\documentclass[12pt]{article}
\usepackage{epsfig,cite}
\usepackage {amsmath}
\usepackage{amssymb}
\include{epsf}

\newenvironment{wide}[2]{%
\begin{list}{}{%
\setlength{\topsep}{0pt}%
\setlength{\leftmargin}{#1}%
\setlength{\rightmargin}{#2}%
\setlength{\listparindent}{\parindent}%
\setlength{\itemindent}{\parindent}%
\setlength{\parsep}{\parskip}}%
\item[]}{\end{list}}

\newcount\timecount
\newcount\hours \newcount\minutes  \newcount\temp \newcount\pmhours
\hours = \time
\divide\hours by 60
\temp = \hours
\multiply\temp by 60
\minutes = \time
\advance\minutes by -\temp
\def\hour{\the\hours}
\def\minute{\ifnum\minutes<10 0\the\minutes
            \else\the\minutes\fi}
\def\clock{
\ifnum\hours=0 12:\minute\ AM
\else\ifnum\hours<12 \hour:\minute\ AM
      \else\ifnum\hours=12 12:\minute\ PM
            \else\ifnum\hours>12
                 \pmhours=\hours
                 \advance\pmhours by -12
                 \the\pmhours:\minute\ PM
                 \fi
            \fi
      \fi
\fi
}

\def\monthname{\relax\ifcase\month 0/\or January\or February\or
   March\or April\or May\or June\or July\or August\or September\or
   October\or November\or December\else\number\month/\fi}

\def\bold#1{\setbox0=\hbox{$#1$}%
     \kern-.025em\copy0\kern-\wd0
     \kern.05em\copy0\kern-\wd0
     \kern-.025em\raise.0433em\box0 }

\def\beq{\begin{equation}}
\def\eeq{\end{equation}}

\def\ss{\scriptscriptstyle}
\def\ga{\mathrel{\raise.3ex\hbox{$>$\kern-.75em\lower1ex\hbox{$\sim$}}}}
\def\la{\mathrel{\raise.3ex\hbox{$<$\kern-.75em\lower1ex\hbox{$\sim$}}}}
\def\gev{{\rm \, Ge\kern-0.125em V}}
\def\tev{{\rm \, Te\kern-0.125em V}}
\def\gyr{{\rm \, G\kern-0.125em yr}}

\def\nl{\hfill\nonumber\\&&}

\def\ttbt{\tan^2 \beta}

\def\gappeq{\mathrel{\rlap {\raise.5ex\hbox{$>$}}
{\lower.5ex\hbox{$\sim$}}}}
\def\lappeq{\mathrel{\rlap{\raise.5ex\hbox{$<$}}
{\lower.5ex\hbox{$\sim$}}}}
\def\Toprel#1\over#2{\mathrel{\mathop{#2}\limits^{#1}}}

\def\stau{\widetilde \tau}

\def\m12{m_{1\!/2}}

\def\mz{m_{\ss Z}}

\def\PL{{Phys.~Lett.} }
\def\PR{{Phys.~Rev.} }
\def\PRL{{Phys.~Rev.~Lett.} }

\def\stau{\tilde{\tau}}

\def\tanb{\tan \beta}

\def\bea{\begin{eqnarray}}
\def\eea{\end{eqnarray}}
\newcommand{\goto}{\rightarrow}
\newcommand{\bmm}{B_s \goto \mu^+ \, \mu^-}

\newcommand{\rmd}{\ensuremath{\mathrm{d}}}
\newcommand{\rmu}{\ensuremath{\mathrm{u}}}
\newcommand{\rms}{\ensuremath{\mathrm{s}}}
\newcommand{\rmc}{\ensuremath{\mathrm{c}}}
\newcommand{\rmb}{\ensuremath{\mathrm{b}}}
\newcommand{\rmt}{\ensuremath{\mathrm{t}}}
\newcommand{\rmpp}{\ensuremath{\mathrm{p}}}
\newcommand{\rmnn}{\ensuremath{\mathrm{n}}}

\newcommand{\Real}{\mathop{\textrm{Re}}}
\newcommand{\diag}{\mathop{\textrm{diag}}}
\newcommand{\ifmulticol}[2]{%
  \ifthenelse{\lengthtest{1.9\columnwidth<\textwidth}}{#1}{#2}%
}
\newcommand{\fTq}[1]{\ensuremath{f_{T_{#1}}}}
\newcommand{\fNTq}[1]{\ensuremath{f_{T_{#1}}^{(N)}}}

\newcommand{\Bq}[1]{\ensuremath{B_{#1}}}
\newcommand{\BNq}[1]{\ensuremath{B_{#1}^{(N)}}}
\newcommand{\Bpq}[1]{\ensuremath{B_{#1}^{(\rmpp)}}}
\newcommand{\Bnq}[1]{\ensuremath{B_{#1}^{(\rmnn)}}}

\newcommand{\DeltaNq}[1]{\ensuremath{\Delta_{#1}^{(N)}}}
\newcommand{\Deltapq}[1]{\ensuremath{\Delta_{#1}^{(\rmpp)}}}
\newcommand{\Deltanq}[1]{\ensuremath{\Delta_{#1}^{(\rmnn)}}}

\newcommand{\mup}{\ensuremath{m_{\rm u}}}

\newcommand{\md}{\ensuremath{m_{\rm d}}}
\newcommand{\ms}{\ensuremath{m_{\rm s}}}

\newcommand{\SigmapiN}{\ensuremath{\Sigma_{\pi\!{\scriptscriptstyle N}}}}

\begin{document}
\begin{titlepage}
\pagestyle{empty}
\baselineskip=21pt
\rightline{CERN-PH-TH/2009-053, UMN--TH--2746/09, FTPI--MINN--09/18, UTTG-03-09, TCC-017-09}
\vskip 0.2in
\begin{center}
{\large{\bf Update on the Direct Detection of Dark Matter in MSSM Models with Non-Universal Higgs Masses}}
\end{center}
\begin{center}
\vskip 0.2in
{\bf John~Ellis}$^1$, {\bf Keith~A.~Olive}$^{2}$ and
{\bf Pearl Sandick}$^{3}$
\vskip 0.1in

{\it
$^1${TH Division, PH Department, CERN, CH-1211 Geneva 23, Switzerland}\\
$^2${William I. Fine Theoretical Physics Institute, \\
University of Minnesota, Minneapolis, MN 55455, USA}\\
$^3${Theory Group and Texas Cosmology Center,\\
The University of Texas at Austin, TX 78712, USA}\\
}

{\bf Abstract}
\end{center}
\baselineskip=18pt \noindent

We discuss the possibilities for the direct detection of neutralino dark matter via elastic
scattering in variants of the minimal supersymmetric extension of the Standard Model
(MSSM) with non-universal supersymmetry-breaking contributions to the Higgs masses,
which may be either equal (NUHM1) or independent (NUHM2). We compare the
ranges found in the NUHM1 and NUHM2 with that found in the MSSM with universal
supersymme- try-breaking contributions to all scalar masses, the CMSSM. We find that
both the NUHM1 and NUHM2 offer the possibility of larger 
spin-independent dark matter scattering cross
sections than in the CMSSM for larger neutralino masses, since they allow the density
of heavier neutralinos with large Higgsino components to fall within the allowed range by
astrophysics. The NUHM1 and NUHM2 also offer more possibilities than the CMSSM for
small cross sections for lower neutralino masses, since they may be suppressed by
scalar and pseudoscalar Higgs masses that are larger than in the CMSSM.

\vfill
\leftline{CERN-PH-TH/2009-053}
\leftline{April 2009}
\end{titlepage}

\section{Introduction}

Supersymmetry is one example of a theory that suggests the possibility of
observing new particles at high-energy particle colliders such as the LHC.
Many supersymmetric models predict that the lightest supersymmetric
particle (LSP) should be stable and hence present in the Universe as a
relic from the Big Bang \cite{EHNOS}. As such, the LSP may provide all or some of the
cold dark matter postulated by astrophysicists and cosmologists. In many
supersymmetric models, the LSP is the lightest neutralino, $\chi$, though
there are other possibilities such as the gravitino.

The most convincing vindication of any model of dark matter would be the direct detection of
astrophysical dark matter particles via their scattering on nuclei in low-background
experiments underground \cite{Goodman:1984dc}. 
The model could then be further verified by comparing
the direct dark matter detection rate with theoretical calculations based on measurements of
the model parameters at colliders. If the LSP is indeed the lightest neutralino $\chi$,
many estimates suggest that supersymmetric dark matter could be detected directly in
present or forthcoming experiments \cite{etal,dd1,EFlO1,EFlOSo,EFFMO,otherMSSMDM,dd2,bot1,bot2,both,eoss8,eosv}. 
On the other hand, if the LSP were the gravitino, the
cross section for its scattering would be so small as to be undetectable.

Within the neutralino LSP scenario, the absence to date of supersymmetric signals
in accelerator experiments imposes constraints on the possible direct detection
rates. In general, these depend on the specific supersymmetric model, and
specifically on the way in which supersymmetry is broken. We consider here the
minimal supersymmetric extension of the Standard Model (MSSM) \cite{martin}, 
and focus on
three possibilities for the pattern of supersymmetry breaking.

In the simplest model the supersymmetry-breaking scalar masses $m_0$ are
constrained to be universal at some input GUT scale, a scenario called the 
CMSSM \cite{cmssm,cmssmwmap}. The absence
of flavour-changing interactions beyond the Standard Model motivates the
assumption that squarks or sleptons with the same quantum numbers should be
(very nearly) degenerate \cite{flavor}, and universality between (some) squarks with
different different quantum numbers and sleptons is suggested by some GUT
models. These also suggest that the SU(3), SU(2) and U(1) gaugino masses
$m_{1/2}$ should be degenerate. The direct detection of neutralino dark
matter within this CMSSM framework has been studied in many papers 
\cite{dd1,EFlO1,EFFMO,dd2,bot1,bot2,both,eoss8,eosv}.

However, there are no good reasons 
to expect universality for the supersymmetry-breaking contributions to the scalar
masses of the MSSM Higgs multiplets. Accordingly, our main focus here is on variants
of the MSSM with non-universal Higgs masses \cite{nonu}, generically called 
NUHM models.
The  supersymmetry-breaking contributions to the two Higgs masses might
either be identical, a scenario called the NUHM1 
\cite{nuhm1,EOS08}, or might be different, a
scenario we call the NUHM2 \cite{nuhm,ehow}. The direct detection of neutralino dark
matter within the NUHM2 has also been studied in several papers \cite{EFlOSo,eoss8,nuhm1}, 
but direct detection within the NUHM1 has been less studied \cite{nuhm1}.

Since the amplitude for spin-independent dark matter scattering on a heavy nucleus
receives important contributions from the exchanges of Higgs bosons, and
since these are less constrained in NUHM models than in the CMSSM, {\it a priori} 
one would expect that the direct dark matter scattering cross section should
exhibit more variability than in the CMSSM. In particular, whereas the present
experimental upper limits on dark matter scattering barely touch the range
expected within the CMSSM, ongoing direct dark matter searches might already
be sensitive to significant samples of NUHM models. Conversely, whereas the
planned direct dark matter searches should be able to explore much of the CMSSM
parameter space, it might be more likely that NUHM models could escape
detection.

In this paper, we first review in Section~2 the relationships between the
CMSSM, NUHM1 and NUHM2 models, as well as basic formulae underlying the
calculation of the spin-independent and -dependent dark matter scattering cross sections.
We then review briefly in Section~3 the expectations for direct dark
matter detection within the CMSSM. Subsequently, we explore in
Sections~4 and 5 the possible scattering cross sections
in the NUHM1 and the NUHM2, respectively. These explorations 
each proceed in two steps: surveys of the variation in the direct dark matter
cross section across some representative planes for fixed values of the
other parameters, followed by a presentation and discussion of the overall
range of possible values throughout the corresponding parameter space.
We concentrate on spin-independent dark matter scattering because this seems 
closer to the present and likely experimental sensitivity than does
spin-dependent scattering. Our conclusions are summarized in Section~6,
where we comment on the possibilities for distinguishing experimentally
between the CMSSM, NUHM1 and NUHM2.

\section{Theoretical Framework}

\subsection{From the CMSSM to the NUHM1 and the NUHM2}

The CMSSM has four continuous parameters and a sign that 
determine the weak-scale observables: the supposedly
universal scalar mass $m_0$, gaugino mass $m_{1/2}$ and
universal trilinear coupling $A_0$; the ratio of the Higgs vacuum
expectation values, $\tanb$, and the sign of the Higgs mixing parameter
$\mu$. The parameters $m_0$, $m_{1/2}$ and $A_0$ 
and the sign of $\mu$ are assumed to be
specified at the GUT scale.  
A useful starting-point for analyzing the
CMSSM is the $(m_{1/2}, m_0)$ plane for some fixed value of $A_0$,
$\tan \beta$ and sign of $\mu$. 
The input supersymmetry-breaking
contributions to the effective Higgs masses-squared, 
$m_{1,2}^2$ for the Higgs fields coupled to the down- and up-type
quarks respectively, are renormalized differently below the GUT
scale, because they are sensitive to different Yukawa couplings, specifically
the $t$-quark coupling in the case of $H_2$, and this renormalization
permits electroweak symmetry breaking at low
energies. In the CMSSM, $|\mu|$ and $m_A$ are calculated from the electroweak vacuum
conditions:
\beq
m_A^2(Q) = m_1^2(Q)+m_2^2(Q)+2\mu^2(Q)+\Delta_A(Q)
\label{eq:mA}
\eeq  
and
\beq
\mu^2=\frac{m_1^2-m_2^2\tan^2\beta+\frac{1}{2}m_Z^2(1-\tan^2\beta)+\Delta_{\mu}^{(1)}}{\tan^2\beta-1+\Delta_{\mu}^{(2)}},
\label{eq:mu}
\eeq
where $\Delta_A$ and $\Delta_\mu^{(1,2)}$ are loop
corrections~\cite{Barger:1993gh,deBoer:1994he,Carena:2001fw},
$Q=(m_{\stau_R}m_{\stau_L})^{1/2}$, and all quantities in
(\ref{eq:mu}) are defined at the electroweak scale, $m_Z$. Unless otherwise noted,
$m_A \equiv m_A(Q)$ and $\mu \equiv \mu(m_Z)$.  The values of the parameters in (\ref{eq:mA})
and (\ref{eq:mu}) are related via
\begin{eqnarray}
m_1^2(Q)=m_1^2+c_1, \nonumber\\
m_2^2(Q)=m_2^2+c_2, \label{eq:radcorr}\\
\mu^2(Q)=\mu^2+c_{\mu}, \nonumber
\end{eqnarray}
where $c_1$, $c_2$ and $c_{\mu}$ are well-known radiative 
corrections~\cite{Barger:1993gh,IL,Martin:1993zk}.

The restriction $m_1(M_{GUT}) = m_2(M_{GUT})$ is retained
in the NUHM1, but these may differ from the otherwise
universal scalar mass, $m_0$. The common GUT-scale value of the
Higgs masses-squared therefore introduces an additional parameter
in the NUHM1.  This can be mapped into an additional free parameter in
the effective low-energy theory, that may be taken to be either
$\mu$ or $m_A$. As in the CMSSM, one uses (\ref{eq:mA}) - (\ref{eq:radcorr}) 
to calculate the relationship between $m_1^2$ and $m_2^2$ at
the weak scale that is required to respect the electroweak boundary conditions,
allowing for the weakened universality condition at $M_{GUT}$.  
If $m_A$ is taken to be the additional free low-energy parameter, then at
$m_Z$ we have \cite{EOS08}
\begin{eqnarray}
m_1^2(\tan^2\beta+1+\Delta_{\mu}^{(2)})
=m_2^2(\tan^2\beta+1-\Delta_{\mu}^{(2)})+m_Z^2(\tan^2\beta-1)-2\Delta_{\mu}^{(1)}\nonumber\\
+\big(m_A^2-(\Delta_A(Q)+c_1+c_2+2c_{\mu})\big)
(\tan^2\beta-1+\Delta_{\mu}^{(2)}).
\end{eqnarray}
Alternatively, if $\mu$ is taken as the free parameter, then at $m_Z$
we have
\beq
m_1^2=m_2^2\tan^2\beta+\mu^2(\tan^2\beta-1+\Delta_{\mu}^{(2)})+\frac{1}{2}m_Z^2(\tan^2\beta-1)-\Delta_{\mu}^{(1)}.
\eeq
In each case, the NUHM1 boundary condition at $M_{GUT}$ is $m_1^2 =
m_2^2$.  Clearly, for some specific input values of $\mu$ and $m_A$, one
finds $m_1^2(M_{GUT}) = m_2^2(M_{GUT}) = m_0^2$, thereby recovering the
CMSSM. The subject of Section~4 is the deviation from the CMSSM
prediction for dark matter scattering that may be found in the NUHM1.
We discuss first planar subspaces of the NUHM1 in which $m_A$ is
taken as the additional free low-energy parameter, and then consider
planes with $\mu$ as a free parameter. 

In the NUHM2, the soft supersymmetry-breaking
contributions to both the Higgs scalar masses $m_{1,2}$ are regarded as free parameters.
Correspondingly, these may be traded for free values of both $\mu$ and $m_A$
at low energies. In this case, we can write \cite{nuhm}
\begin{eqnarray}
m_1^2(1+ \tan^2 \beta) &=& m_A^2(Q) \tan^2 \beta - \mu^2 (\tan^2 \beta + 1 -
\Delta_\mu^{(2)} ) 
- (c_1 + c_2 + 2 c_\mu) \ttbt \nl - \Delta_A(Q) \ttbt 
- \frac{1}{2} \mz^2 (1 - \ttbt) - \Delta_\mu^{(1)} 
\label{m1}
\end{eqnarray}
and 
\begin{eqnarray}
m_2^2(1+ \tan^2 \beta) &=& m_A^2(Q) - \mu^2 (\tan^2 \beta + 1 +
\Delta_\mu^{(2)} )
- (c_1 + c_2 + 2 c_\mu) \nl
- \Delta_A(Q) + \frac{1}{2} \mz^2 (1 - \ttbt) + \Delta_\mu^{(1)}.
\label{m2}
\end{eqnarray}
 Each point in a CMSSM
$(m_{1/2}, m_0)$ plane can be `blown up' into a $(\mu, m_A)$ plane.
The NUHM1 subspace may be represented as a line in such a $(\mu, m_A)$ plane, 
and the CMSSM as one or two points on this line.

\subsection{Dark Matter Scattering}

The neutralino LSP $\chi$ is the
lowest-mass eigenstate combination of the Bino ${\tilde B}$, Wino 
$\tilde W$ and neutral Higgsinos ${\tilde H}_{1,2}$. Its mass matrix $N$ is
diagonalized by a matrix $Z$: $\diag(m_{\chi_{1,..,4}}) = Z^* N Z^{-1}$, and
the composition of the lightest neutralino may be written as
\begin{equation} \label{eqn:chi}
  \chi = Z_{\chi 1}\tilde{B} + Z_{\chi 2}\tilde{W} +
         Z_{\chi 3}\tilde{H_{1}} + Z_{\chi 4}\tilde{H_{2}}.
\end{equation}
The amplitude for $\chi$ scattering on quarks depends on the squark
mass-squared matrix, which is diagonalized by a matrix $\eta$:
$diag(m^2_1, m^2_2) \equiv \eta M^2 \eta^{-1}$, which can be
parameterized for each flavour $f$ by an angle $\theta_{f}$~\footnote{We ignore
here all possible CP-violating phases.}.  The diagonalization matrix can
be written as
\begin{equation} \label{eqn:etamatrix}
  \left( \begin{array}{cc}
    \cos{\theta_{f}} & \sin{\theta_{f}} e^{i \gamma_{f}} \\
    -\sin{\theta_{f}} e^{-i \gamma_{f}} & \cos{\theta_{f}}
  \end{array} \right) 
  \hspace{0.5cm}
  \equiv 
  \hspace{0.5cm}
  \left( \begin{array}{cc}
    \eta_{11} & \eta_{12} \\
    \eta_{21} & \eta_{22}
  \end{array} \right).
\end{equation}
The magnitudes of $\mu$ and the pseudoscalar Higgs mass $m_A$ are
calculated using the appropriate electroweak vacuum conditions in the CMSSM,
NUHM1 and NUHM2, as discussed in the previous Subsection.

The effective four-fermion Lagrangian relevant for relic dark matter
scattering is~\cite{Falk:1998xj}:
\begin{equation} \label{eqn:lagrangian}
  {\cal L}
    = \alpha_{2i} \bar{\chi} \gamma^\mu \gamma^5 \chi \bar{q_{i}} 
    \gamma_{\mu} \gamma^{5}  q_{i}
+ \alpha_{3i} \bar{\chi} \chi \bar{q_{i}} q_{i}.
\end{equation}
This Lagrangian is to be summed over the quark generations, and the 
subscript $i$ labels up-type quarks ($i=1$) and down-type quarks
($i=2$).   The coefficients are given by:
\begin{eqnarray} \label{eqn:alpha2}
  \alpha_{2i}
    &=& \frac{1}{4(m^{2}_{1i} - m^{2}_{\chi})}
        \left[ \left| X_{i} \right|^{2} + \left| Y_{i} \right|^{2} \right] 
        {}+ \frac{1}{4(m^{2}_{2i} - m^{2}_{\chi})}
            \left[ \left| W_{i} \right|^{2}
                   + \left| V_{i} \right|^{2} \right]
        \nonumber \\
    & & {}- \frac{g^{2}}{4 m_{Z}^{2} \cos^{2}{\theta_{W}}}
            \left[ \left| Z_{\chi_{3}} \right|^{2}
                   - \left| Z_{\chi_{4}} \right|^{2} \right]
            \frac{T_{3i}}{2}
\end{eqnarray}
and
\begin{eqnarray} \label{eqn:alpha3}
  \alpha_{3i}
    &=& {}- \frac{1}{2(m^{2}_{1i} - m^{2}_{\chi})}
            \Real \left[ \left( X_{i} \right)
                         \left( Y_{i} \right)^{\ast} \right] 
        {}- \frac{1}{2(m^{2}_{2i} - m^{2}_{\chi})}
            \Real \left[ \left( W_{i} \right)
                         \left( V_{i} \right)^{\ast} \right]
        \nonumber \\
    & & {}- \frac{g m_{q_{i}}}{4 m_{W} B_{i}}
        \bigg\{
          \left( \frac{D_{i}^{2}}{m^{2}_{H_{2}}}
                 + \frac{C_{i}^{2}}{m^{2}_{H_{1}}}
          \right)
          \Real \left[ \delta_{2i}
                  \left( g Z_{\chi 2} - g' Z_{\chi 1}\right)
                \right] \nonumber \\
         & & {}+ D_{i} C_{i} \left( \frac{1}{m^{2}_{H_{2}}}
                                 - \frac{1}{m^{2}_{H_{1}}} \right)
              \Real \left[ \delta_{1i}
                      \left( g Z_{\chi 2} - g' Z_{\chi 1} \right)
                    \right]
        \bigg\},
\end{eqnarray}
where
\begin{eqnarray} \label{eqn:XYWV}
  X_{i} &\equiv&
    \eta^{\ast}_{11} \frac{g m_{q_{i}}Z_{\chi 5-i}^{\ast}}{2 m_{W} B_{i}}
    - \eta_{12}^{\ast} e_{i} g' Z_{\chi 1}^{\ast} ,
    \nonumber \\
  Y_{i} &\equiv&
    \eta^{\ast}_{11} \left( \frac{y_{i}}{2} g' Z_{\chi 1}
                            + g T_{3i} Z_{\chi 2} \right)
    + \eta^{\ast}_{12} \frac{g m_{q_{i}} Z_{\chi 5-i}}{2 m_{W} B_{i}} ,
    \nonumber \\
  W_{i} &\equiv&
    \eta_{21}^{\ast} \frac{g m_{q_{i}}Z_{\chi 5-i}^{\ast}}{2 m_{W} B_{i}}
    - \eta_{22}^{\ast} e_{i} g' Z_{\chi 1}^{\ast} ,
    \nonumber \\
  V_{i} &\equiv&
    \eta_{21}^{\ast}\left( \frac{y_{i}}{2} g' Z_{\chi 1}
                           + g T_{3i} Z_{\chi 2} \right)
    + \eta_{22}^{\ast} \frac{g m_{q_{i}} Z_{\chi 5-i}}{2 m_{W} B_{i}},
\end{eqnarray}
where $y_i, T_{3i}$ denote hypercharge and isospin, and
\begin{eqnarray} \label{eqn:deltas}
  \delta_{1i} = Z_{\chi 3} (Z_{\chi 4}),
  \qquad
  \delta_{2i} = Z_{\chi 4} (-Z_{\chi 3}),
\end{eqnarray}
\begin{equation} \label{eqn:CD}
  B_{i}             =     \sin{\beta} \, (\cos{\beta}), \;
  C_{i} 
  {=} \sin{\alpha} (\cos{\alpha}), \;
  D_{i}              =      \cos{\alpha} \, (-\sin{\alpha}) 
\end{equation}
for up (down) type quarks.  We denote by $m_{H_2} < m_{H_1}$ the masses
of the two neutral scalar Higgs bosons, and $\alpha$ denotes the
neutral Higgs boson mixing angle.

In the NUHM1 and NUHM2, the greater freedom in the choice(s) of the
input values of $m_1^2$ and $m_2^2$ translates into greater freedom
for $m_{H_2}$ and $m_{H_1}$ than in the CMSSM and hence, {\it a priori}, 
more variation in the dark matter scattering amplitude.

The elastic cross section for $\chi$ scattering on a nucleus can be
decomposed into a scalar (spin-independent) part obtained from
$\alpha_{3i}$ (\ref{eqn:alpha3}), and a spin-dependent
part obtained from $\alpha_{2i}$ (\ref{eqn:alpha2}).  Each of these can be
written in terms of the cross sections for elastic scattering off individual nucleons, as we now review.

The scalar, or spin-independent (SI), part of the cross section can be
written as~\footnote{This expression is valid in the
         zero-momentum-transfer limit.  For non-zero momentum exchange,
         the expression must include a form factor due to the finite
         size of the nucleus.  See, for example, Ref.~\cite{SmithLewin}.}%
\begin{equation} \label{eqn:sigmaSI}
  \sigma_{\rm SI} = \frac{4 m_{r}^{2}}{\pi}
                    \left[ Z f_{p} + (A-Z) f_{n}  \right]^{2},
\end{equation}
where $m_r$ is the $\chi$-nuclear reduced mass and
\begin{equation} \label{eqn:fN}
  \frac{f_N}{m_N}
    = \sum_{q=\rmu,\rmd,\rms} \fNTq{q} \frac{\alpha_{3q}}{m_{q}}
      + \frac{2}{27} f_{TG}^{(N)}
        \sum_{q=\rmc,\rmb,\rmt} \frac{\alpha_{3q}}{m_q}
\end{equation}
for $N$ = p or n.  The parameters $\fNTq{q}$ are defined by
\begin{equation} \label{eqn:Bq}
  m_N \fNTq{q}
  \equiv \langle N | m_{q} \bar{q} q | N \rangle
  \equiv m_q \BNq{q} ,
\end{equation}
where~\cite{Shifman:1978zn,Vainshtein:1980ea}
\begin{equation} \label{eqn:fTG}
  f_{TG}^{(N)} = 1 - \sum_{q=\rmu,\rmd,\rms} \fNTq{q} .
\end{equation}
We take the central ratios of the light quark masses from
\cite{Leutwyler:1996qg}:
\begin{equation} \label{eqn:mqmd}
  \frac{\mup}{\md} = 0.553, \qquad
  \frac{\ms}{\md}  = 18.9.
\end{equation}
We take the other quark masses from~\cite{rpp}, except for the
top mass, which is taken from the combined CDF and D0 result
\cite{mtop}.

Following~\cite{Cheng:1988im}, we introduce the quantity:
\begin{equation} \label{eqn:z}
  z \equiv \frac{\Bpq{\rmu} - \Bpq{\rms}}{\Bpq{\rmd} - \Bpq{\rms}}
    = 1.49 ,
\end{equation}
which has an experimental error that is negligible compared with others
in this calculation, and the strange scalar density
\begin{equation} \label{eqn:y}
  y \equiv \frac{2 \BNq{\rms}}{\BNq{\rmu} + \BNq{\rmd}}.
\end{equation}
In terms of these, one may write
\begin{equation} \label{eqn:BdBu}
  \frac{\Bpq{\rmd}}{\Bpq{\rmu}}
   = \frac{2 + ((z - 1) \times y)}{2 \times z - ((z - 1) \times y)} \; .
\end{equation}
Proton and neutron scalar matrix elements are related by an interchange
of $B_{\rmu}$ and $B_{\rmd}$, i.e.,
\begin{equation} \label{eqn:Bn}
  \Bnq{\rmu} = \Bpq{\rmd} , \quad
  \Bnq{\rmd} = \Bpq{\rmu} , \quad \text{and} \quad
  \Bnq{\rms} = \Bpq{\rms} .
\end{equation}
The $\pi$-nucleon sigma term, $\SigmapiN$, may be written as
\begin{equation} \label{eqn:SigmapiN}
  \SigmapiN \equiv \frac{1}{2} (\mup + \md)
                   \times \left( \BNq{\rmu} + \BNq{\rmd} \right) \; ,
\end{equation}
and the coefficients $\fTq{q}$ may be written in the forms \cite{eosv}:
\begin{alignat}{3}
  \label{eqn:fNTu}
  \fTq{\rmu}
    & \ = \ & \frac{\mup \Bq{\rmu}}{m_N}
    & \ = \ & \frac{2 \SigmapiN}{m_N (1+\frac{\md}{\mup})
                                 (1+\frac{\Bq{\rmd}}{\Bq{\rmu}})}
    \; , \\
  \label{eqn:fNTd}
  \fTq{\rmd}
    & \ = \ & \frac{\md \Bq{\rmd}}{m_N}
    & \ = \ & \frac{2 \SigmapiN}{m_N (1+\frac{\mup}{\md})
                                 (1+\frac{\Bq{\rmu}}{\Bq{\rmd}})}
    \; , \\
  \label{eqn:fNTs}
  \fTq{\rms}
    & \ = \ & \frac{\ms \Bq{\rms}}{m_N}
    & \ = \ & \frac{(\frac{\ms}{\md}) \SigmapiN \, y}%
                   {m_N (1+\frac{\mup}{\md})}
    \; ; \quad\quad\quad
\end{alignat}
where we have dropped the $(N)$ superscript from $\fTq{q}$ and
$\Bq{q}$.

The value of $y$ is related to the $\pi$-nucleon sigma term $\SigmapiN$ by
\begin{equation} \label{eqn:y2}
  y = 1 - \sigma_0/\SigmapiN \; .
\end{equation}
The central value for $\sigma_0$  is estimated on the
basis of octet baryon mass differences to be 
$\sigma_0 = 36$~MeV~\cite{Borasoy:1996bx,Gasser:1990ce,Knecht:1999dp,Sainio:2001bq},
and the latest determination of $\SigmapiN = 64$~MeV.
These are the values assumed in our analyses of the CMSSM, NUHM1 and NUHM2:
the effect of varying these assumptions are discussed in the context of the CMSSM
in~\cite{eoss8,eosv}. {\it We take this opportunity to reiterate the importance of
measuring $\SigmapiN$ as accurately as possible~\cite{eosv}~\footnote{Lattice calculations are now reaching the stage where they may also provide useful information on $\Sigma_{\pi N}$~\cite{0901.3310}.}.}

The spin-dependent (SD) part of the elastic $\chi$-nucleus cross
section can be written in the zero momentum transfer limit as
\begin{equation} \label{eqn:sigmaSD}
  \sigma_{\rm SD} = \frac{32}{\pi} G_{F}^{2} m_{r}^{2}
                    \Lambda^{2} J(J + 1) \; ,
\end{equation}
where $m_{r}$ is again the reduced neutralino mass, $J$ is the spin 
of the nucleus, and
\begin{equation} \label{eqn:Lambda}
  \Lambda \equiv \frac{1}{J} \left(
                 a_{\rmpp} \langle S_{\rmpp} \rangle
                 + a_{\rmnn} \langle S_{\rmnn} \rangle
                 \right) \; ,
\end{equation}
where
\begin{equation} \label{eqn:aN}
  a_{\rmpp} \equiv \sum_{q} \frac{\alpha_{2q}}{\sqrt{2} G_{f}}
              \Deltapq{q} , \qquad
  a_{\rmnn} \equiv \sum_{i} \frac{\alpha_{2q}}{\sqrt{2} G_{f}}
              \Deltanq{q} \; .
\end{equation}
The factors $\DeltaNq{q}$ parametrize the quark spin content of the
nucleon and are only significant for the light (u,d,s) quarks.
For definiteness, we assume
\begin{equation}
  \Deltapq{\rmu}
    = 0.84
    \; ,
  \Deltapq{\rmd}
    = -0.43
    \; ,
  \Deltapq{\rms} = -0.09.
  \label{deltas}
  \end{equation}
The effects of varying these assumptions are also discussed in the context of the CMSSM
in~\cite{eosv}.

\section{CMSSM Models}

\begin{figure}[p]
\begin{wide}{-1in}{-1in}
\begin{center}
\vskip -1.4in
\hskip .6in
\resizebox{0.55\textwidth}{!}{\includegraphics{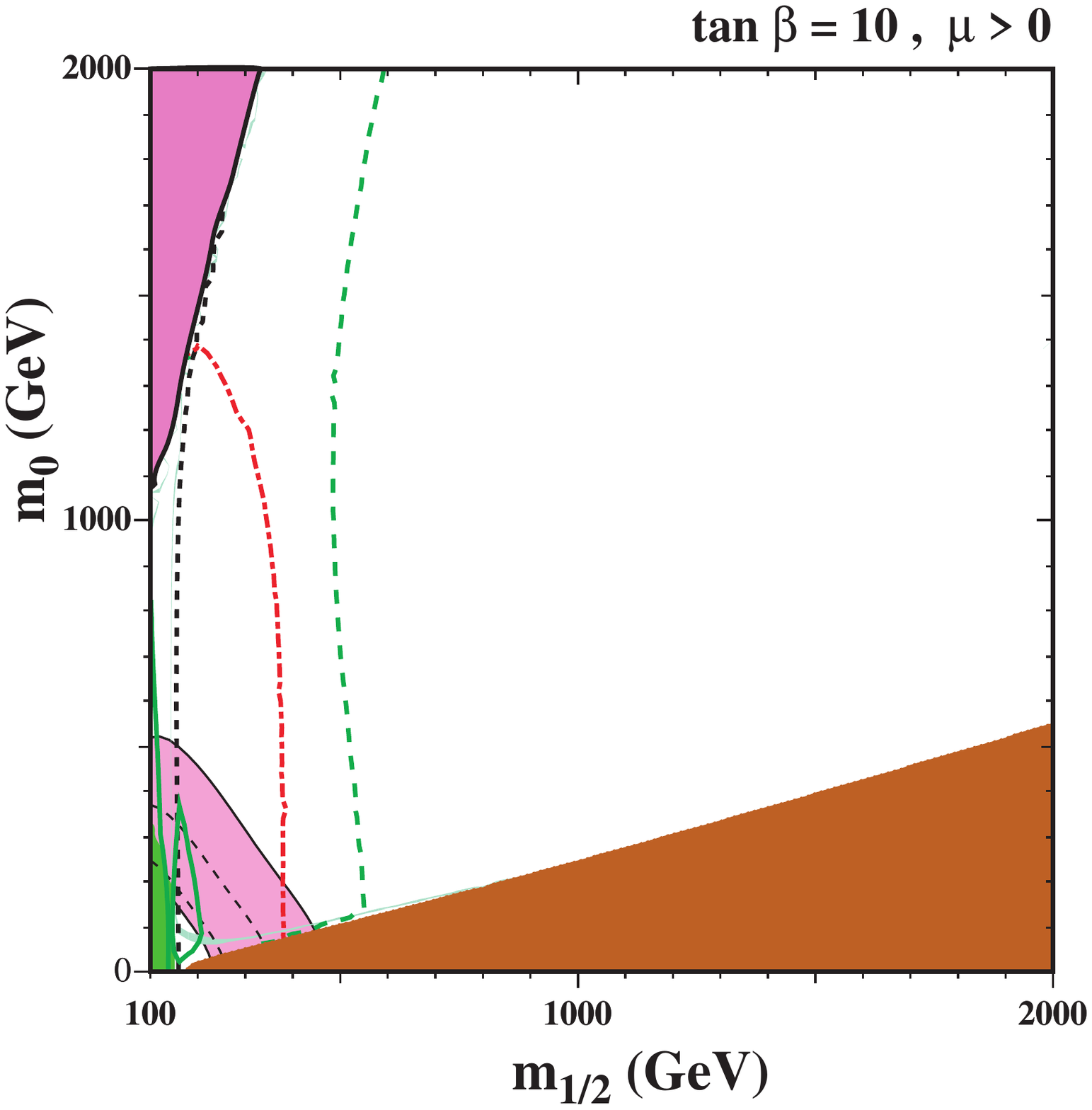}}
\hskip -.6in
\resizebox{0.55\textwidth}{!}{\includegraphics{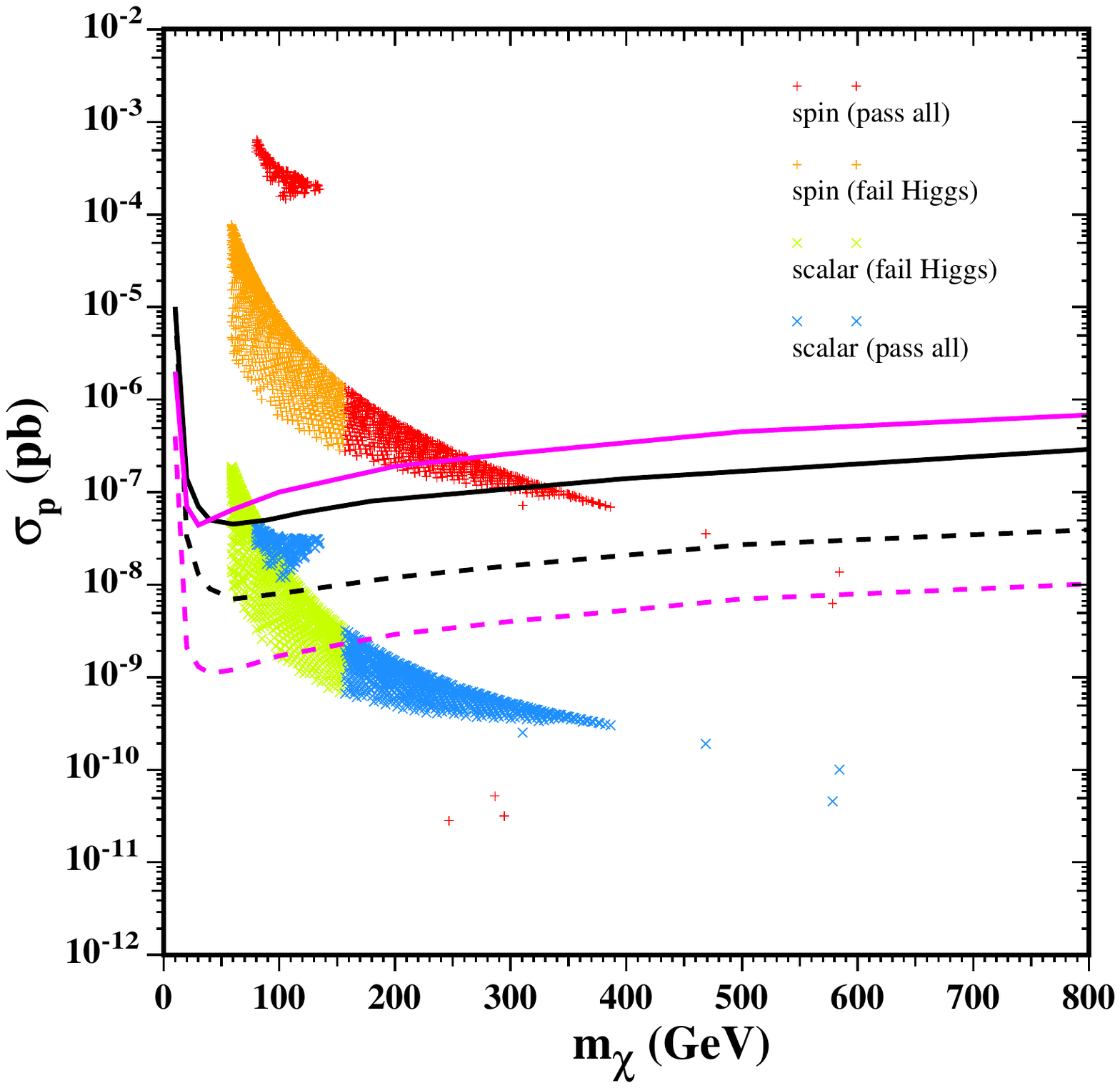}}
\vskip -1.8in
\hskip .6in
\resizebox{0.55\textwidth}{!}{\includegraphics{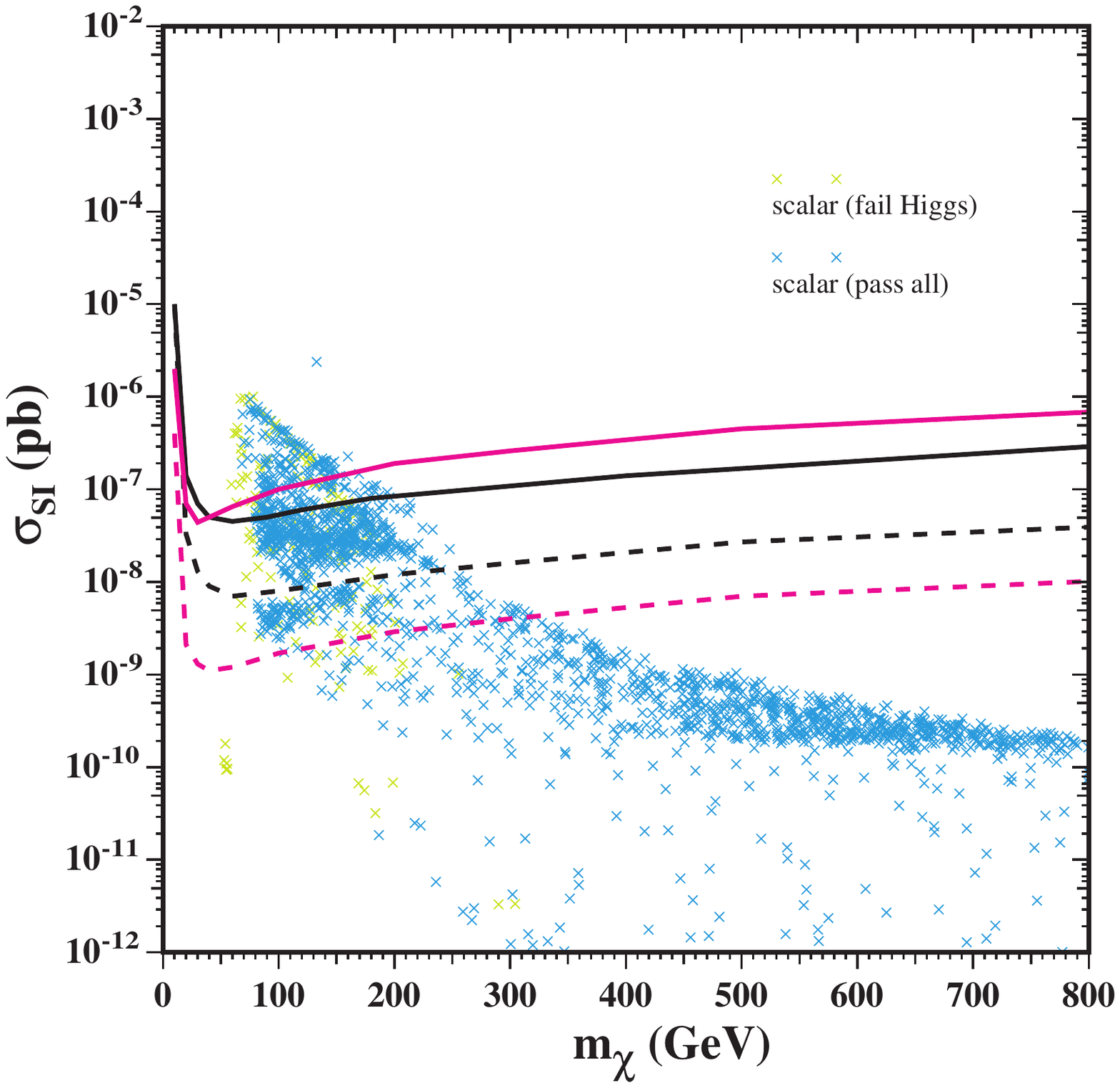}}
\hskip -.6in
\resizebox{0.55\textwidth}{!}{\includegraphics{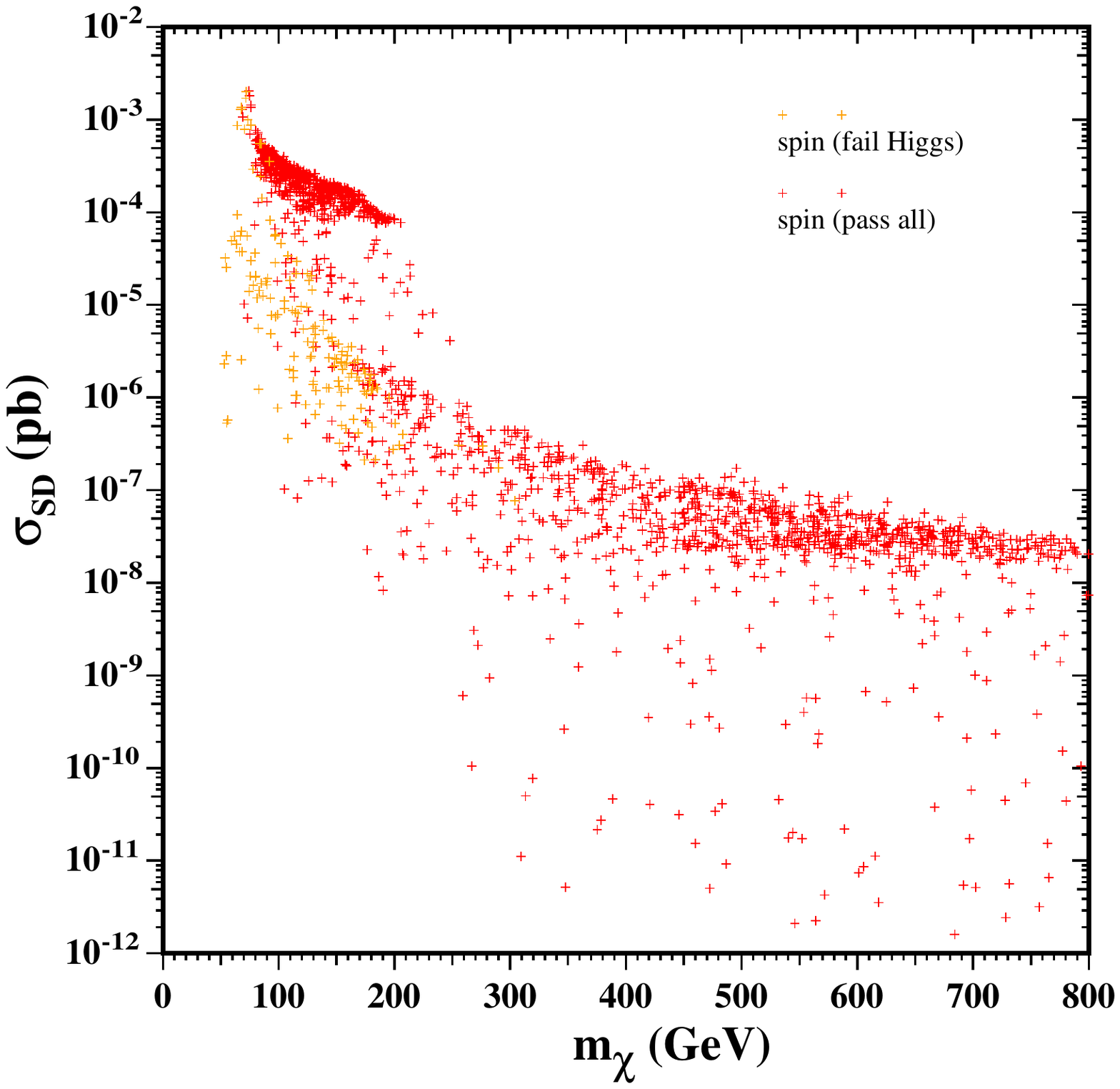}}
\end{center}
\end{wide}
\caption{\it Panels (a) and (b) show the CMSSM $(m_{1/2},m_0)$ plane and the corresponding cosmologically viable neutralino-nucleon elastic scattering cross sections as functions of neutralino mass for $\tanb=10$ and $A_0 =0$.  Panel (c) shows the entire potential range of neutralino-nucleon cross sections as functions of neutralino mass for the CMSSM, with $5 \leq \tanb \leq 55$, 0 $\leq m_{1/2} \leq 2000$ GeV, 100 GeV $\leq m_0 \leq 2000$ GeV, and $-3 m_{1/2} \leq A_0 \leq 3 m_{1/2}$  We consider $\mu<0$ only for $\tanb<30$. Also shown are upper limits on the spin-independent 
dark matter scattering cross section from CDMS~II~\protect\cite{cdmsII} (solid black line) and
XENON10~\cite{XENON10} (solid pink line), as well as the expected sensitivities  
for XENON100~\cite{XENON100} (dashed pink line) and SuperCDMS at the Soudan 
Mine~\protect\cite{superCDMS} (dashed black line). 
\label{fig:cmssm}}
\end{figure}

We begin with a brief discussion of detection prospects in the CMSSM.  In panel (a) of 
Fig.~\ref{fig:cmssm}, we show the $(m_{1/2},m_0)$ plane for $\tanb=10$, $A_0=0$, 
and $\mu > 0$.  The region excluded because the LSP is a charged $\stau$ is shaded 
brown, and that where electroweak symmetry breaking cannot be obtained, resulting
in unphysical $\mu^2 < 0$, in dark pink. The red dot-dashed contour corresponds to a
Higgs mass of 114 GeV.  At lower $m_{1/2}$ the Higgs boson would be lighter, which is 
excluded by its non-observation at LEP~\cite{LEPsusy}.  We also plot a black dashed 
contour for $m_{\chi^{\pm}}=104$ GeV, the region at lower $m_{1/2}$ also being 
excluded by LEP.  The green shaded region at very low $m_{1/2}$ and $m_0$ is 
disfavored by the measured branching ratio for $b \goto s \gamma$~\cite{bsgex}, 
while the light pink shaded region is favored by the measurement of the muon 
anomalous magnetic moment at the 2-$\sigma$ level~\cite{g-2}. Finally, in the turquoise 
shaded regions, the relic density of neutralinos falls within the WMAP range~\cite{WMAP}.  
For the value $\tanb = 10$ used here, the only cosmologically-preferred regions are the 
coannihilation strip, bordering the $\stau$-LSP region, and the focus-point region at large $m_0$, 
where $\mu$ is small and the LSP is a mixed bino-Higgsino state.  Over the bulk of the 
plane, the relic density of neutralinos exceeds the WMAP range by more than 2 $\sigma$.  
There are, however, slim strips where $\Omega_{\chi} h^2$ is below the WMAP 
range, which lie between the strips of good relic density and the excluded regions they 
border.  These portions of the plane are not forbidden, as there could be some
additional source of cold dark matter.

In panel (b) of Fig.~\ref{fig:cmssm} we show the neutralino-nucleon elastic scattering cross sections as
functions of neutralino mass for the regions of panel (a) that are cosmologically viable 
(i.e., those where the upper limit on the relic density of neutralinos is respected), 
and are not excluded by constraints from colliders.  We also plot the limits on the 
spin-independent cross section from CDMS~II~\cite{cdmsII} (solid black line)
and XENON10~\cite{XENON10} (solid red line), as well as the sensitivities projected  
for XENON100~\cite{XENON100} (or a similar 100-kg liquid noble-gas detector such as LUX, 
dashed red line) and SuperCDMS at the Soudan Mine~\cite{superCDMS} (dashed black line). 
We do not show the experimental bound for the spin-dependent scattering
of a neutralino on a proton as the current best limit is 0.2 pb (at $m_\chi = 100$ GeV) \cite{kims}
and is off the scale of our plot.  The limit for the scattering on neutrons is stronger
but is still only at the level of 0.02 pb \cite{cdmsII}.

When stating limits or projections, direct detection experiments assume as a prior that 
$\Omega_{WIMP} = \Omega_{CDM}$, where $\Omega_{CDM}$ is that measured by WMAP. 
If $\Omega_{WIMP} < \Omega_{CDM}$, we assume that the rest of the cold dark matter
is due to some other source, and we rescale the plotted neutralino-nucleon cross sections 
by a factor $\Omega_{\chi} / \Omega_{CDM}$, so that comparison with direct searches is 
possible. 

In panel (b), we display both the spin-dependent and spin-independent (scalar) 
cross sections for neutralino-nucleon elastic scattering.  In each case, there are two distinct 
regions in the $(M_{\chi},\sigma)$ plane, that arising from the focus-point region
at $m_{\chi} \lesssim 150$ GeV and relatively large $\sigma$, and that from the 
coannihilation strip.  In the coannihilation strip, 50 GeV $< m_{\chi} <$ 400 GeV, 
where the lower limit on $m_{\chi}$ is a result of the LEP constraint on the chargino mass,
and the upper limit on $m_{\chi}$ corresponds to the end-point of the coannihilation strip 
for $\tan \beta = 10$.  In contrast, the end point of the focus-point region shown is due 
only to the cut-off $m_0 < 2$ TeV that we assume.  

In addition, for $m_{1/2} \lesssim 380$~GeV 
in the coannihilation strip ($m_{\chi} \lesssim 160$ GeV), the nominal calculated mass of the lighter
scalar MSSM Higgs boson is less than the LEP lower bound.  These points are indicated
by lighter shadings: lime green for the spin-independent scattering cross sections, 
and ochre for the spin-dependent scattering cross sections~\footnote{Note that in the
focus-point region the calculated Higgs mass always exceeds the LEP lower limit.}. 
In general, such points may 
not be strictly forbidden, as the calculated mass\footnote{We use FeynHiggs \cite{FeynHiggs} 
to calculate the Higgs mass.} has a theoretical uncertainty estimated
as 1.5~GeV, and the MSSM Higgs has slightly different couplings from the SM Higgs for 
which the LEP limit was set. However, in the CMSSM the couplings are generally very 
close and (up to theoretical and experimental uncertainties) the limit should hold in this case. 

At very low $m_{1/2}$, CDMS~II and XENON10 have definitively excluded some of the 
region where $m_h$ is below the LEP limit. We show this explicitly in panel (a), 
by plotting the reach of current and future direct detection experiments in the supersymmetric 
parameter space.  Here and in subsequent parameter space scans, we display contours of the scalar 
neutralino-nucleon cross section, scaled by $\Omega_{\chi} / \Omega_{CDM}$ if necessary, 
of $5 \times 10^{-8}$ pb (solid green lines) and $10^{-9}$ pb (dashed green lines).  
A cross section of $5 \times 10^{-8}$ pb is currently excluded by XENON10 for 
$m_{\chi}=30$ GeV and by CDMS~II for $m_{\chi}=60$ GeV, and will be probed by 
SuperCDMS for $m_{\chi}$ up to $\sim 1000$ GeV. Tonne-scale liquid noble-gas 
detectors such as the proposed XENON1T or a similar detector mass for LUX/ZEP 
will be sensitive to scalar cross sections below $10^{-9}$ pb for all neutralino masses 
in the range 10 GeV $\lesssim m_{\chi} \lesssim$ a few TeV~\cite{XENON100}.  Indeed, they will be sensitive 
to cross sections below $10^{-10}$ pb over much of the preferred mass range
$m_{\chi} \sim O(100)$ GeV.

The choices $\tanb=10$ and $A_0=0$ do not yield viable direct detection 
cross sections that are completely representative of the range of possibilities within the CMSSM.
Therefore, in panels (c) and (d) of Fig.~\ref{fig:cmssm} we show CMSSM spin-independent and 
spin-dependent 
neutralino-nucleon cross sections, respectively, as obtained in a scan over all 
CMSSM parameters with $5 \leq \tanb \leq 55$, 0 $\leq m_{1/2} \leq 2000$ GeV, 
100 GeV $\leq m_0 \leq 2000$ GeV, and $-3 m_{1/2} \leq A_0 \leq 3 m_{1/2}$.  
We also allow both positive and negative $\mu$, except for large $\tanb > 30$, 
where convergence becomes difficult in the $\mu < 0$ case~\footnote{We recall 
that models with $\mu < 0$ are generally disfavoured by $g_\mu - 2$, but we do
not use this as a restriction on our parameter scans.}.  

For future reference, we note the ranges of CMSSM cross 
sections for different given neutralino masses. At low $m_{\chi} < 300$~GeV, cross sections 
generally exceed $10^{-9}$~pb, and the largest scalar cross sections, 
which occur for $m_{\chi} \sim 100$ GeV, are already excluded by XENON10 and/or 
CDMS~II~\footnote{These results are consistent with those presented in \cite{eoss8}.}. 
These exclusions occur primarily in the focus-point region at large $\tanb$.  
On the other hand, for $m_{\chi} \gtrsim 400$ GeV scalar cross sections are well 
below $10^{-9}$ pb, and come from the coannihilation strip or the rapid-annihilation 
funnel that appears at large $\tanb$ in the CMSSM~\footnote{Bordering these regions, the relic density of neutralinos can be quite low with respect to $\Omega_{CDM}$ such that some of the cross sections in panels (c) and (d) are highly scaled.}. Moreover, the effective cross sections are
suppressed for points with $\Omega_\chi \ll \Omega_{CDM}$, and there may be cancellations
at larger $m_{\chi}$ that suppress the cross sections substantially. 
These regions of parameter space will not be probed by direct detection experiments in the near future.

\section{NUHM1 Models}

In NUHM1 models, either $\mu$ or $m_A$ may be taken as a free parameter, 
in addition to the four parameters and $\mu$ sign choice in the CMSSM.  To 
examine the parameter space, we may choose to fix either $m_{1/2}$ or $m_0$, 
along with $\tan{\beta}$, $A_0$, and the sign of $\mu$, and perform a scan over 
the remaining two parameters~\cite{EOS08}.  

Before exploring the NUHM1
parameter space, we first make some general comments. The LSP may be either bino-like or
a mixed bino-Higgsino state.  If the LSP is bino-like, its mass is nearly equal to the
bino mass, $M_1$, which is proportional to $m_{1/2}$: $m_\chi \sim 0.42 m_{1/2}$ 
for a bino-like neutralino.  Scans with fixed $m_{1/2}$ can therefore provide only limited 
information on the accessibility of the full parameter 
space with direct dark matter detection experiments. When $\mu$ becomes less than
$M_1$, however, the LSP becomes increasingly Higgsino-like, and its mass is strongly 
influenced by the value of $\mu$.  Specifically, one finds that the mass drops off as the 
Higgsino component becomes more substantial. 

\subsection{Exploring Large $m_0$ in Sample NUHM1 Planes}

\begin{figure}[p]
\begin{wide}{-1in}{-1in}
\begin{center}
\vskip -1.4in
\hskip .6in
\resizebox{0.55\textwidth}{!}{\includegraphics{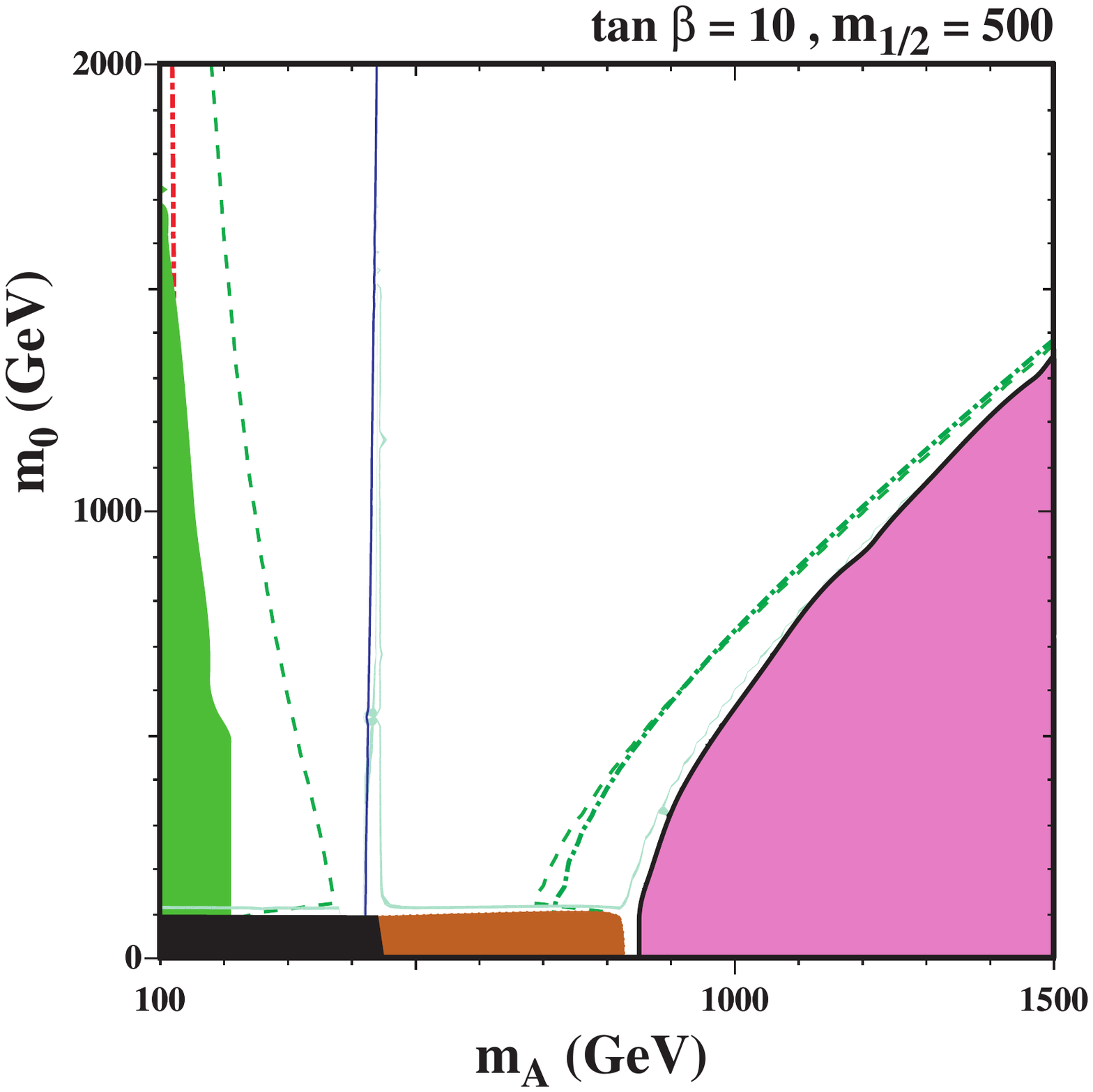}}
\hskip -.6in
\resizebox{0.55\textwidth}{!}{\includegraphics{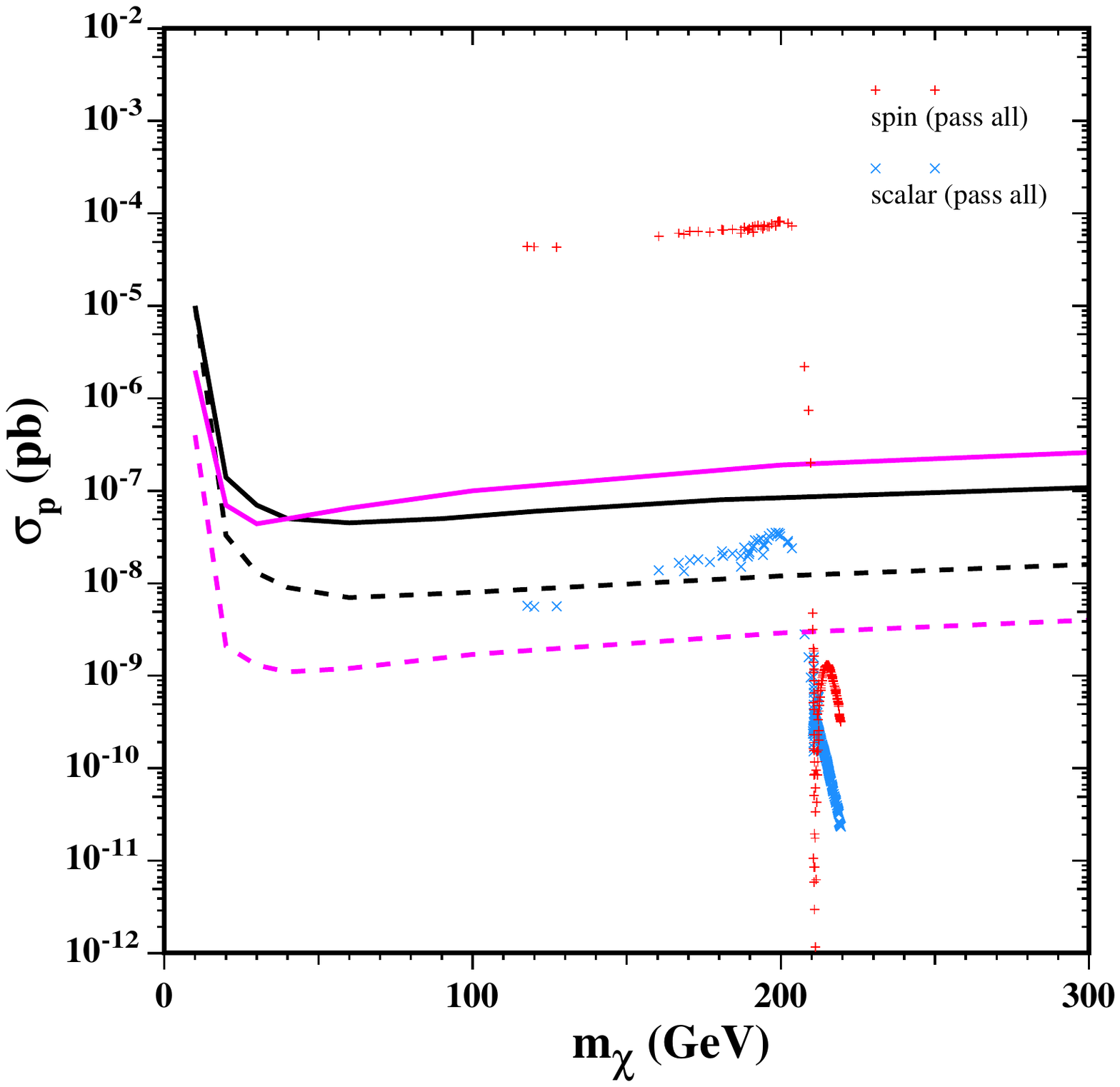}}
\vskip -1.8in
\hskip .6in
\resizebox{0.55\textwidth}{!}{\includegraphics{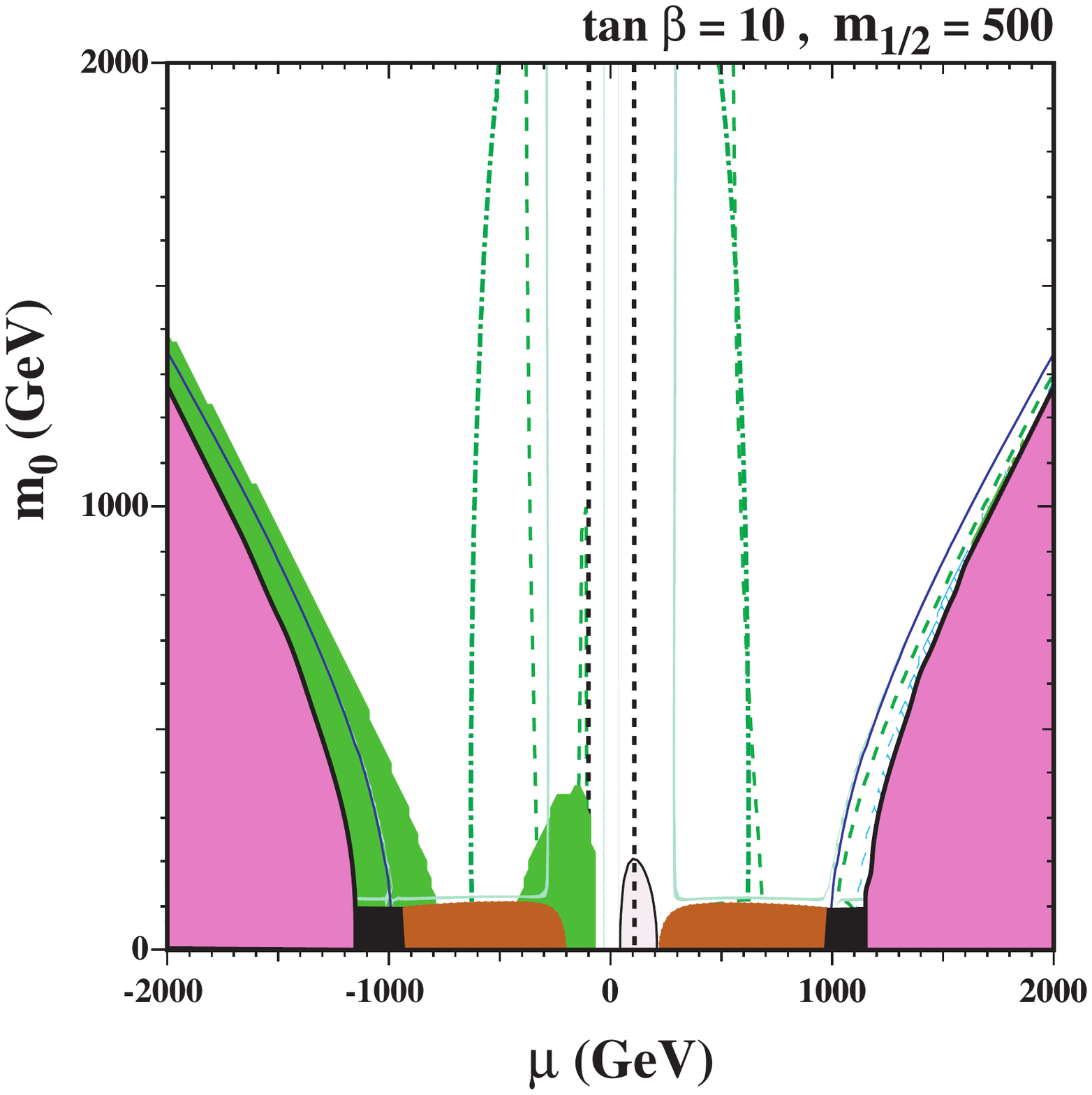}}
\hskip -.6in
\resizebox{0.55\textwidth}{!}{\includegraphics{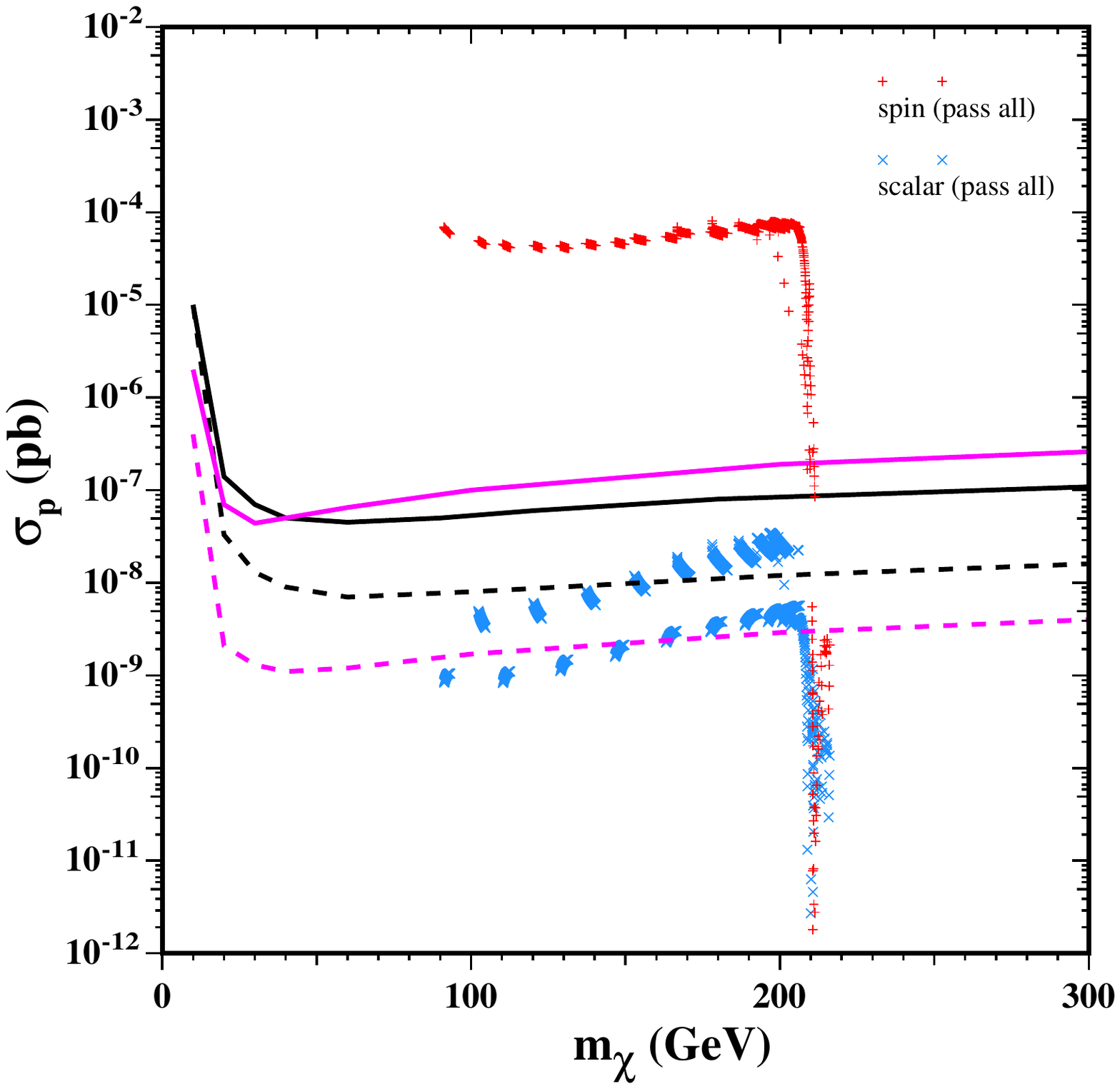}}
\end{center}
\end{wide}
\caption{\it Panels (a) and (c) show the NUHM1 $(m_A,m_0)$ and $(\mu,m_0)$ 
planes for $m_{1/2}=500$ GeV, $\tanb=10$, and $A_0=0$ ($\mu > 0$ in panel (a)). 
Panels (b) and (d) show the corresponding neutralino-nucleon elastic scattering 
cross sections for cosmologically-viable models as functions of neutralino mass.
\label{fig:fixm0}}
\end{figure}

We display in Fig.~\ref{fig:fixm0} representative 
NUHM1 $(m_A,m_0)$ and $(\mu,m_0)$ planes in panels (a) and (c), and the 
corresponding neutralino-nucleon elastic scattering cross sections as functions of the
neutralino mass for cosmologically viable regions of the NUHM1 plane in panels (b) and 
(d), respectively.   
Although fixing $m_{1/2}$ limits the range of neutralino masses, as already
commented, and hence also the cross sections, 
the $(m_A,m_0)$ and $(\mu,m_0)$ planes in panels (a) and (c) of Fig.~\ref{fig:fixm0}
exhibit many of the constraints and features that provide the basis for understanding 
direct detection prospects in models with non-universal Higgs masses discussed in the 
following sections.  In panel (a), for example, regions of the plane at very low $m_A$ are 
excluded because the lighter scalar Higgs mass falls below the LEP bound 
(red dot-dashed line) and/or BR($b \goto s \gamma$) is too large (green shading). 
At large $m_A$ relative to $m_0$, it becomes impossible to satisfy the electroweak 
vacuum conditions with a real value of $\mu$, leading to the pink unphysical region.
The brown and black regions at very low $m_0$ are forbidden because they have 
a stau or selectron/smuon LSP, respectively.   

There are four distinct regions in panel (a) of Fig.~\ref{fig:fixm0} where the relic density 
of neutralinos falls within the WMAP range. At low $m_A$ and $m_0$, the selectron and 
stau are nearly degenerate, and both are lighter than the lightest 
neutralino, with the selectron being the lighter. Close to this forbidden region,
there is an allowed turquoise strip where {\it coannihilations} of neutralinos with 
selectrons and staus bring the relic density down into the WMAP range.  
Near $m_A = 420$ GeV, where $m_A = 2 m_\chi$ as indicated by the solid
blue line, a {\it rapid-annihilation funnel} rises out of the coannihilation strip.
Between the two funnel walls where $2 m_{\chi} \approx m_A$, the relic density of 
neutralinos is brought below the WMAP range by an enhanced annihilation rate through 
the direct-channel $A$ pole. On the other side of the funnel, we see a continuation of the coannihilation 
strip, near the border of the region where the stau is the LSP. The final region where the 
relic density of neutralinos falls within the WMAP range occurs along the border of the 
unphysical region at large $m_A$ where electroweak symmetry breaking cannot be 
obtained, in the {\it focus-point region}.  Here, as $\mu \goto 0$, the LSP becomes increasingly 
Higgsino-like, and annihilations to gauge bosons are enhanced. Since $m_{1/2}=500$~GeV 
in this plane, the LSP is bino-like with a mass of just over 200 GeV in all cosmologically 
viable regions except the focus-point region, where the mass of the lightest neutralino 
depends on $\mu$ and may be smaller.  

As a result, panel (b) shows cross sections possible in this plane as a 
strip at $m_\chi \sim 200$ GeV, including values in both the coannihilation strip and the
rapid-annihilation funnel, where the LSP is bino-like and $m_\chi \sim 0.42 m_{1/2}$.  
We note that there are points with $m_\chi \sim 210$~GeV that have very low cross
sections because they lie in the rapid-annihilation funnel at large $m_0$, a possibility
offered in the NUHM1 because of the freedom in the Higgs masses, that was absent in the CMSSM.
On the other hand, the focus-point region yields the largest cross sections. However, 
as the LSP becomes increasingly Higgsino-like, its mass decreases with $\mu$, and the 
relic density eventually falls below the WMAP range. Thus, we rescale the cross sections 
at lower $m_{\chi}$ by a factor of $\Omega_{\chi} / \Omega_{WMAP}$, which decreases 
as the Higgsino fraction increases and $\Omega_{\chi}$ decreases.  Had a larger (smaller) 
value of $m_{1/2}$ been chosen, these features in panel (b) would have been shifted to 
larger (smaller) $m_{\chi}$.

The green dot-dashed line in panel (a) marks the CMSSM subspace within this
NUHM1 plane. It crosses the
cosmologically viable regions in two places: at large $(m_A,m_0)$ in the focus-point region,
and at low $m_0$ in the stau coannihilation strip.  The largest CMSSM cross sections of a 
few $\times 10^{-8}$ pb come from the focus-point region, as already seen in Fig.~\ref{fig:cmssm},
whereas cross sections from CMSSM points along the coannihilation strip peak at $\sim 10^{-9}$~pb.
Moreover, they are rescaled in the portion of the CMSSM contour that extends into the area below the 
coannihilation strip, where the relic density of neutralinos can be as much as an order of 
magnitude below the WMAP range.

Panel (c) shows an NUHM1 $(\mu,m_0)$ plane, again with $m_{1/2}$ fixed to be 500 GeV.  
Many of the features mentioned above are also present in panel (c). Note that there is
a small region with 100~GeV$ < \mu <$200~GeV and $m_0 < 200$~GeV that
may be favoured by $g_\mu - 2$ (light pink shading, bounded by solid black line). There are 
rapid-annihilation funnels and coannihilation strips at both positive and negative $\mu$.
Also, at large $|\mu|$ there are unphysical regions where the electroweak 
vacuum conditions imply $m_A^2 < 0$, and the rapid-annihilation funnels, 
where $m_A \sim 420$ GeV, run close to these borders. 
There is also an analogue of the focus-point region 
in the form of {\it crossover strips} of good relic density at $|\mu| \sim 300$ GeV,
where the LSP has a substantial Higgsino component.  
The BR($b \goto s \gamma$) constraint is important primarily for $\mu<0$, shown
in the left half-plane of (c), in which case it excludes the funnel and parts of the coannihilation and crossover strips.  The LEP constraint on the Higgs mass is insignificant in panel (c), 
as is the upper limit on BR($\bmm$), which becomes important only at larger $\tanb$.
The green dot-dashed CMSSM contours run roughly vertically at $|\mu| \sim$ 500-600 GeV,
and in the visible part of the plane they intersect only the cosmologically preferred coannihilation 
strips at low $m_0$. However, at larger $m_0$ the CMSSM contours would join together
to form a parabola, and intersect the crossover strips at both positive and negative $\mu$,
outside the visible part of the plane.
In the area between the crossover strips, the 
relic density of neutralinos is below the WMAP range, resulting in rescaled cross sections in 
panel (d) that decrease for the increasingly Higgsino-like LSPs at lower  $m_{\chi}$. 
There are two distinct strips of scalar cross sections in panel (d) for Higgsino-like LSPs:
the upper one corresponding to $\mu>0$, and the lower to $\mu<0$.  Cross sections for 
$\mu<0$ are suppressed by cancellations in the scattering matrix element due to
sign differences in the neutralino eigenvectors.

We illustrate in panels (b) and (d)
the reaches of current and future direct detection experiments in the NUHM1 
parameter space.  As can be seen from the green dashed contours of $\sigma_{SI}$
drawn in panel (a), probing this plane will require a sensitivity below 
$5 \times 10^{-8}$~pb. In particular, to probe the coannihilation strip at low $m_A$ and all 
of the focus-point region would require a direct detection experiment with a sensitivity of 
$10^{-9}$ pb. For the $(\mu,m_0)$ plane in panel (c), 
$10^{-9}$ pb would again suffice to probe the phenomenologically viable region between the crossover strips~\footnote{Within the region at very low $|\mu|$ where the chargino mass is below the LEP bound, we omit the contours of $\sigma_{SI} = 10^{-9}$ pb. The relic density of neutralinos can be quite low in this region, such that 10$^{-9}$ pb is not sufficient to probe the entirety of this already-excluded strip.} as well as 
some of the coannihilation strip.  Unfortunately, scalar neutralino-nucleon elastic scattering cross sections in the rapid-annihilation funnel regions of both panels (a) and (c)
may be considerably lower, as these points may have very large values
of $m_0$ - a feature not found in the CMSSM. 
Only a detector like LUX/ZEP with 20 tonnes would have 
sufficient sensitivity to probe 
also the funnel. 

\begin{figure}[ht!]
\begin{wide}{-1in}{-1in}
\begin{center}
\vskip -1.4in
\hskip .6in
\resizebox{0.55\textwidth}{!}{\includegraphics{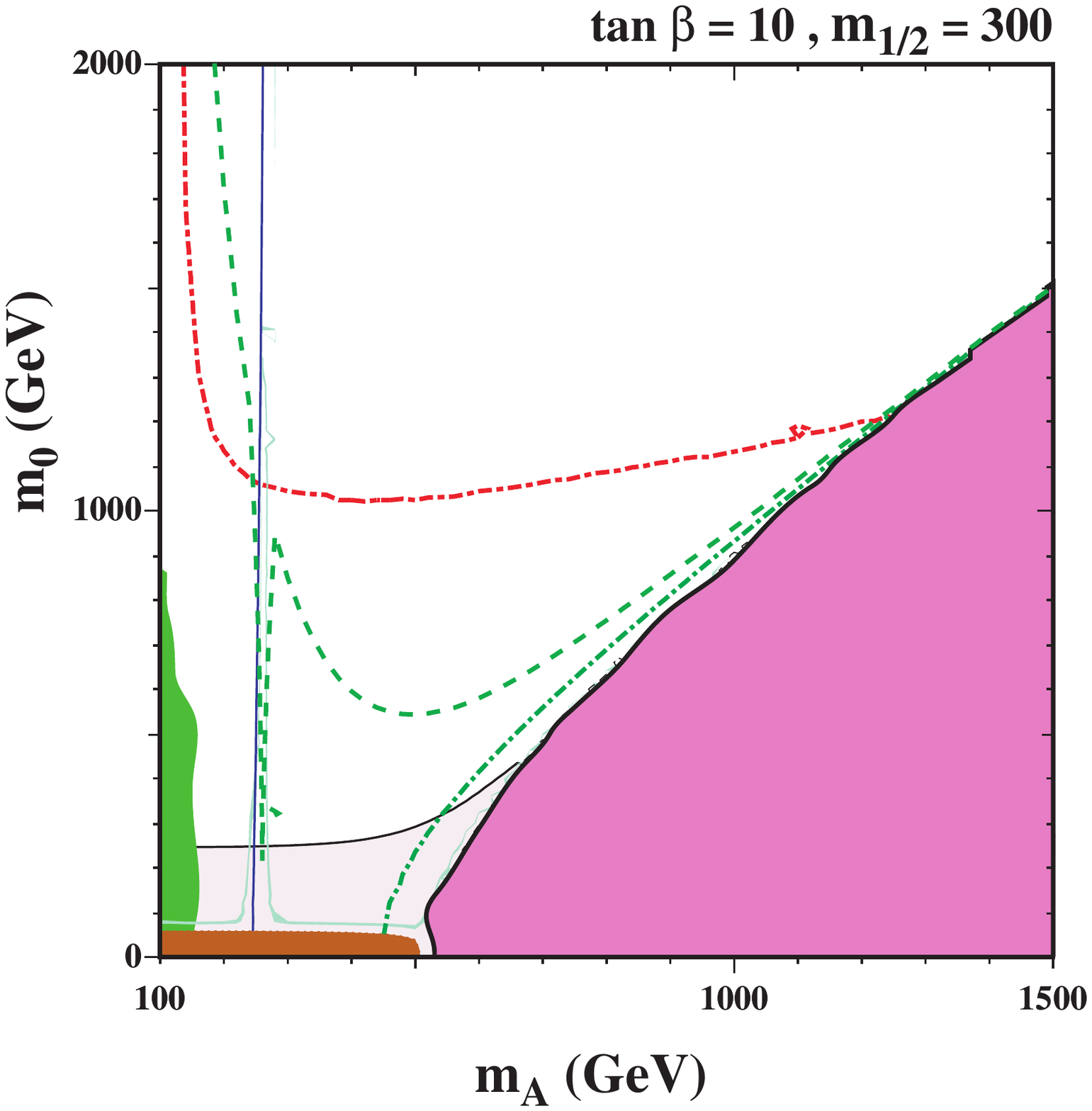}}
\hskip -.6in
\resizebox{0.55\textwidth}{!}{\includegraphics{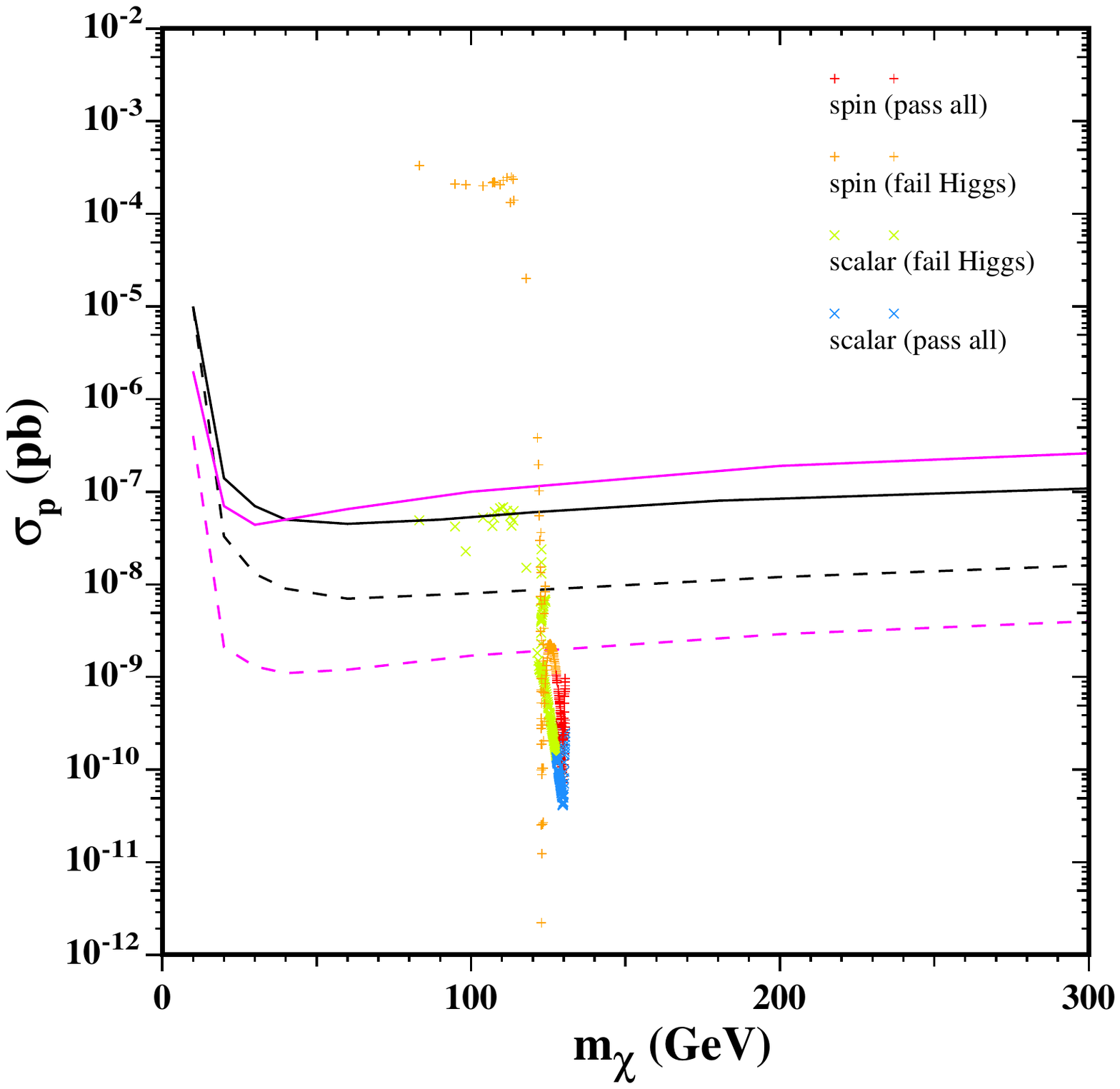}}
\vskip -1.8in
\hskip .6in
\resizebox{0.55\textwidth}{!}{\includegraphics{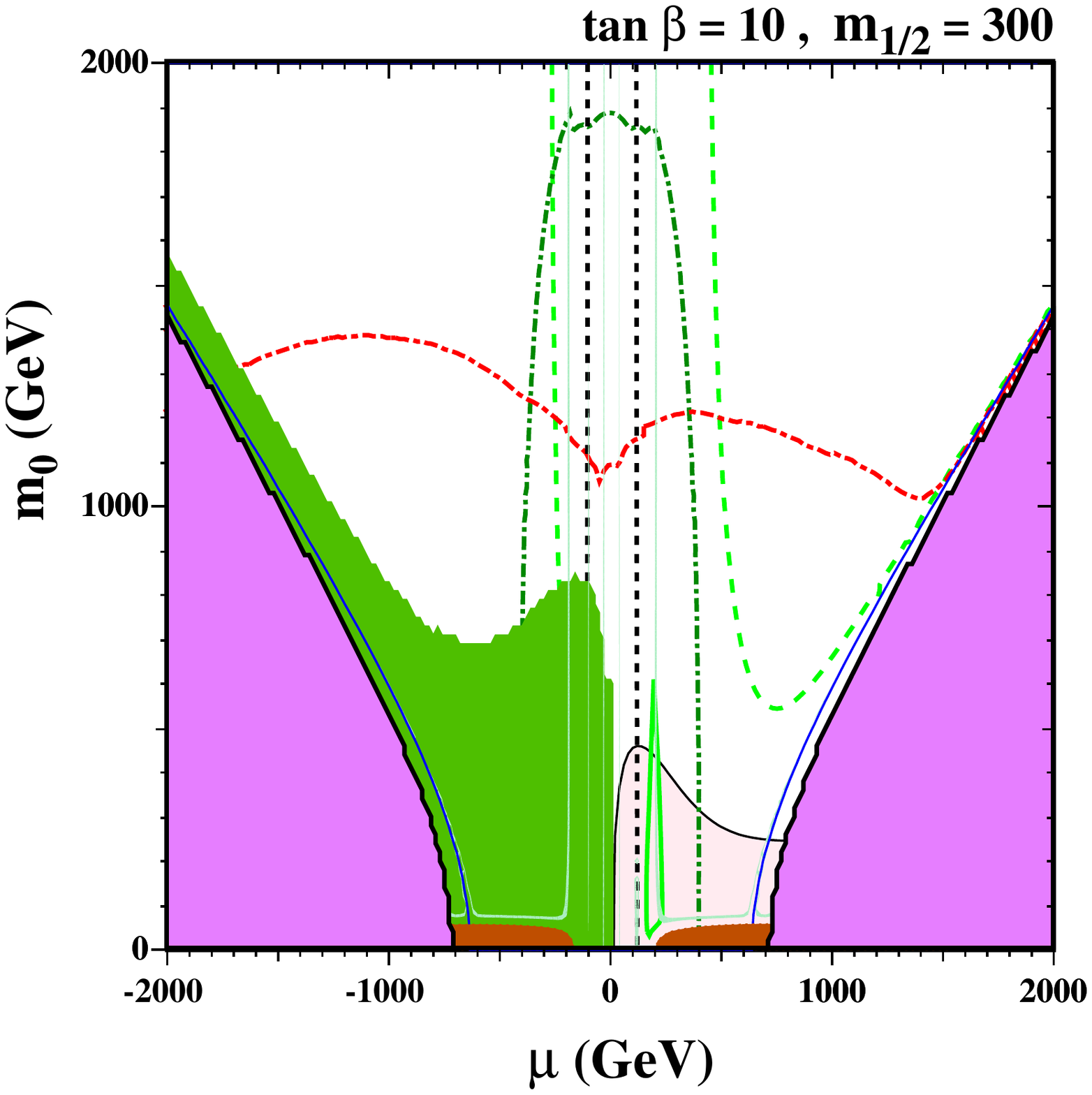}}
\hskip -.6in
\resizebox{0.55\textwidth}{!}{\includegraphics{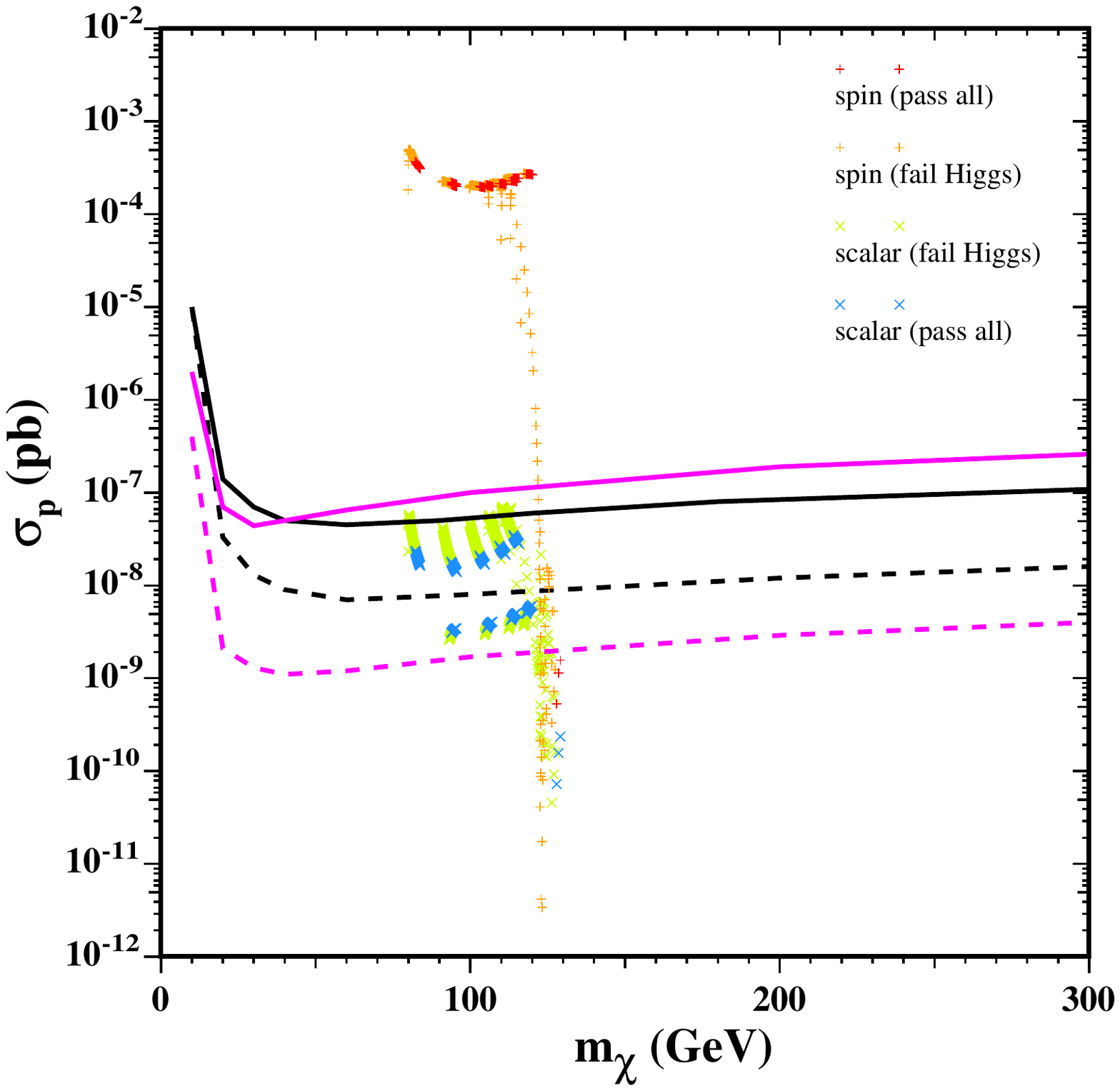}}
\end{center}
\end{wide}
\caption{\it As in Fig. \protect\ref{fig:fixm0}, but for $m_{1/2} = 300$~GeV.
\label{fig:fixm0-2}}
\end{figure}

Fig.~\ref{fig:fixm0-2} shows corresponding $(m_A, m_0)$ and $(\mu, m_0)$ planes
with $\tan \beta = 10$ and a lower value of $m_{1/2} = 300$~GeV close to the best
fit value for the NUHM1 found in \cite{mc2}. Comparing its panel (a) 
with that of Fig.~\ref{fig:fixm0}, we see that the
electroweak symmetry breaking constraint is more important, as is the Higgs mass
constraint, whereas the $b \to s \gamma$ constraint is less important. A new feature is that
a region at low $m_0$ is apparently favoured by $g_\mu - 2$. We also note that the 
prospective reach of future searches for spin-independent dark matter scattering in
the $(m_A, m_0)$ plane is 
considerably greater for $m_{1/2} = 300$~GeV than for $m_{1/2} = 500$~GeV, as seen
by the shift in the green dashed contour between the corresponding panels (a)~\footnote{The 
dip around $m_A = 280$~GeV in Fig.~\ref{fig:fixm0-2}(a) is due to the rescaling of the
cross section in the underdense band between the rapid-annihilation strips.}. However,
as seen in panel (b), there is no large increase in the spin-independent dark matter 
scattering cross section in the regions of the plane allowed by WMAP, and the points
with larger cross sections, most of which have $m_\chi \sim 120$~GeV or less,
generally have $m_h < 114$~GeV. We also note that there are points with $m_\chi \sim
120$~GeV and very low cross sections, due again to their locations at large $m_0$
along the rapid-annihilation funnel in panel (a).

Comparing the $(\mu, m_0)$ plane in Fig.~\ref{fig:fixm0-2}(c) with that in
in panel (c) of Fig.~\ref{fig:fixm0}, we again see increased importance
for the electroweak vacuum and LEP Higgs constraints, and the $b \to s \gamma$
constraint for $\mu < 0$ is also more important. We also note the appearance of
a region at low $m_0$ that is apparently favoured by $g_\mu - 2$, and that  the 
prospective reach of future searches for spin-independent dark matter scattering in
the $(m_A, m_0)$ plane is again considerably greater for $m_{1/2} = 300$~GeV 
than for $m_{1/2} = 500$~GeV. As in panel (b) of Fig.~\ref{fig:fixm0-2}, only a few
points in panel (d) escape the LEP Higgs constraint, but some of these are in the
focus-point region with $m_\chi < 120$~GeV, and have spin-independent
scattering cross sections close to the present experimental upper limits.
We again see points with $m_\chi \sim
120$~GeV and very low cross sections, that lie along the rapid-annihilation 
funnel in panel (c), at large $m_0$. These points have no counterparts in the CMSSM.

After these exploratory studies introducing some properties of characteristic
dark matter regions, we next study more systematically NUHM1 planes
with $m_{1/2}$ and either $m_A$ or $\mu$ as free parameters, keeping $m_0$ fixed.

\subsection{The NUHM1 with $m_A$ as a free parameter}

\begin{figure}[ht!]
\begin{wide}{-1in}{-1in}
\begin{center}
\vskip -1.4in
\hskip .6in
\resizebox{0.55\textwidth}{!}{\includegraphics{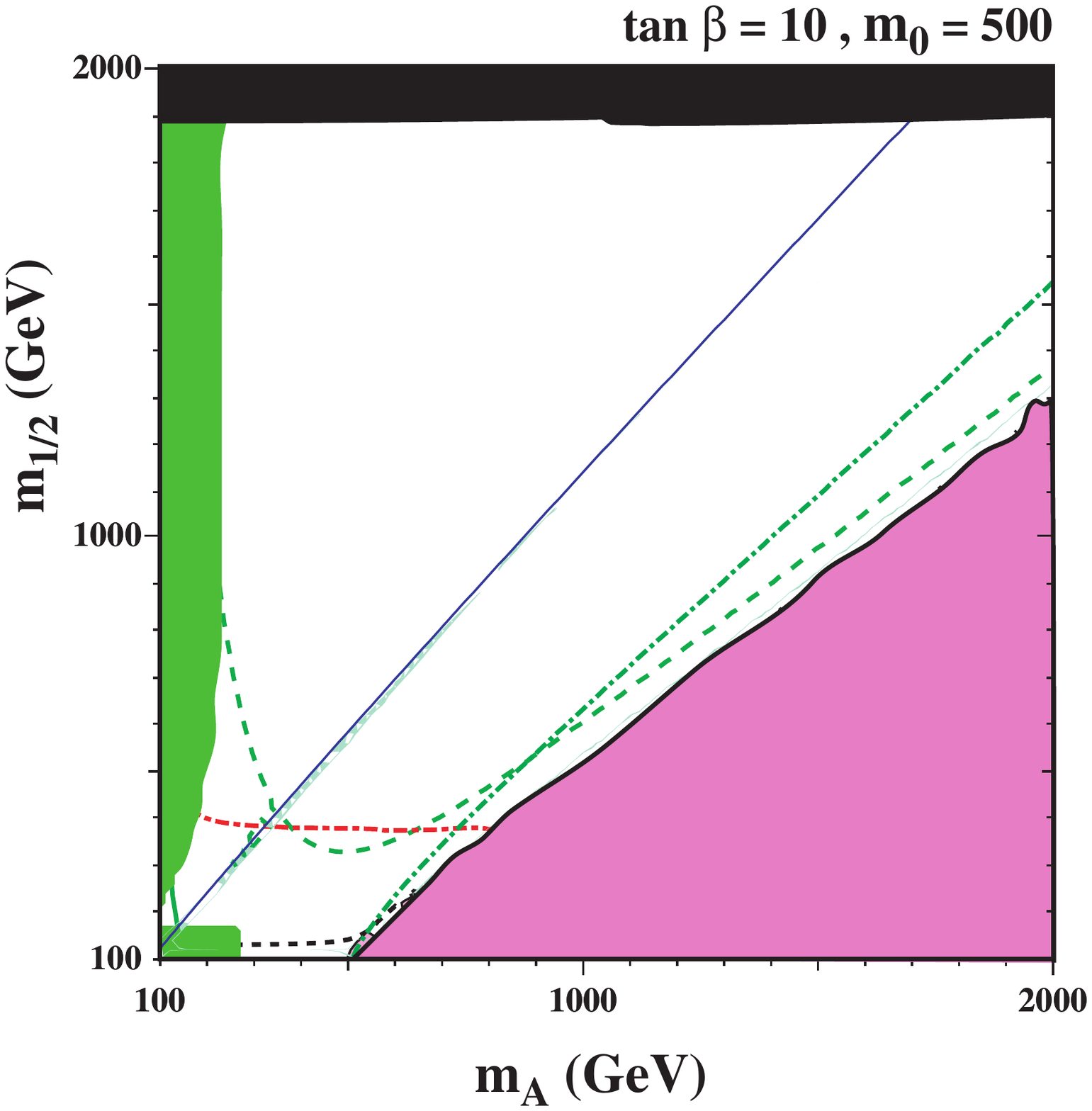}}
\hskip -.6in
\resizebox{0.55\textwidth}{!}{\includegraphics{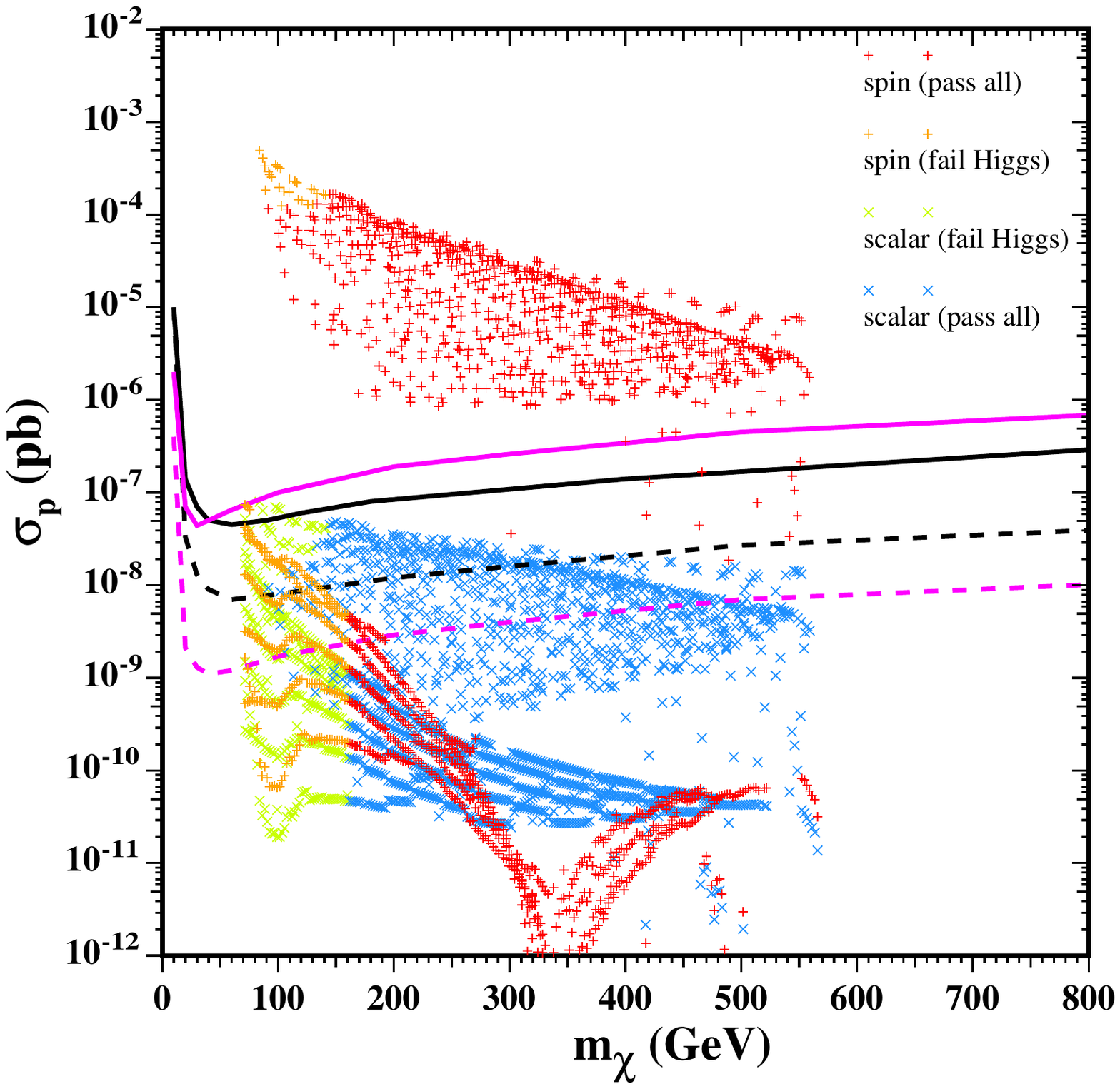}}
\vskip -1.8in
\hskip .6in
\resizebox{0.55\textwidth}{!}{\includegraphics{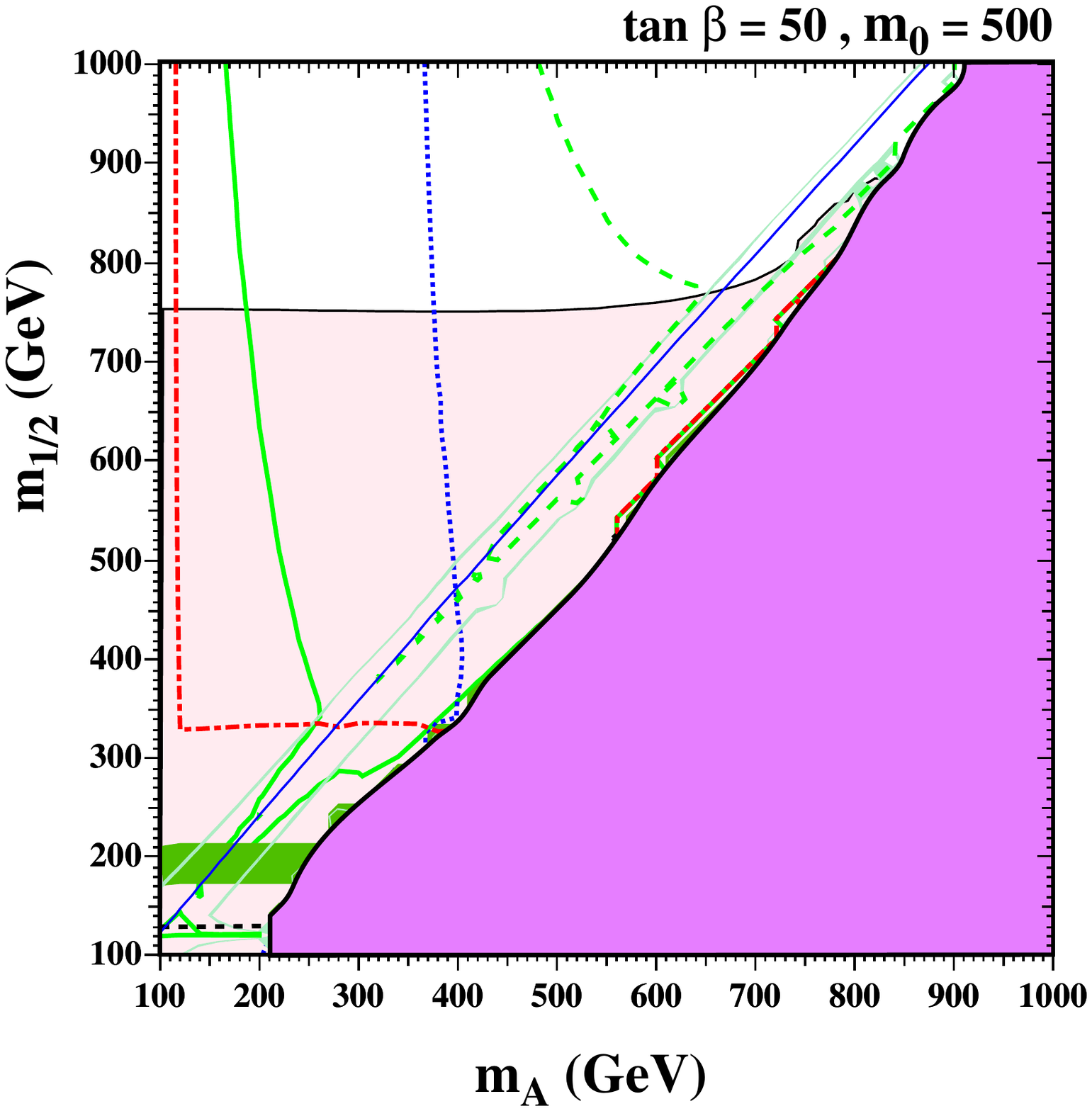}}
\hskip -.6in
\resizebox{0.55\textwidth}{!}{\includegraphics{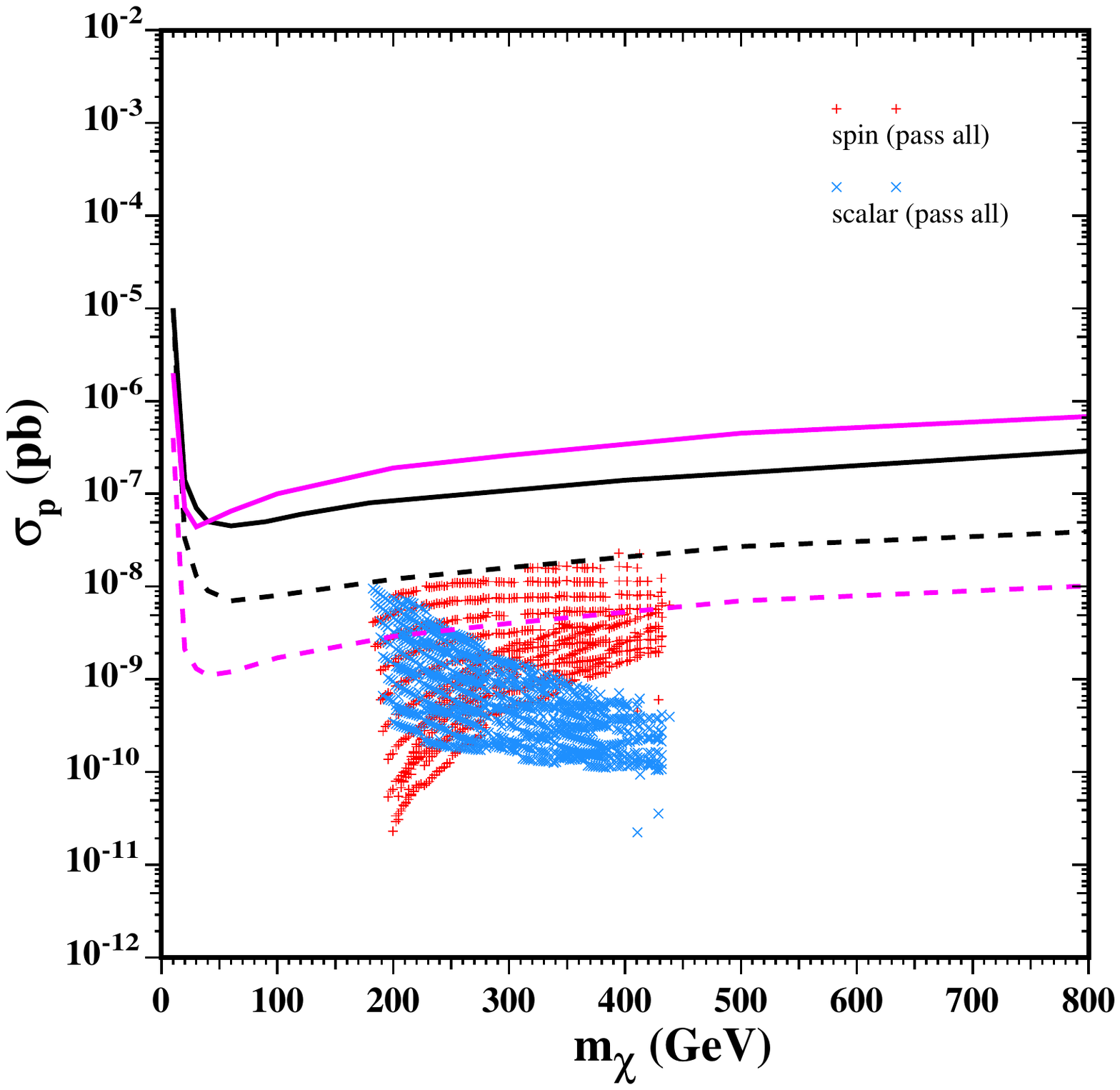}}
\end{center}
\end{wide}
\caption{\it Panels (a) and (c) show the NUHM1 $(m_A,m_{1/2})$ planes for 
$m_0=500$ GeV, $\tanb=10$ and 50. Panels (b) and (d) show the corresponding 
neutralino-nucleon elastic scattering cross sections as functions of neutralino mass.
\label{fig:1mAvMlowtb}}
\end{figure}

Fig.~\ref{fig:1mAvMlowtb}(a) displays a $(m_A, m_{1/2})$ plane with fixed 
$\tanb = 10, m_0 = 500$~GeV. The triangular allowed region is bounded by $b \to s \gamma$
at small $m_A$, the appearance of a slepton LSP at large $m_{1/2}$, and the absence of a
consistent electroweak vacuum at larger $m_A$ and smaller $m_{1/2}$.
The diagonal blue line indicates where $m_\chi = m_A/2$: on either side there is a
narrow rapid-annihilation funnel strip where the relic density falls within the WMAP
range. This funnel extends only to $m_{1/2} \approx 1200$ GeV, and therefore 
$m_{\chi} \lesssim 550$ GeV for this region. There is another WMAP strip in the 
focus-point region close to the electroweak vacuum boundary, 
where the LSP is more Higgsino-like. The displayed part of the focus-point 
strip is cut off at $m_A = 2000$ GeV, corresponding to $m_{\chi} \lesssim 600$ GeV:
larger values of $m_\chi$ would be allowed if one considered larger values of $m_A$. 
Apart from the region between this strip and
the boundary, and between the funnel strips, the relic LSP density would exceed the WMAP
range. The green dot-dashed CMSSM contour runs only through regions excluded 
either by excessive $\Omega_{\chi} h^2$ or by the LEP chargino mass limit. 
Whilst the BR($b \goto s \gamma$) limit is also important at very low $m_A$ and $m_0$, 
it is the constraint on the Higgs mass (shown by the red 
dot-dashed curve that is roughly horizontal at $m_{1/2} = 400$ GeV) 
that places a lower limit on the expected LSP mass 
for the funnel region of $m_\chi \sim 160$ GeV, where the LSP is bino-like.

We see in panel (a) that the green dashed line indicating future sensitivity to
spin-independent dark matter scattering is mainly in the region where the
relic density exceeds the WMAP upper limit.
SuperCDMS at Soudan would be sensitive to much of the focus-point region 
shown in panel (a), but a more advanced detector would be required if the relic density 
of neutralinos is below the WMAP range. Unfortunately, even if neutralinos
make up all the dark matter in the universe, much of the funnel region will remain out of 
reach, even to next-generation direct detection experiments.
Points associated with the funnel with cross sections larger than $10^{-9}$ pb
fail the Higgs mass constraint.

Panel (b) of Fig.~\ref{fig:1mAvMlowtb} displays the scattering cross sections 
for the allowed points in panel (a). The fact that scalar cross sections in the funnel region are 
generally smaller than those in the focus-point region is reflected in the lower cutoff on 
the LSP mass, $m_{\chi} \lesssim 550$ GeV, seen in panel (b) for the points with
$\sigma_{SI} \lesssim$ few $\times 10^{-10}$ pb.  In the focus-point region
the LSP is in a mixed state $m_{\chi} \lesssim 600$ GeV, and the cross section may be 
much larger: by two orders of magnitude for the spin-independent cross section, 
and four orders of magnitude for the spin-dependent cross section. These large
cross sections at large $m_\chi$ have no counterparts in the CMSSM, since in that model 
mixed states in the focus-point region have much lower masses. Furthermore,
in the CMSSM, the focus point is reached only at large $m_0$, whereas here,
$m_0 = 500$ GeV is fixed. 
The large scalar cross sections in the focus-point region are already beginning 
to be probed by CDMS~II and XENON10, with a few points at very low $m_{1/2}$ 
already being excluded, as one can see directly in panel (b)~\footnote{We note
these points also have $m_h < 114$~GeV and are within the solid green contour in panel (a)
at very low values of $m_{1/2}$ and $m_A$.}. We also note that there are points in
panel (b) with very low cross sections even though they have $m_\chi < 150$~GeV,
which also have no counterparts in the CMSSM. These points associated with the
funnel and their cross sections have been scaled down due to the low
relic density in that region.

A new feature seen in this plot is the near-vanishing of the spin-dependent 
cross section when $m_\chi = 350$ GeV.  This feature
is associated with the funnel region and this group of points all have 
cross sections substantially lower than those of the focus-point region
as remarked above.  However, near $m_\chi = 350$ GeV there is the possibility
for a complete cancellation in Eq. (\ref{eqn:Lambda}) when the spin
contribution from up quarks cancels that due to down and strange quarks
for the values in Eq. (\ref{deltas}). While the exact position of the 
cancellation is sensitive to the values of the spin matrix elements adopted,
the existence of the cancellation is quite robust. Of course, in this case the 
cross section for the scattering on neutrons will not exhibit a cancellation at the same place.

Panels (c) and (d) of Fig.~\ref{fig:1mAvMlowtb} show a corresponding analysis
for $\tanb = 50$. We first observe that there is no longer any substantial focus-point 
region, and we do not display in panel (c) the CMSSM contour,
which lies very close to the boundary of the region where electroweak symmetry
breaking is possible. We see that much of the plane is consistent with the apparent
anomaly in the value of $g_\mu - 2$. At the larger value of $\tanb$, 
the branching ratio for $\bmm$ is substantially larger, and the 
current limit \cite{bmm} on the $\bmm$ branching ratio of $5.8 \times 10^{-8}$ excludes
points with $m_A < 370 - 400$ GeV (depending on the value of $m_{1/2}$),
as shown by the blue dotted curve.
In this case, this limit is stronger than that due to the Higgs mass and 
excludes the lower part of the funnel and, as a consequence, 
neutralino masses $\la 200$ GeV.

Even though spin-independent cross sections from the funnel region are slightly 
boosted at large $\tanb$, much of the funnel remains out of reach of
direct detection experiments, as this region is above
the solid green contour corresponding to $\sigma_p = 5 \times 10^{-8}$ pb. 
Moreover, the absence of a focus-point region
for $\tan \beta = 50$ removes a prospective source of points with larger
cross sections. Complementary collider searches and confirmation 
from astrophysical indirect dark matter observations may be needed to
discover dark matter if the relic 
density of neutralinos obtains a cosmologically-acceptable value through 
enhanced annihilations near a direct-channel pole. However, a good portion of the 
funnel (corresponding to acceptable $g_\mu - 2$) is within the dashed green contour, 
and hence has a cross section above $10^{-9}$ pb.

\begin{figure}[h]
\begin{wide}{-1in}{-1in}
\begin{center}
\vskip -1.4in
\hskip .6in
\resizebox{0.55\textwidth}{!}{\includegraphics{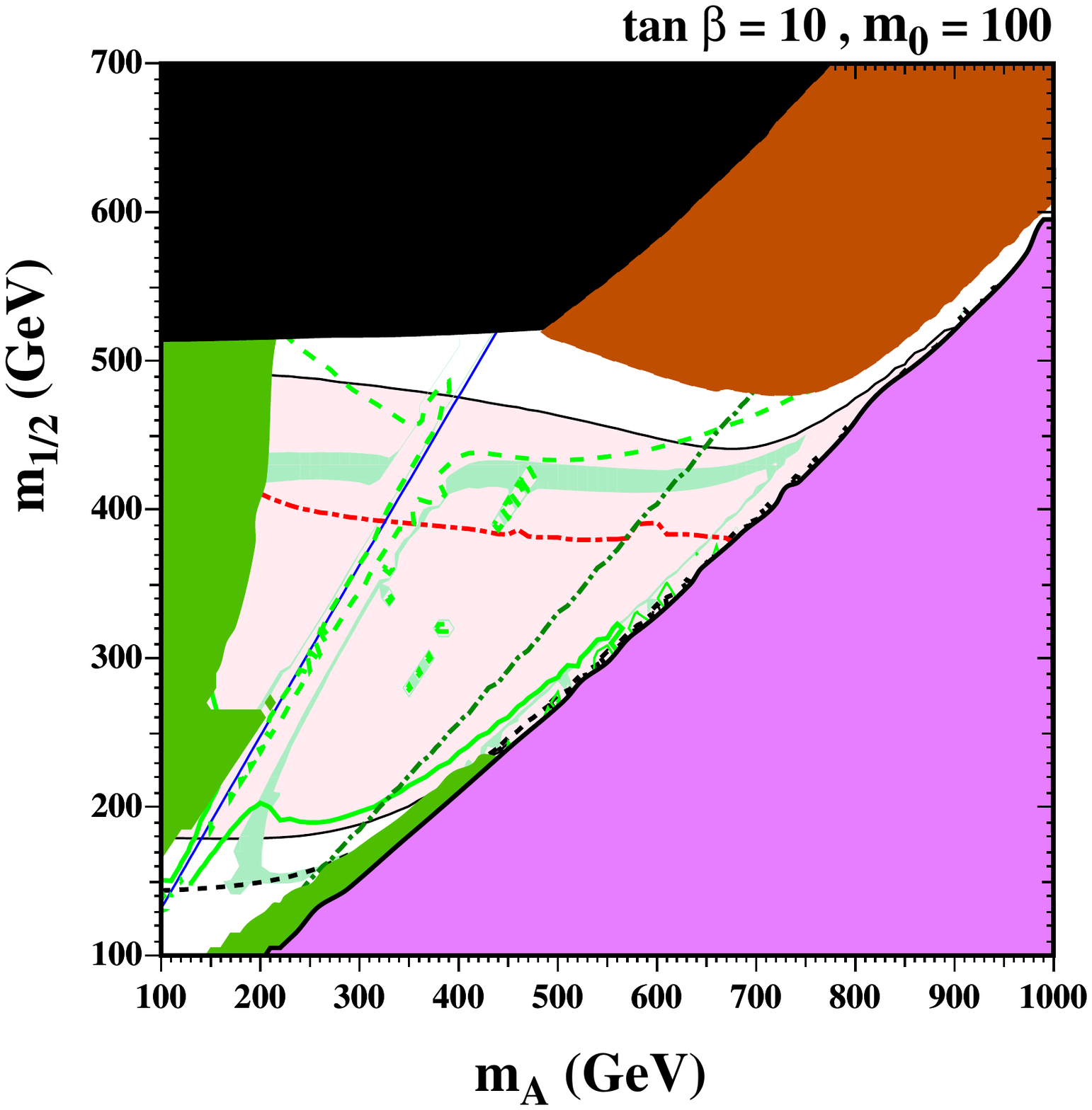}}
\hskip -.6in
\resizebox{0.55\textwidth}{!}{\includegraphics{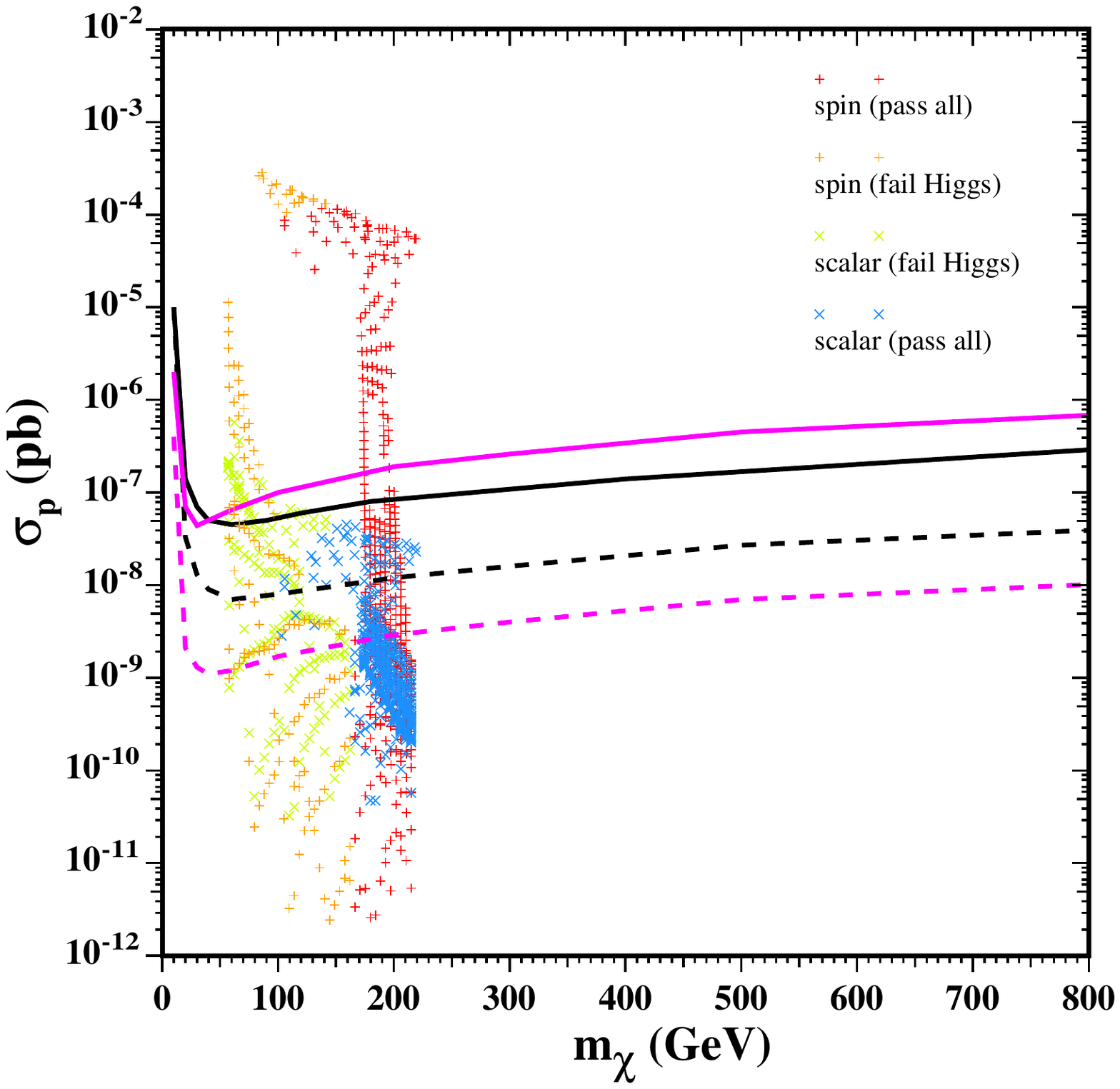}}
\end{center}
\end{wide}
\caption{\it Panel (a) shows the NUHM1 $(m_A,m_{1/2})$ planes for $m_0=100$ GeV and 
$\tanb=10$. Panel (b) shows the corresponding neutralino-nucleon elastic scattering cross 
sections as functions of neutralino mass.
\label{fig:1mAvMlowm0}}
\end{figure}

Fig.~\ref {fig:1mAvMlowm0} shows how the $(m_A, m_{1/2})$ plane changes
if we decrease $m_0$ to 100~GeV, which is close to the best fit found for the NUHM1 \cite{mc2}, from the value $m_0 = 500$~GeV studied in 
Fig.~\ref{fig:1mAvMlowtb}. The most obvious effects in panel (a) are that the region where the 
stau and/or selectron/smuon is lighter than the lightest neutralino descends from
$m_{1/2} \sim 1900$~GeV to $m_{1/2} \sim 500$~GeV. The $b \to s \gamma$
and $m_h$ constraints also become more aggressive, with the result that most of the
points allowed by WMAP are in a coannihilation strip where $m_{1/2} \sim 420$~GeV, 
corresponding to $m_\chi \sim 200$~GeV - which is favoured by $g_\mu - 2$,
moreover. There are, however, some points at
both larger and smaller $m_{1/2}$ in the rapid-annihilation funnel, and also
a few points in a focus-point region close to the boundary of consistent
electroweak symmetry breaking. We see in panel (b) of Fig.~\ref {fig:1mAvMlowm0}
that there is a wide range of possible values of the spin-independent cross section,
particularly for small $m_\chi$ where $m_h < 114$~GeV. Points with $m_\chi < 150$~GeV
and such low cross sections which are associated with the funnel have no analogues in the CMSSM.

\subsection{The NUHM1 with $\mu$ as a free parameter}

\begin{figure}[ht!]
\begin{wide}{-1in}{-1in}
\begin{center}
\vskip -1.4in
\hskip .6in
\resizebox{0.55\textwidth}{!}{\includegraphics{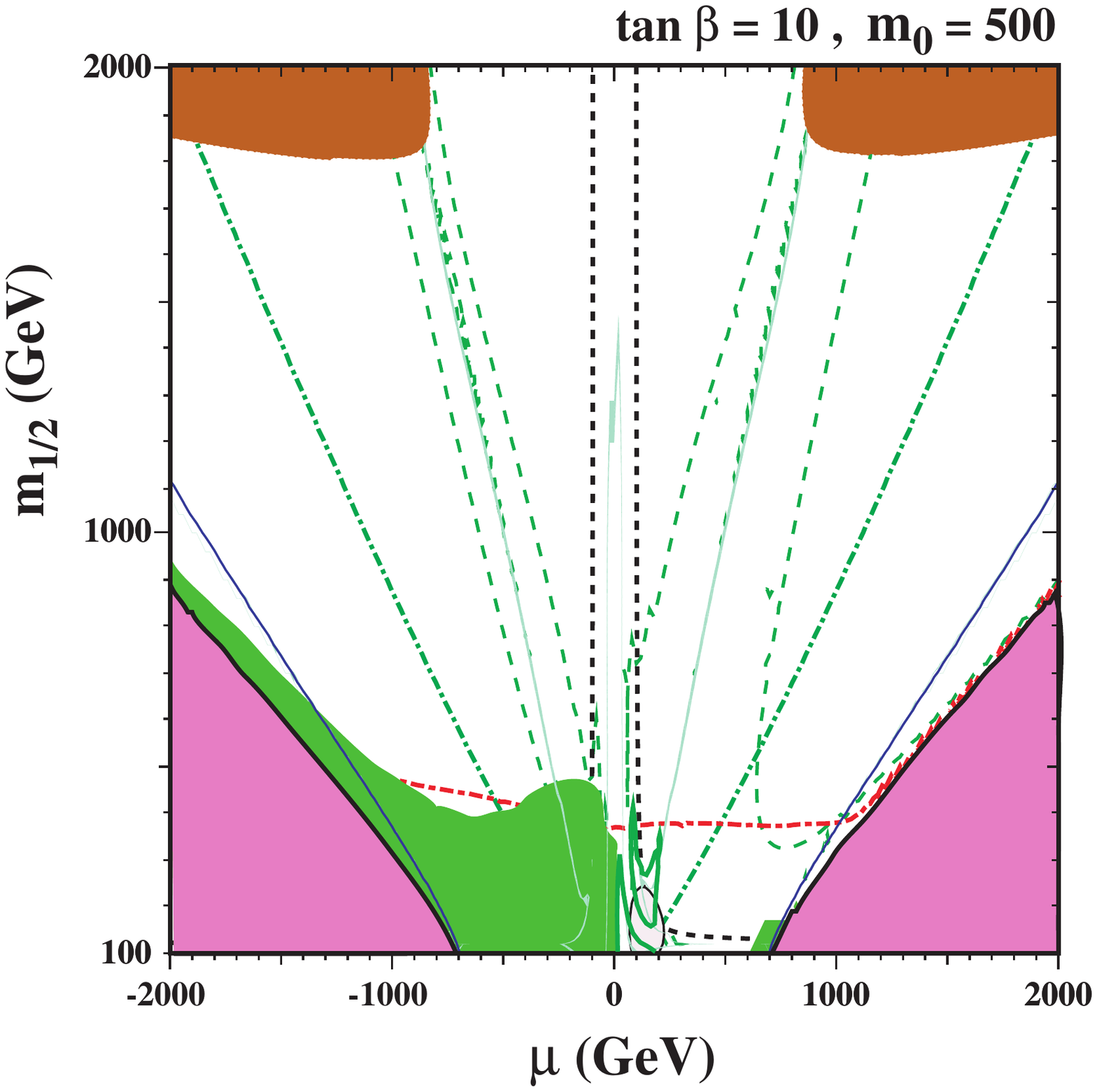}}
\hskip -.6in
\resizebox{0.55\textwidth}{!}{\includegraphics{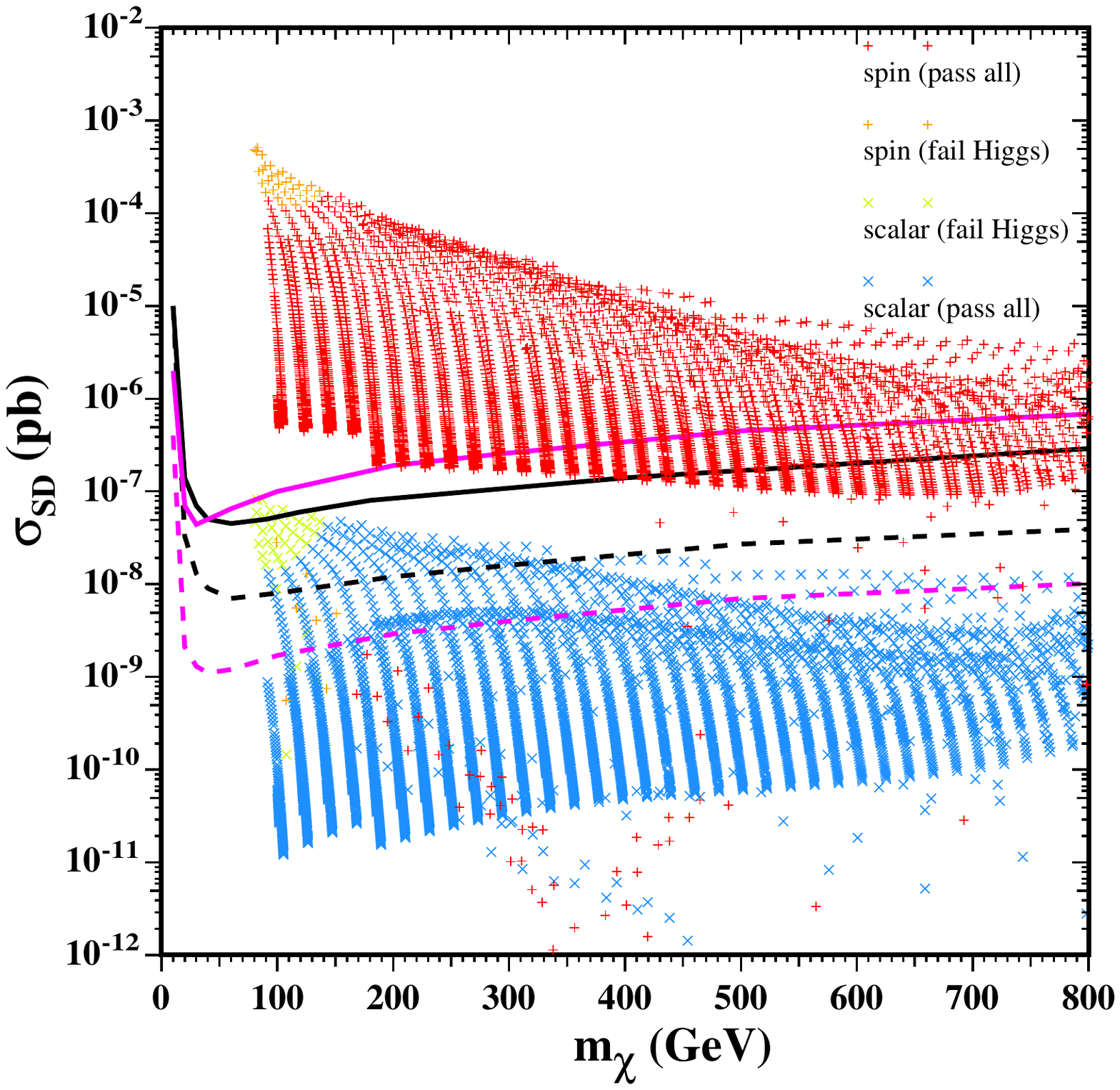}}
\vskip -1.8in
\hskip .6in
\resizebox{0.55\textwidth}{!}{\includegraphics{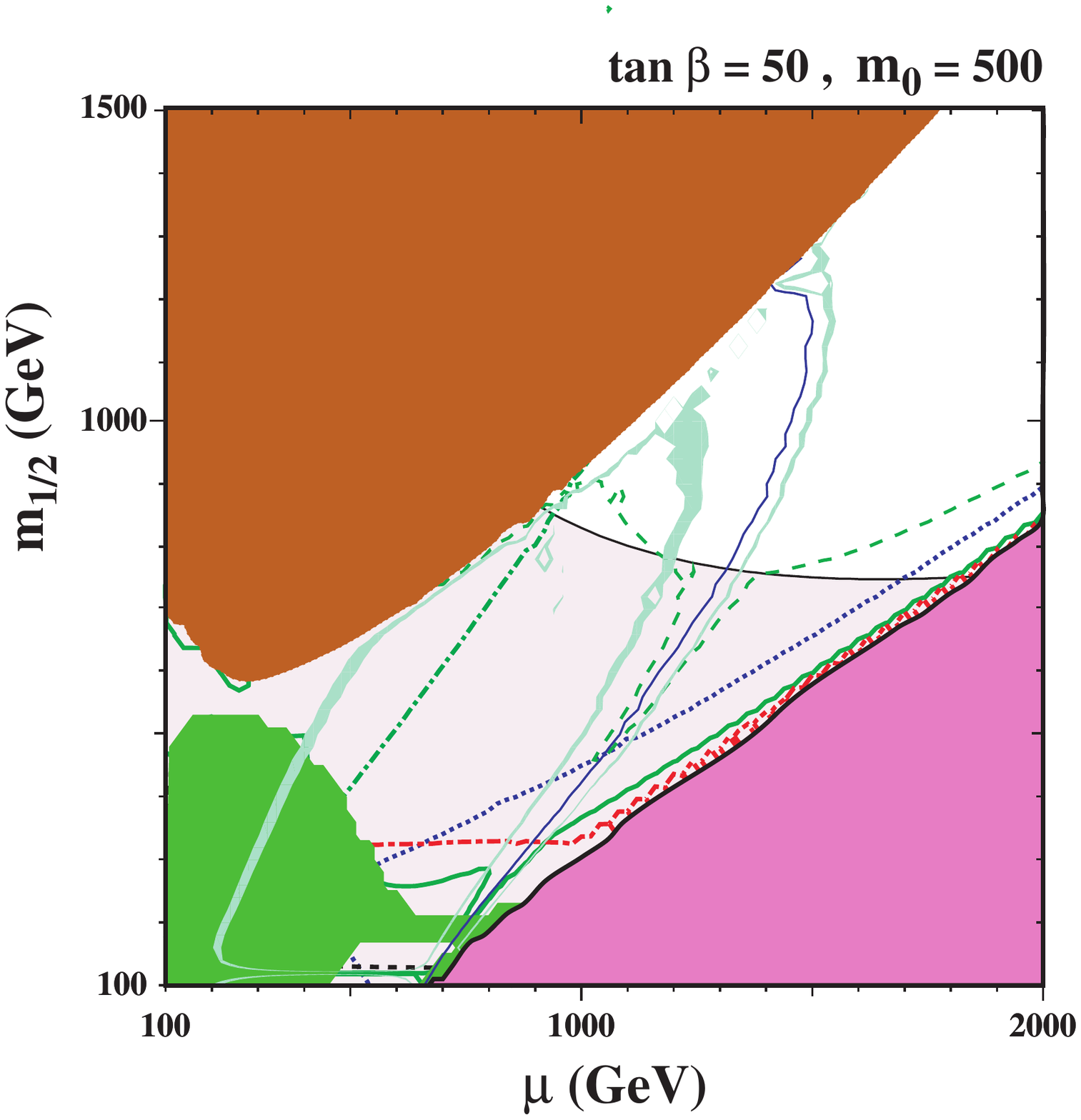}}
\hskip -.6in
\resizebox{0.55\textwidth}{!}{\includegraphics{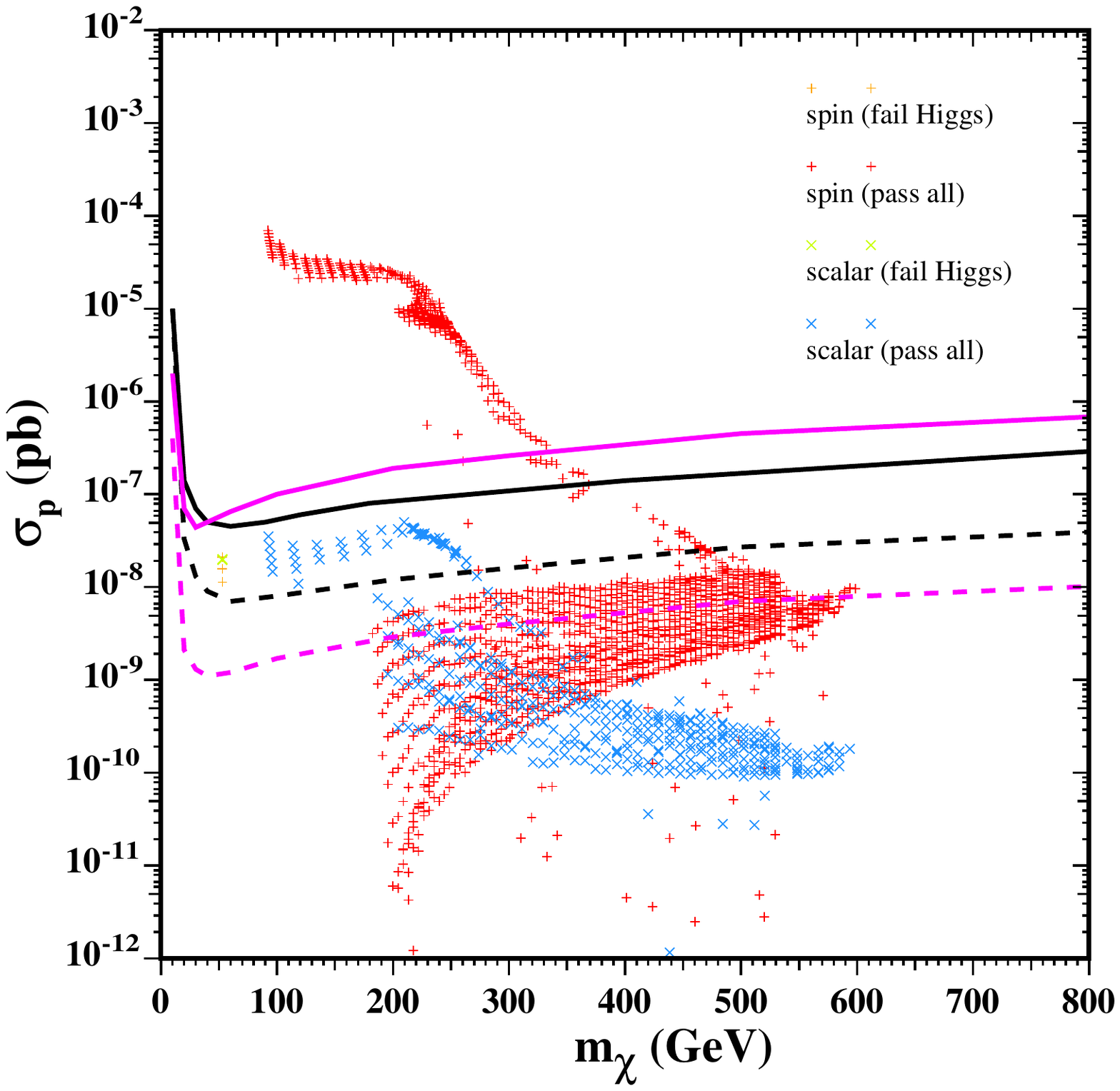}}
\end{center}
\end{wide}
\caption{\it Panels (a) and (c) show the NUHM1 $(\mu,m_{1/2})$ planes for 
$m_0=500$ GeV, $\tanb=10$ and 50. Panels (b) and (d) show the corresponding 
neutralino-nucleon elastic scattering cross sections as functions of neutralino mass.
\label{fig:1muvMlowtb}}
\end{figure}

Fig.~\ref{fig:1muvMlowtb}(a) shows a $(\mu, m_{1/2})$ plane for $\tanb = 10$
and $m_0 = 500$~GeV. We see in panel (a) that $b \to s \gamma$ excludes
a region at small $m_{1/2}$ for $\mu < 0$, and that the Higgs constraint excludes
points with low $m_{1/2}$ and $\mu > 0$ in both the crossover strip and the 
rapid-annihilation funnel (which lies close to the electroweak vacuum boundary in
this case). The intersection of the CMSSM contour with regions of acceptable relic 
density occurs only at very low $m_{1/2}$, in regions already excluded by the 
Higgs constraint (and also by $b \goto s \gamma$ for $\mu<0$). 

The largest cross sections excluded by BR($b \goto s \gamma$) come from 
points in or between the vee-shaped crossover strips. 
As the $\chi$-nucleon scattering cross sections 
are smaller for $\mu<0$ than for $\mu>0$~\footnote{For this reason, the upper limit on the scalar cross section is larger for $\mu>0$ than for $\mu<0$, resulting in the lower ``shadow'' edge in panel (b).}, the cross sections in these excluded regions are lower 
by more than an order of magnitude than those in regions with $\mu>0$ that are still 
allowed. Even between the crossover strips, at lower $| \mu |$ where the 
relic density is below the WMAP range, the scaled cross sections are typically a few 
$\times 10^{-8}$ pb. The largest cross sections shown in panel (b) of 
Fig.~\ref{fig:1muvMlowtb} come from points where the relic density is within the WMAP 
range (and therefore need no rescaling) and at low $m_{1/2}$, and a few of these points 
are in fact already excluded by CDMS~II. The funnel region, as has been previously noted, 
gives rise to rather low spin-independent neutralino-nucleon cross sections, even at low $m_\chi$.
A XENON100-type detector would be sensitive to the entire crossover strip region shown 
here for $\mu>0$, and much of that for $\mu<0$, if neutralinos make up all the 
dark matter. We note in Fig.~\ref{fig:1muvMlowtb}(b) the existence of points with
100~GeV$ < m_\chi < $200~GeV with spin-independent cross sections $\sim 10^{-11}$~pb,
much smaller than those found in the CMSSM for this range of masses.

In panels (c) and (d) of Figure~\ref{fig:1muvMlowtb}, for $\tanb = 50$ and 
$m_0 = 500$~GeV, the stau LSP region has intruded so far into the parameter space 
that the coannihilation strip at its boundary joins the crossover strip to the rapid-annihilation 
funnel.  The funnel itself is deformed and broadened significantly, leaving a significant 
region within the funnel walls where the relic density of neutralinos is below the WMAP 
range. Much of the funnel is favoured by $g_\mu - 2$.
The CMSSM contour crosses regions of acceptable relic density at very low $\mu$ 
and $m_{\chi}$, in an area excluded by $b \goto s \gamma$ and the Higgs mass constraint, 
and again near the boundary of the stau LSP region, where $m_{\chi} \sim 400$ GeV.
The largest cross sections, which occur at low $m_{\chi}$, are excluded both by
$b \goto s \gamma$ and by present limits on direct detection cross sections~\footnote{The
range of $m_\chi$ in panel (d) is restricted by the range of $m_{1/2}$ shown in panel (c).}.
While some of 
the coannihilation strip may be probed in the future, cross sections in the focus-point 
region become as low as $\sim 10^{-10}$ pb when the relic density is within the WMAP 
range. In panel (c), the large value of $\tan \beta$ again leads to a sizeable contribution to 
$\bmm$.  Points below the blue dotted curve are above the current experimental
constraint of $5 \times 10^{-8}$ \cite{bmm}. 

\begin{figure}[ht!]
\begin{wide}{-1in}{-1in}
\begin{center}
\vskip -1.4in
\hskip .6in
\resizebox{0.55\textwidth}{!}{\includegraphics{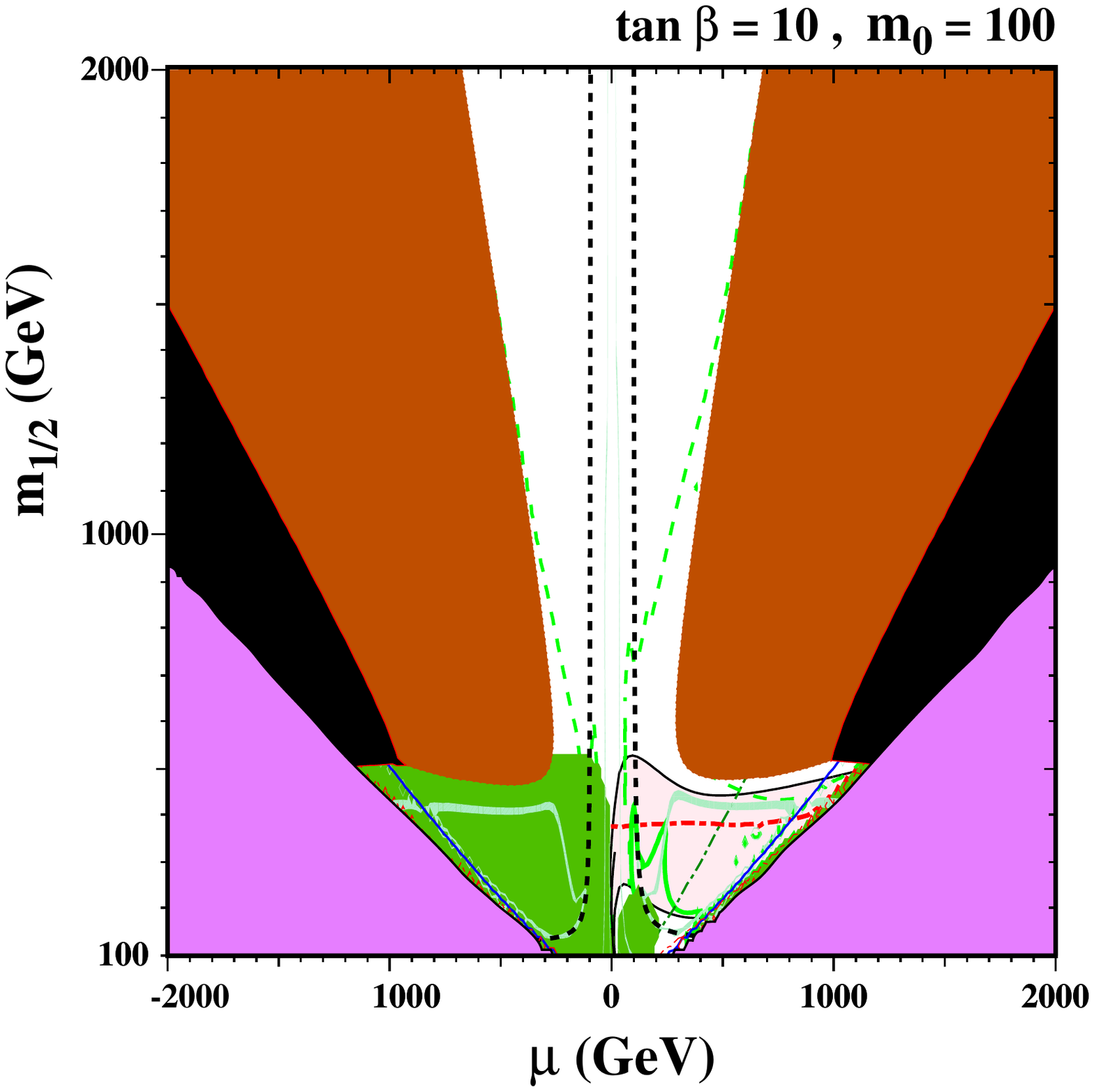}}
\hskip -.6in
\resizebox{0.55\textwidth}{!}{\includegraphics{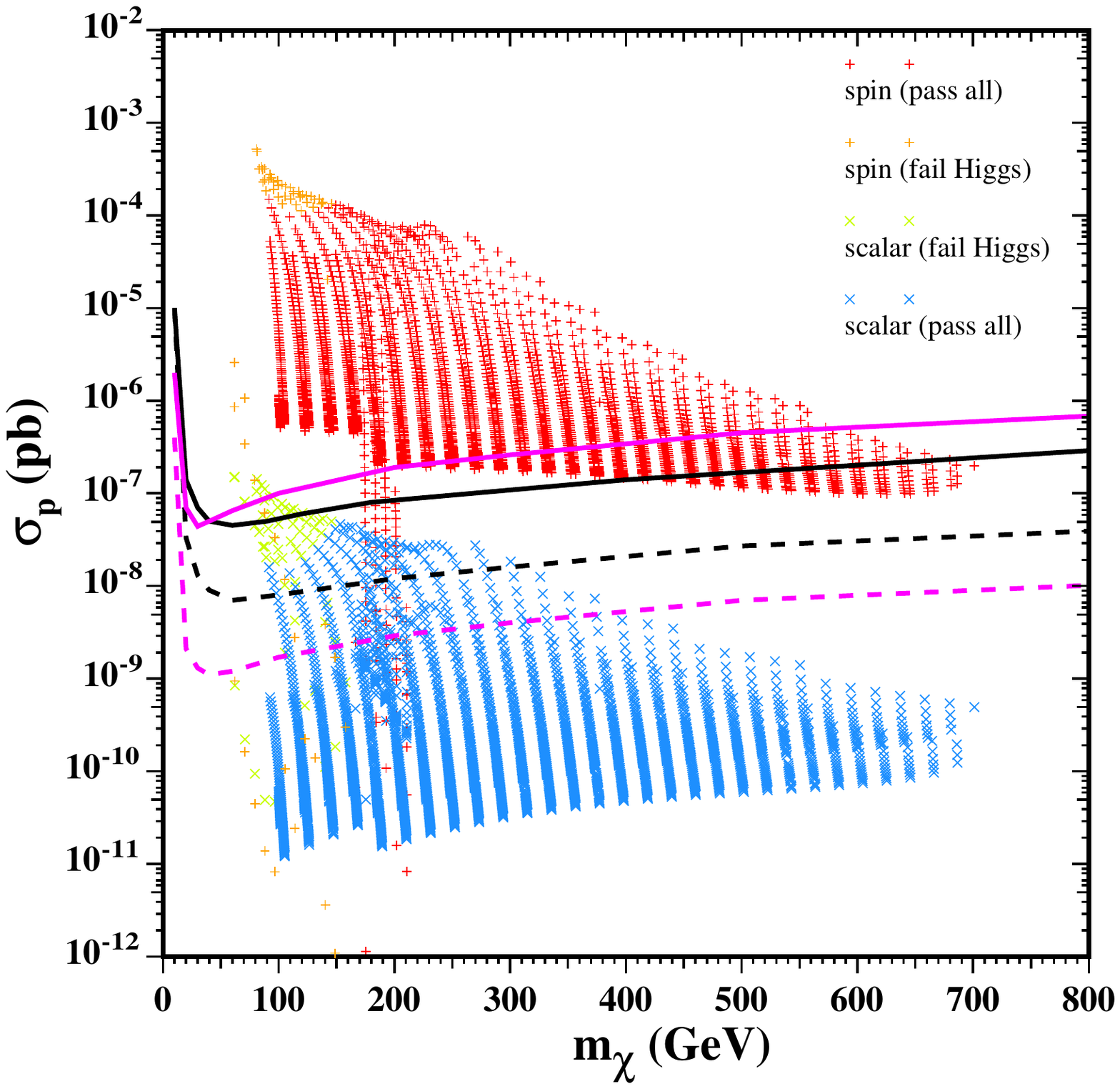}}
\end{center}
\end{wide}
\caption{\it Panel (a) shows the NUHM1 $(\mu,m_{1/2})$ plane for $m_0=100$ GeV and
$\tanb=10$. Panel (b) shows the corresponding neutralino-nucleon elastic scattering 
cross sections as functions of neutralino mass.
\label{fig:1muvMlowm0}}
\end{figure}

In the $(\mu, m_{1/2})$ plane for $\tanb = 10$ and $m_0 = 100$~GeV, 
shown in Fig.~\ref{fig:1muvMlowm0}(a),
a combination of the electroweak vacuum constraint, the $\chi$ LSP constraint and
$b \to s \gamma$ excludes almost all the half-plane with $\mu < 0$. On the other hand,
more points survive for $\mu > 0$, in the region bounded by the $\chi$ LSP
constraint, the LEP Higgs constraint and the LEP chargino constraint. 
The surviving points have spin-independent cross sections below the current
upper limits, though many points with $m_\chi < 400$~GeV should be accessible to
future experiments. We note that there are a few points with $m_h < 114$~GeV
that are already excluded by XENON10 and particularly CDMS~II. On the other
hand, there are some points in panel (b) with very low scalar cross sections
which arise from regions where the relic density is rather low (eg. in most
of the region above the WMAP strip found at $m_{1/2} = 450$ GeV) and the 
cross section has been scaled.  There is also a very sharp downturn 
in the spin-dependent cross section
at $m_\chi \sim 200$~GeV with values much smaller than in the CMSSM.
This is due to a cancellation 
similar to the one seen in Fig.~\ref{fig:1mAvMlowtb} between the up, down
and strange contributions.
 
\subsection{Summary}

\begin{figure}[h!]
\begin{wide}{-1in}{-1in}
\begin{center}
\vskip -1.4in
\hskip .6in
\resizebox{0.55\textwidth}{!}{\includegraphics{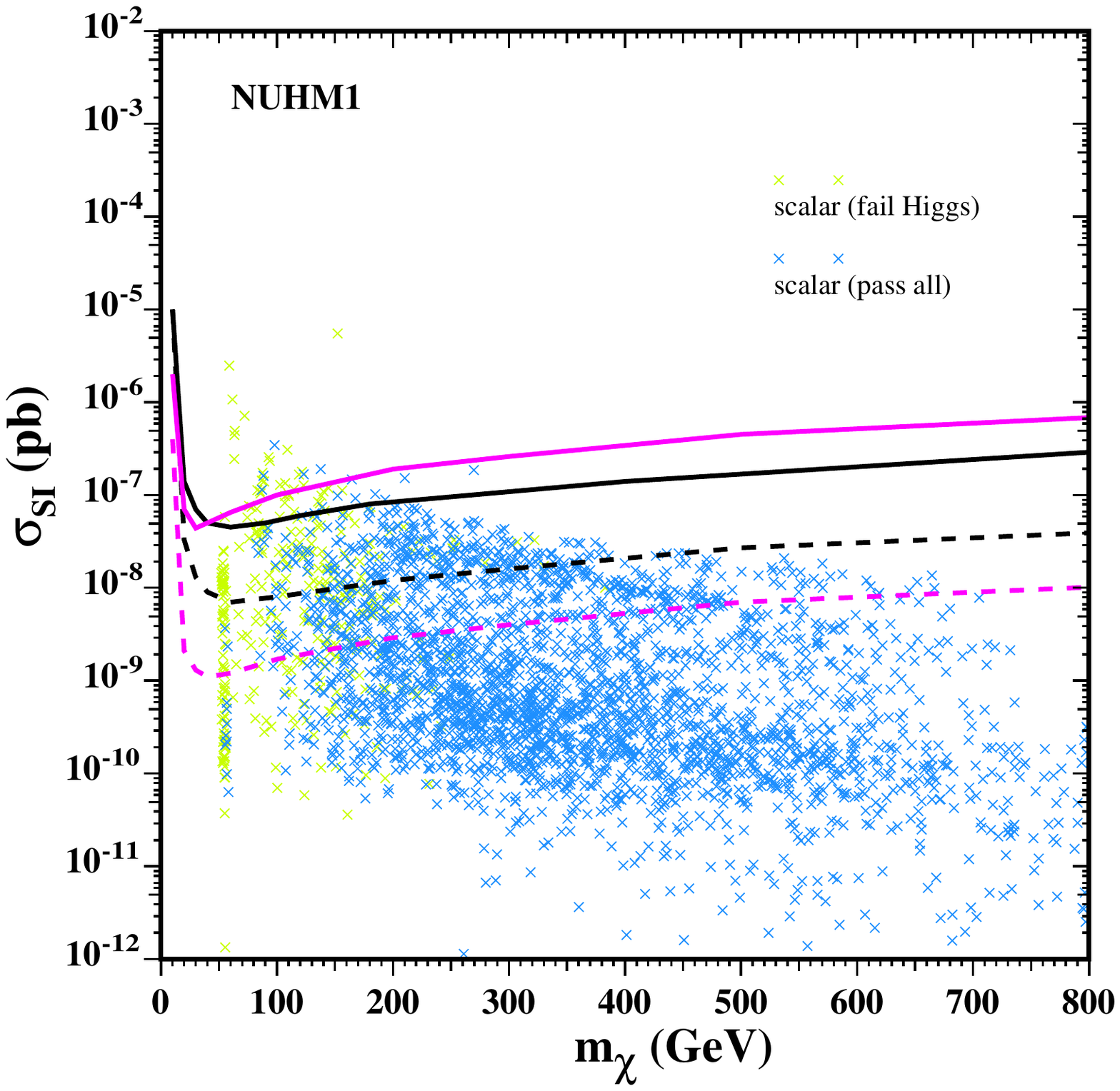}}
\hskip -.6in
\resizebox{0.55\textwidth}{!}{\includegraphics{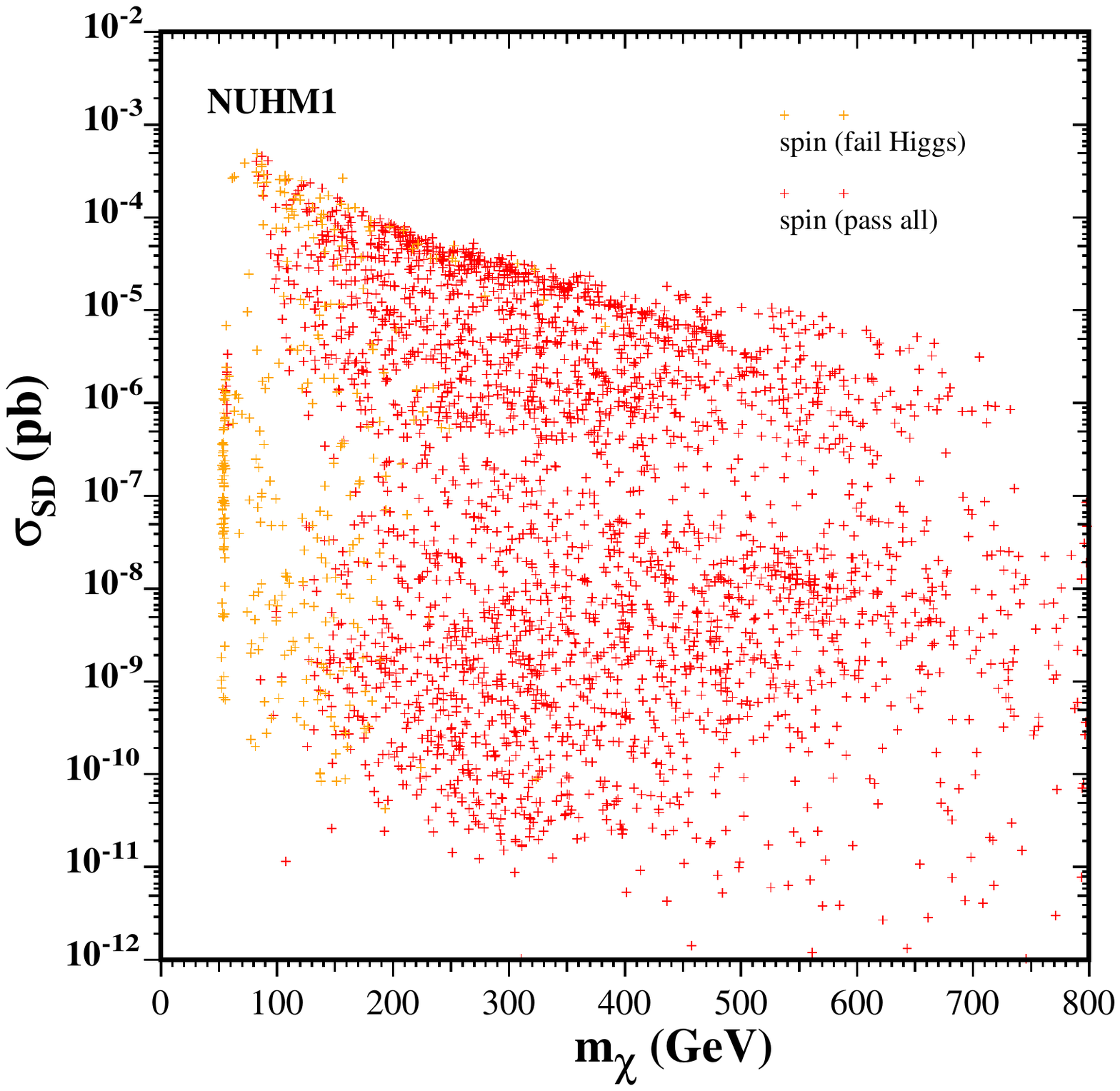}}
\end{center}
\end{wide}
\caption{\it Panels (a) and (b) show the entire potential ranges in the NUHM1 of 
the scalar and spin-dependent neutralino-nucleon cross sections, respectively, 
as functions of neutralino mass.  In both plots, we scan $5 \leq \tanb \leq 55$, 
0 $\leq m_{1/2} \leq 2000$ GeV, 100 GeV $\leq m_0 \leq 2000$ GeV, and
$-3 m_{1/2} \leq A_0 \leq 3 m_{1/2}$. The common GUT-scale 
value of $m_1 = m_2$ is in the range $(-2000,2000)$ GeV. 
\label{fig:NUHM1summary}}
\end{figure}

Fig.~\ref{fig:NUHM1summary} is a pair of scatter plots displaying the potential ranges
of (a) the spin-independent and (b) the spin-dependent dark matter cross sections in
the NUHM1. Comparing with the corresponding plots for the CMSSM in
Fig.~\ref{fig:cmssm}, we note that the spin-independent cross section in the NUHM1 may be
up to an order of magnitude larger for $m_\chi > 300$~GeV. We also note the
appearance of points in the NUHM1 with $m_\chi < 200$~GeV and low spin-independent
cross sections $\sim 10^{-10}$~pb. Similar features are present for the spin-dependent
cross sections: in the NUHM1 this may be $\sim 10^{-6}$~pb for $m_\chi > 500$~GeV,
whereas values in the CMSSM are an order of magnitude lower. Also, the NUHM1
allows the possibility of much lower spin-dependent cross sections for 
$m_\chi < 300$~GeV than are attained in the CMSSM.

The points with large cross sections at large $m_\chi$ are generally those
with relatively large Higgsino components. In the CMSSM, this is possible
only in focus-point regions that have relatively small $m_\chi$. However, in
the NUHM1 there are focus-point regions extending to larger $m_\chi$,
and there are also other cross-over regions where the LSP has a relatively
large Higgsino component. Relatively low cross sections may occur in the
NUHM1 at points with relatively large values of $m_0$ or in the funnel
regions due to the scaling applied when the relic density is small, as was illustrated
in some of the sample planes discussed earlier.

\section{NUHM2 Models}

As already discussed, the NUHM2 has two parameters in addition to those
already present in the CMSSM, which may be transposed into free choices
of both the quantities $m_A$ and $\mu$. The relatively large number renders
complex a systematic survey of the NUHM2 parameter space. We restrict
ourselves here to studies of a few parameter planes whose
features we compare with the CMSSM and NUHM1.  See \cite{EFlOSo},
for further discussions of direct detection cross sections in the NUHM2.

\subsection{Sample $(m_{1/2}, m_0)$ Plane}

\begin{figure}[h]
\begin{wide}{-1in}{-1in}
\begin{center}
\vskip -1.4in
\hskip .6in
\resizebox{0.55\textwidth}{!}{\includegraphics{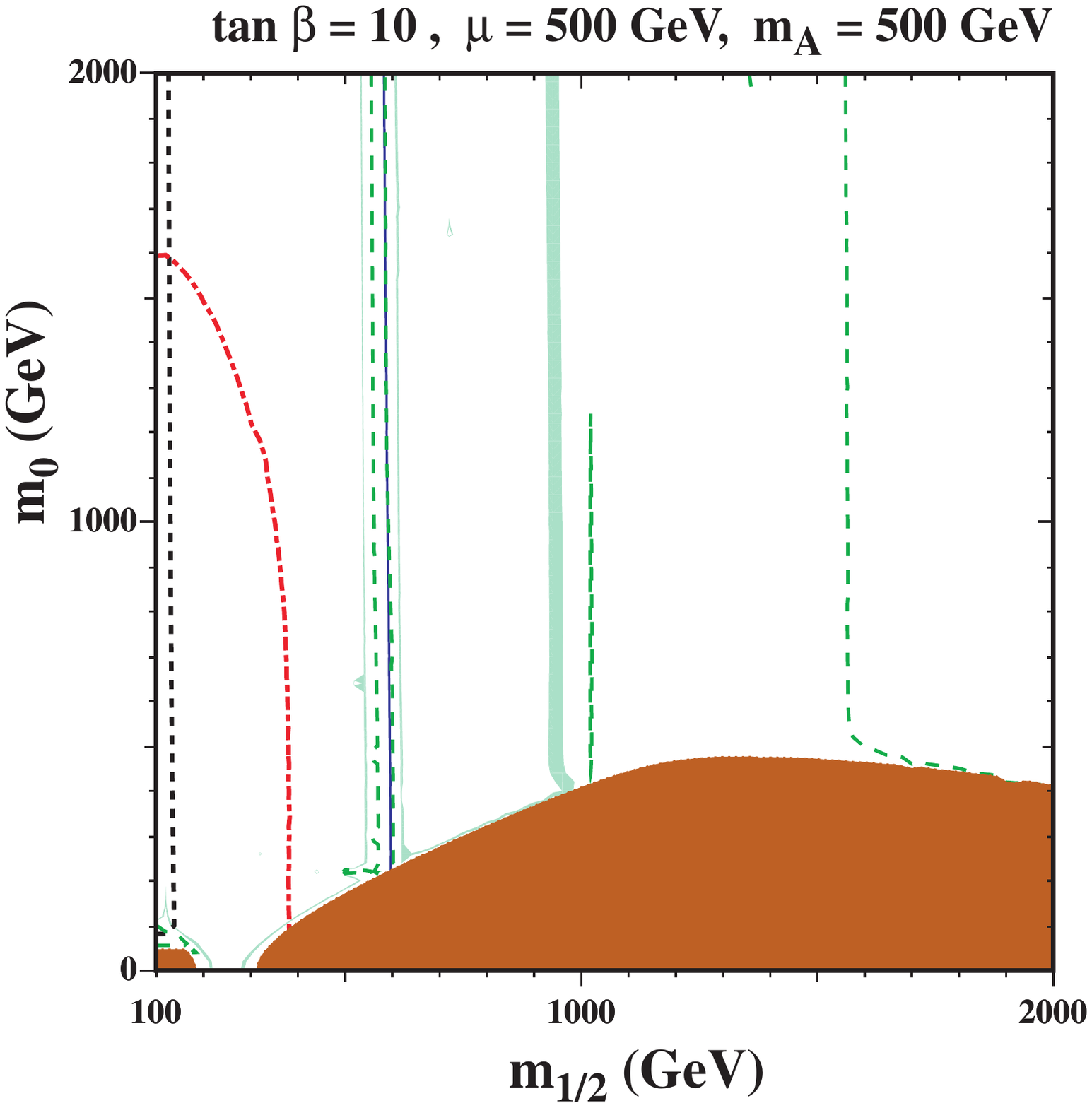}}
\hskip -.6in
\resizebox{0.55\textwidth}{!}{\includegraphics{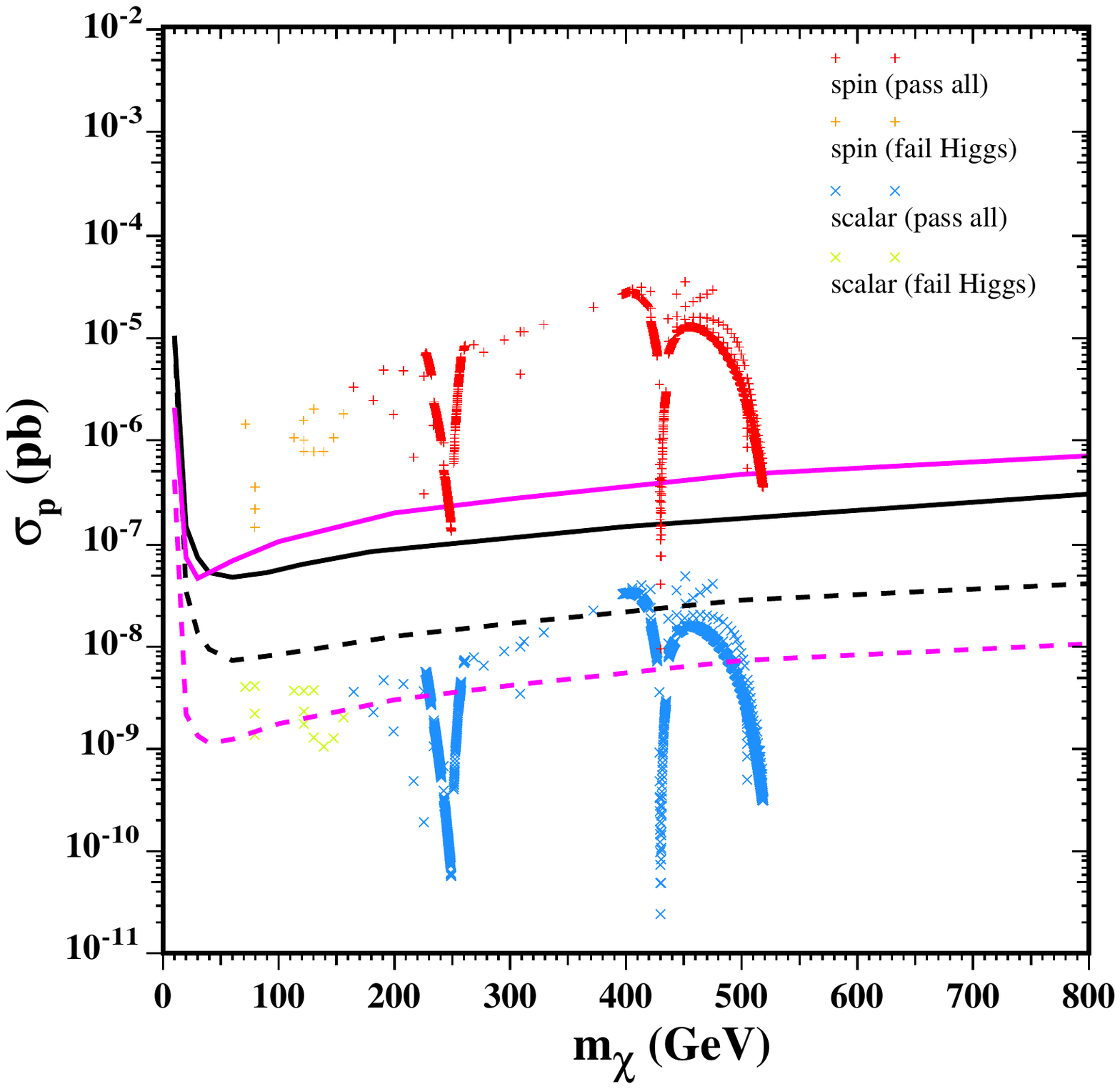}}
\end{center}
\end{wide}
\caption{\it Panel (a) shows the NUHM2 $(m_{1/2},m_0)$ plane for $m_A=500$ GeV, 
$\mu=500$ GeV and $\tanb=10$. Panel (b) shows the corresponding neutralino-nucleon 
elastic scattering cross sections as functions of neutralino mass.
\label{fig:NUHM2m12m0}}
\end{figure}

We first display in Fig.~\ref{fig:NUHM2m12m0} a sample $(m_{1/2}, m_0)$ plane
with $\tanb = 10$ and fixed $m_A = 500$~GeV and $\mu = 500$~GeV, 
which reveals a couple of new features. As in the CMSSM, there is a region in panel (a)
at large $m_{1/2}$ and small $m_0$ which is forbidden because the lighter stau
is the LSP. Just above this forbidden region, as in the CMSSM, there is a 
stau-coannihilation strip. However, jutting up from this strip at $m_{1/2} \sim 600$~GeV
and $\sim 950$~GeV, there are vertical strips where the relic $\chi$ density falls within
the WMAP range. The double strips at $m_{1/2} \sim 600$~GeV form a rapid-annihilation
funnel on either side of the line (indicated in solid blue) where $m_A = 2 m_\chi$.
Such funnels appear only at large $\tanb$ in the CMSSM, but the freedom to choose
different values of $m_A$ in the NUHM2 permits the appearance of a rapid-annihilation
funnel also at the low value $\tanb = 10$ shown here. The other vertical WMAP strip
appears because, as $m_{1/2}$ increases relative to $\mu$ which is fixed here, 
the Higgsino fraction
in the lightest neutralino $\chi$ increases, which in turn increases the
annihilation rate and hence decreases the relic density, resulting in this case in 
a crossover region when $m_{1/2} \sim 950$~GeV.
At slightly higher $m_{1/2} = 1020$ GeV, there is a rapid coannihilation pole
(through $Z$-exchange)
between the lightest and next-to-lightest neutralinos resulting in a narrow region 
with suppressed relic density and hence a suppressed scalar cross section.

These novel regions are clearly visible in panel (b) of Fig.~\ref{fig:NUHM2m12m0}.
The elastic scattering cross sections do not vary rapidly as $m_\chi$ increases, until
the funnel region at $m_\chi \sim 250$~GeV is reached. The vee-shaped suppression
in the cross section arises from the increasing value of $m_0$ as one rises up the funnel.
The two sides of the funnel approach each other as $m_0$ increases, eventually
joining together and resulting in a minimum value $\sigma_{\rm SI} \sim 5 \times 10^{-11}$~pb
where the two sides of the funnel meet (at a value of $m_0 > 2000$~GeV, and hence
invisible in panel (a) of Fig.~\ref{fig:NUHM2m12m0}). After this excitement, the cross
section continues to rise gradually as one follows the coannihilation strip, until the
crossover strip is reached at $m_\chi \sim 400$~GeV. Here the cross section decreases
again as $m_0$ increases, to values even smaller than in the rapid-annihilation funnel,
before rising again and finally declining towards the end of the coannihilation strip.
Note that the entire region to the right of the transition strip is viable, albeit with
a relic density below the WMAP range.  The cross section will thus be reduced
due to scaling.  Because the neutralino is predominantly a Higgsino here, 
its mass is given by $\mu$ rather than $m_{1/2}$ and so points at large $m_\chi$ 
end in panel (b) because of our choice of fixed $\mu$.

\subsection{Sample $(m_A, m_{1/2})$ Planes}

\begin{figure}[ht!]
\begin{wide}{-1in}{-1in}
\begin{center}
\vskip -1.4in
\hskip .6in
\resizebox{0.55\textwidth}{!}{\includegraphics{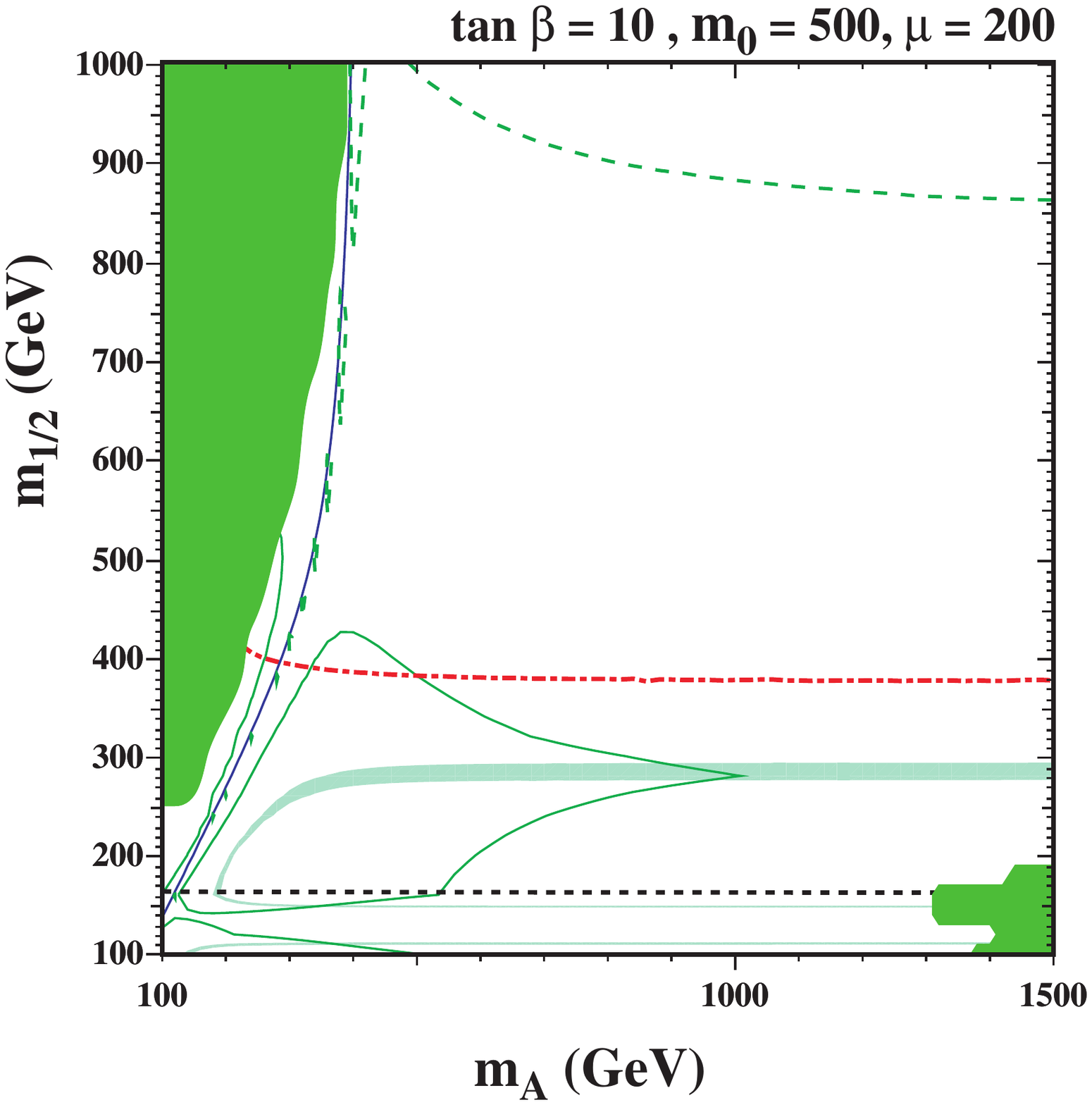}}
\hskip -.6in
\resizebox{0.55\textwidth}{!}{\includegraphics{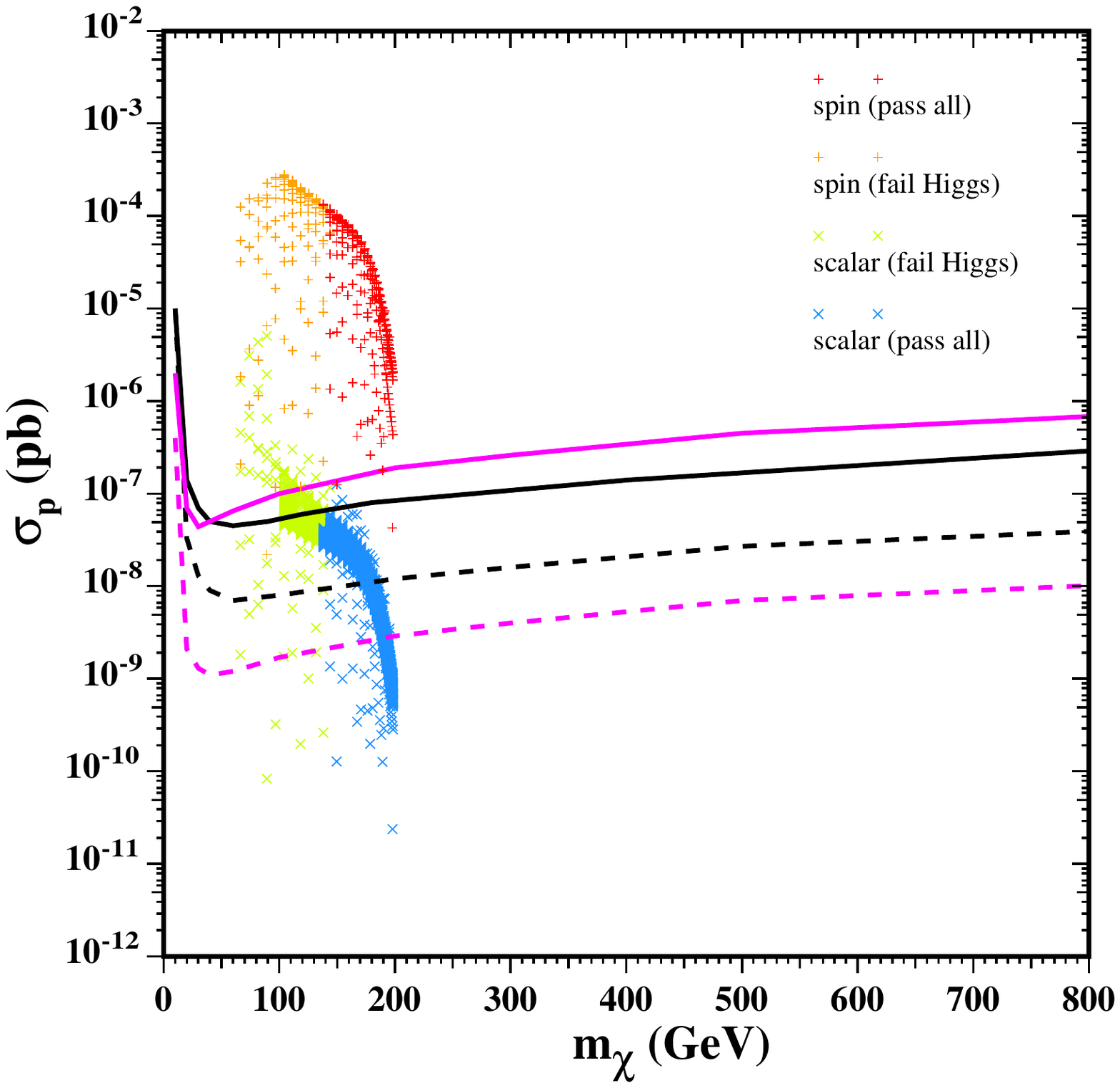}}
\vskip -1.8in
\hskip .6in
\resizebox{0.55\textwidth}{!}{\includegraphics{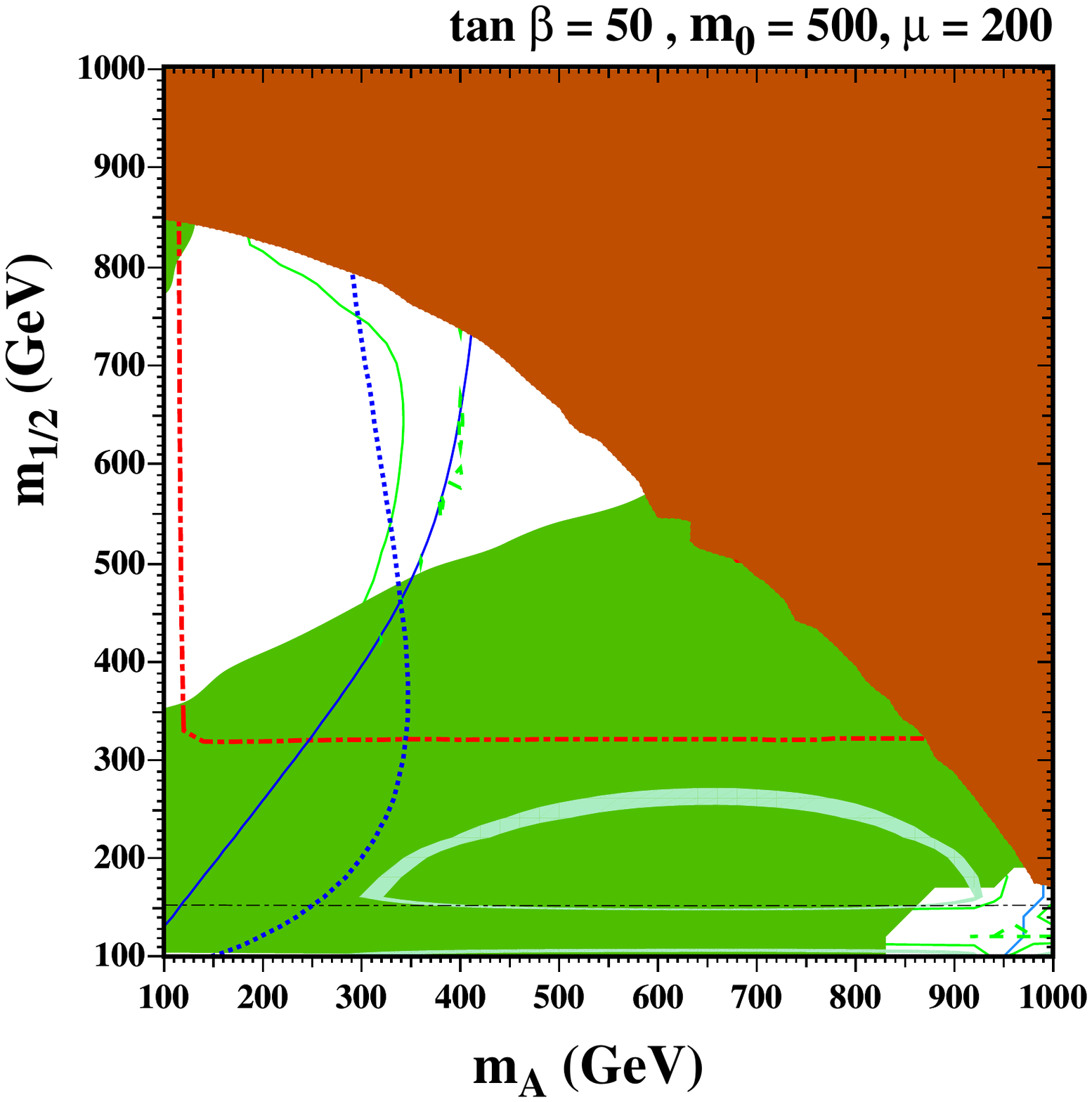}}
\hskip -.6in
\resizebox{0.55\textwidth}{!}{\includegraphics{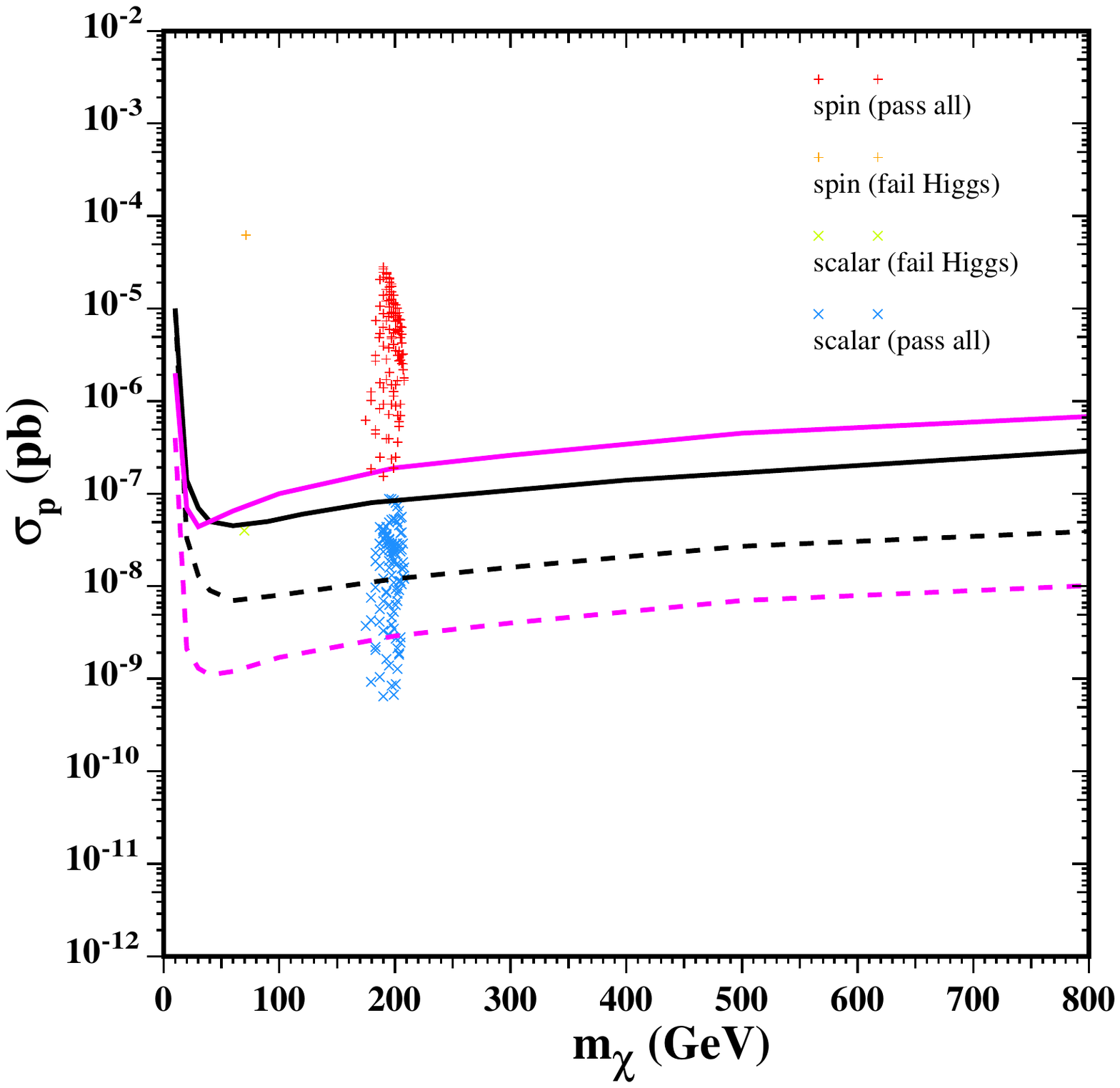}}
\end{center}
\end{wide}
\caption{\it Panels (a) and (c) show the NUHM2 $(m_A,m_{1/2})$ planes for $m_0=500$ GeV, $\mu=200$ GeV, $\tanb=10$ and 50. Panels (b) and (d) show the corresponding neutralino-nucleon elastic scattering cross sections as functions of neutralino mass.
\label{fig:NUHM2mA500200}}
\end{figure}

Fig.~\ref{fig:NUHM2mA500200} shows two NUHM2 $(m_A,m_{1/2})$ planes. We see in
panel (a) for $m_0 = 500$~GeV, $\mu = 200$~GeV and $\tanb = 10$ a strip with the
relic density in the WMAP range that extends to large $m_A$ at $m_{1/2} \sim 290$~GeV.
This is the transition strip, and above it, the relic density is always below the 
WMAP range. There are other strips below this, but they fall in a region 
where the chargino mass is below 104 GeV.
As seen from the location of the dash-dotted red line,
all of these cosmologically-preferred strips have $m_h < 114$~GeV.
In this case, the funnel region shown by the solid blue line where $m_A = 2 m_\chi$, 
occurs past the transition region and there 
rapid annihilation further suppresses the relic density. It is clear from the
shape of this line that $m_\chi \sim \mu = 200$~GeV for larger $m_{1/2}$.
We see in panel (b) that the low-$m_h$ points typically have spin-independent
cross sections $\sim 10^{-7}$~pb, and are largely excluded by the XENON10 and
CDMS~II experiments. On the other hand, essentially all the points with $m_h > 114$~GeV
survive the direct dark matter search experiments, so far, in particular because the
effective cross section is suppressed by the small relic density.

Turning to the corresponding
$(m_A,m_{1/2})$ plane for $\tanb = 50$, we see a large region at large $m_A$ and/or 
$m_{1/2}$ that is excluded because the LSP is charged. There is also a large region
at smaller $m_{1/2}$ that is excluded by the $b \to s \gamma$ constraint . 
At large $\tanb$, the constraint from $\bmm$ is important, excluding values
of $m_A \la 300$ GeV in this case, as shown by the dotted blue curve.
Once again,
although the funnel region, shown by
the solid blue line where $m_A = 2 m_\chi$,  has $m_h > 114$~GeV in between the shaded regions and to the right of the $\bmm$ constraint, 
the relic density is small as this occurs past the transition region (which here is
excluded by $b \to s \gamma$) and the neutralino has a large Higgsino component.
As seen in panel (d) of Fig.~\ref{fig:NUHM2mA500200}, all the allowed points have 
$m_\chi \sim \mu = 200$~GeV. Typical spin-independent cross sections are somewhat
larger than in the case $\tanb = 10$, shown previously in panel (b).

\begin{figure}[ht!]
\begin{wide}{-1in}{-1in}
\begin{center}
\vskip -1.4in
\hskip .6in
\resizebox{0.55\textwidth}{!}{\includegraphics{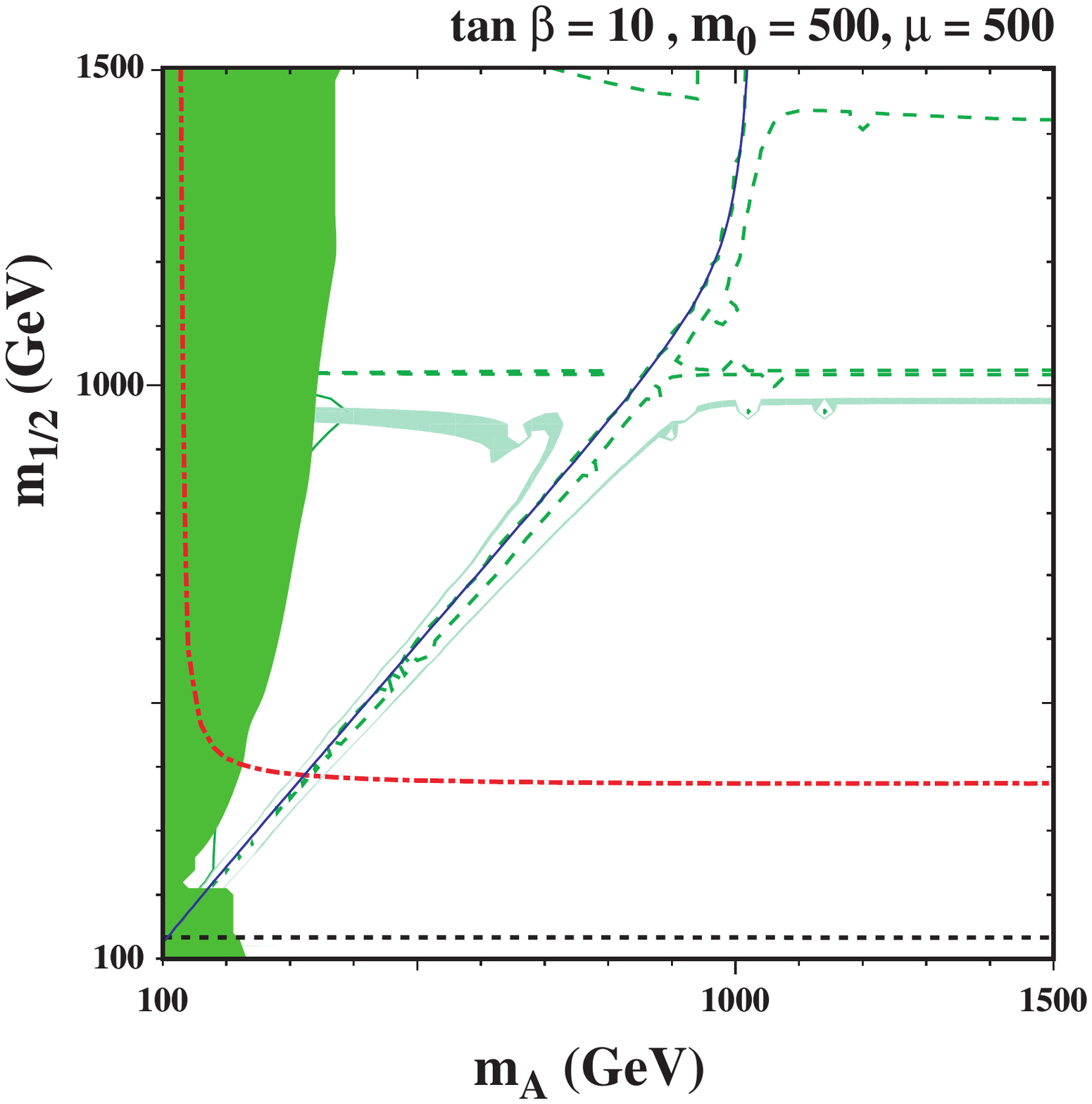}}
\hskip -.6in
\resizebox{0.55\textwidth}{!}{\includegraphics{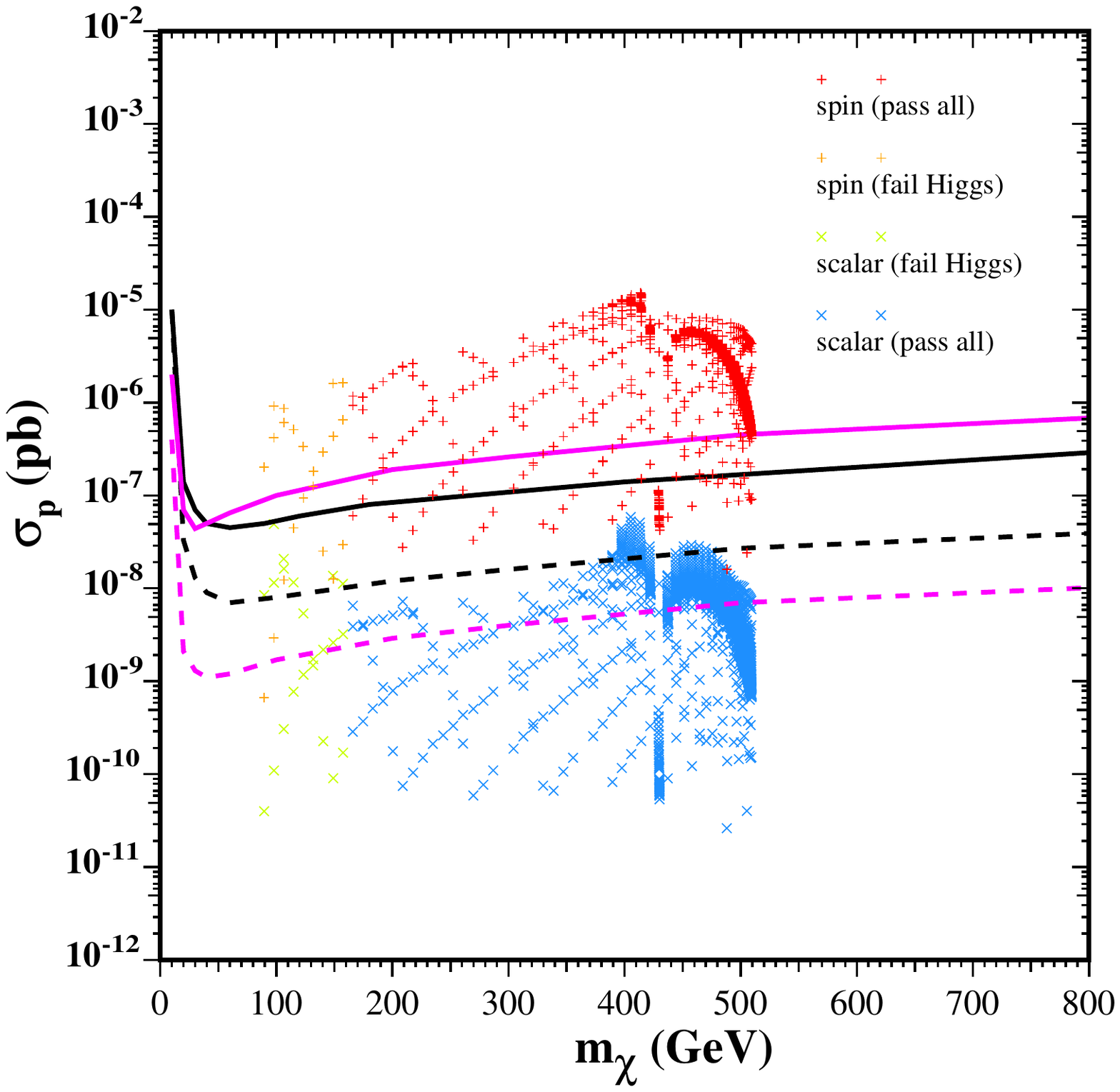}}
\vskip -1.8in
\hskip .6in
\resizebox{0.55\textwidth}{!}{\includegraphics{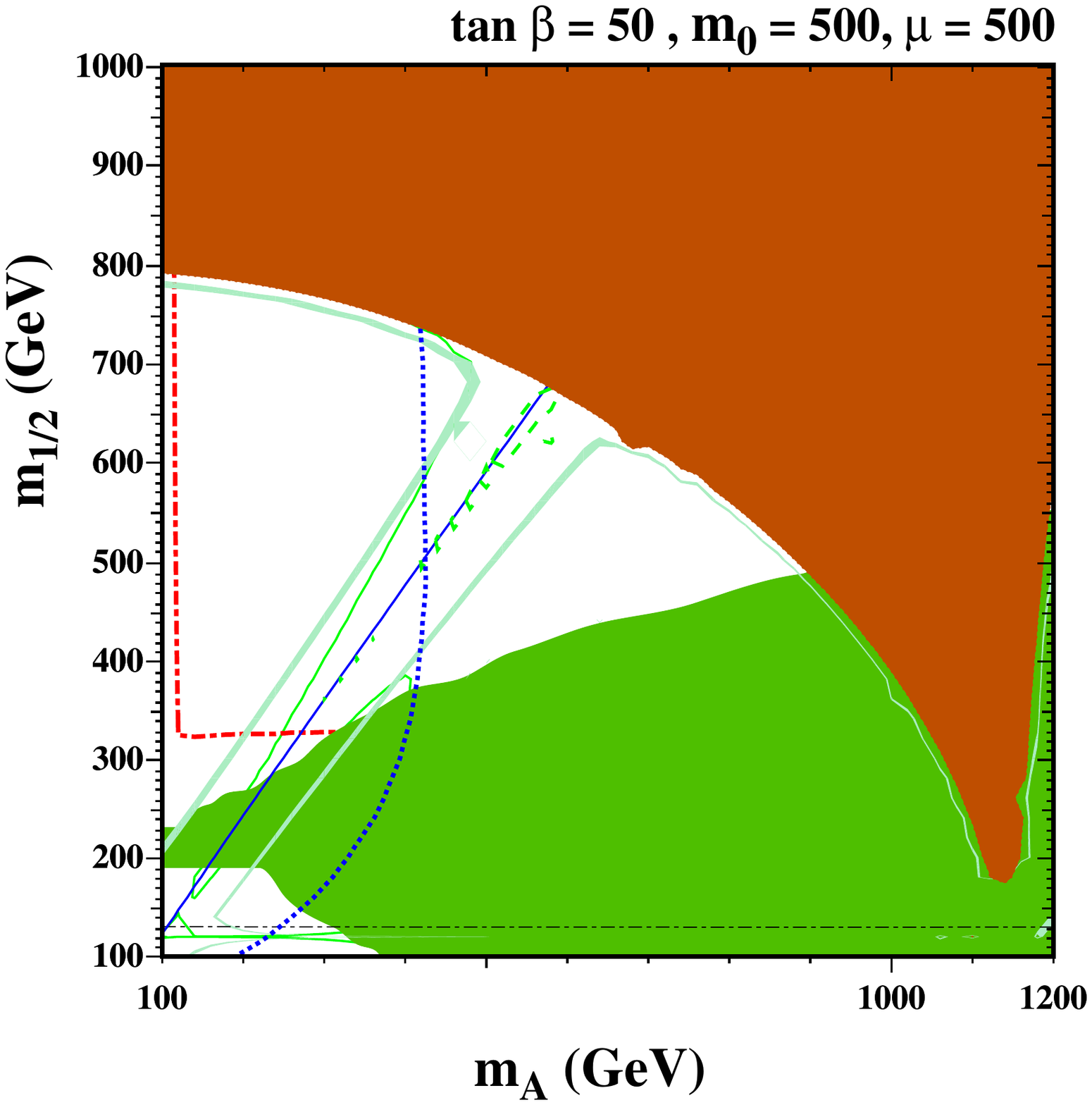}}
\hskip -.6in
\resizebox{0.55\textwidth}{!}{\includegraphics{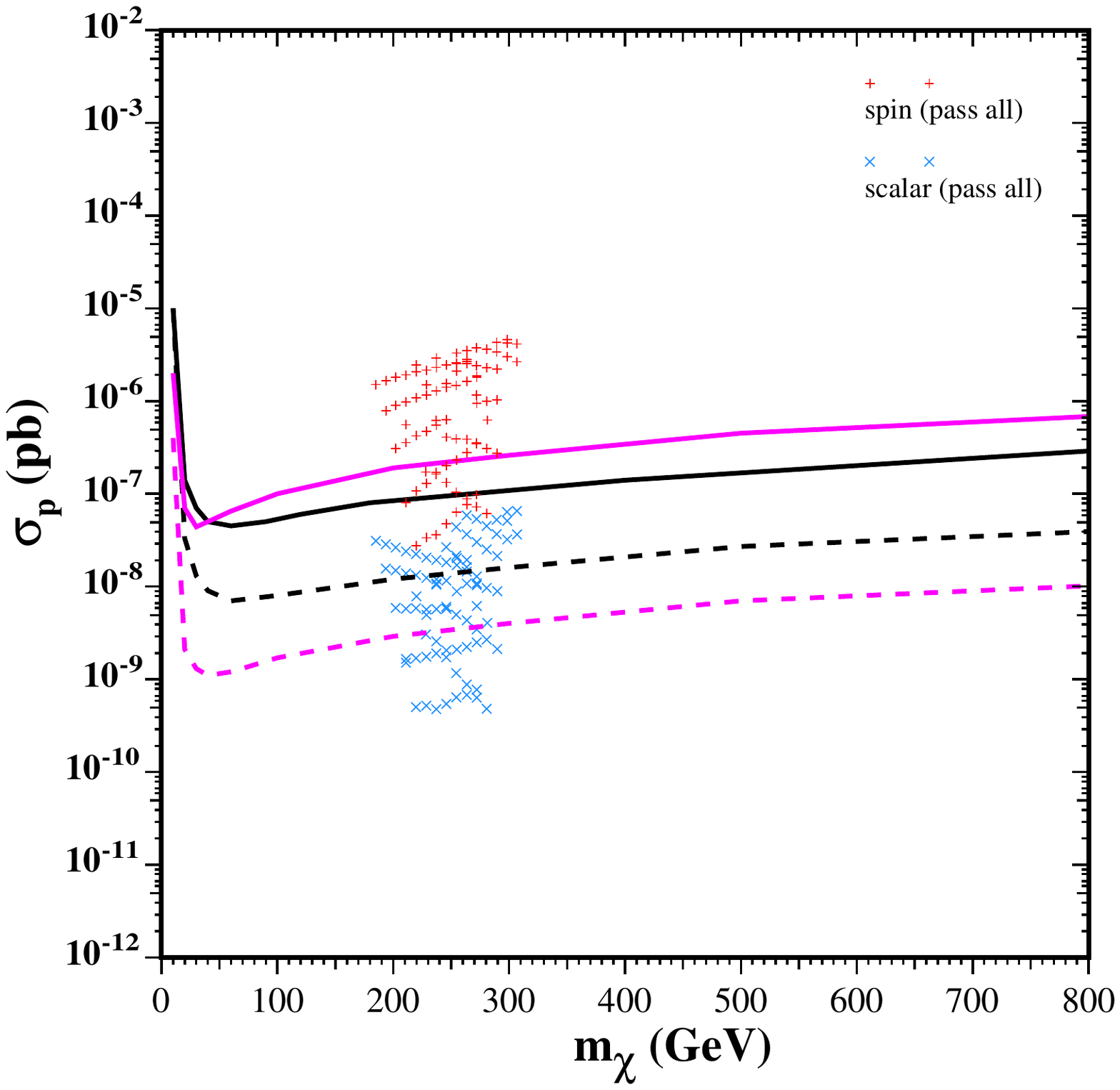}}
\end{center}
\end{wide}
\caption{\it Panels (a) and (c) show the NUHM2 $(m_A,m_{1/2})$ planes for $m_0=500$ GeV, $\mu=500$ GeV, $\tanb=10$ and 50. Panels (b) and (d) show the corresponding neutralino-nucleon elastic scattering cross sections as functions of neutralino mass.
\label{fig:NUHM2mA500500}}
\end{figure}

We see in Fig.~\ref{fig:NUHM2mA500500} the evolution of these features when 
$\mu = 500$~GeV, with the other parameters left unchanged. For $\tanb = 10$
in panel (a), we see that the transition strip at $m_{1/2} \approx 950$ GeV is now
split  by the two halves of a rapid-annihilation funnel. 
In contrast to Fig.~\ref{fig:NUHM2mA500200}, a portion of the funnel now
lies below the transition strip and the WMAP density can be realized.
Above the crossover strips, at larger $m_{1/2}$, the LSP becomes Higgsino-like.
Only small portions of the WMAP strips have $m_h < 114$~GeV.
We see in panel (b) that the scatter plot of the spin-dependent and spin-independent cross sections has a
feature at $m_\chi \sim 430$~GeV, corresponding to the opening out of the
rapid-annihilation funnel. The points at larger $m_\chi$ have relatively large
spin-independent cross sections, higher than in the CMSSM for similar values
of $m_\chi$, reflecting the Higgsino nature of the LSP in these cases.

Turning now to panel (c) of Fig.~\ref{fig:NUHM2mA500500}, for $\tanb = 50$,
we see again the increased importance of the neutralino LSP and $b \to s \gamma$
constraints and the constraint from $\bmm$ excluding small values of $m_A$. 
We also see the WMAP strips on either side of the solid blue line 
where $m_A = 2 m_\chi$, and notice that although almost all of them have $m_h > 114$~GeV
only the upper part of the funnel is allowed by $\bmm$.
The would-be transition strip has been pushed into the region with a charged LSP.
The corresponding values of $m_\chi \sim 250$~GeV, as seen in panel (c) of
Fig.~\ref{fig:NUHM2mA500500}, and the spin-independent cross section may be
as large as $\sim 10^{-7}$~pb, namely considerably larger than in the CMSSM,
but decreasing for points with a low relic density.

\subsection{Sample $(\mu,m_{1/2})$ Planes}

\begin{figure}[htb]
\begin{wide}{-1in}{-1in}
\begin{center}
\vskip -1.4in
\hskip .6in
\resizebox{0.55\textwidth}{!}{\includegraphics{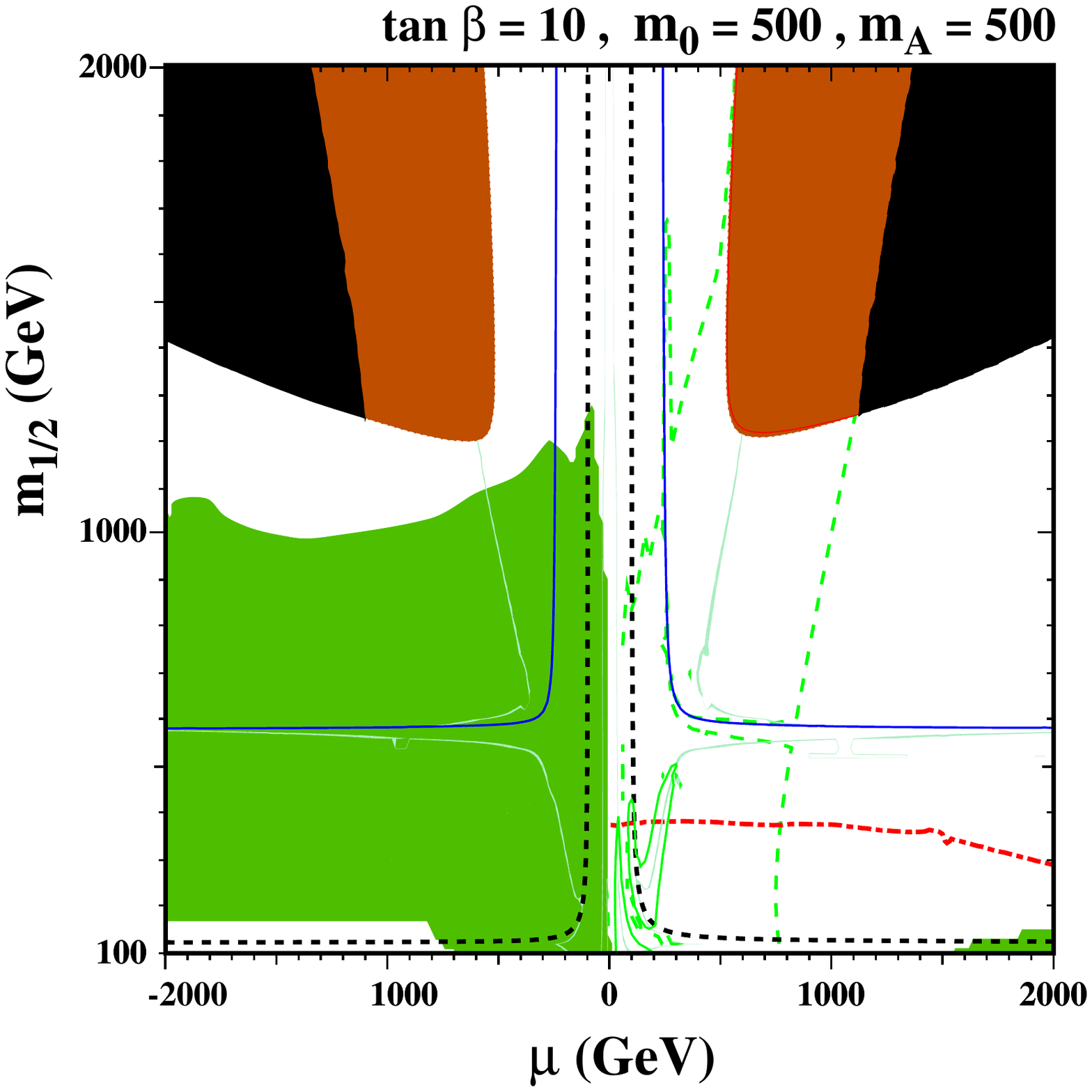}}
\hskip -.6in
\resizebox{0.55\textwidth}{!}{\includegraphics{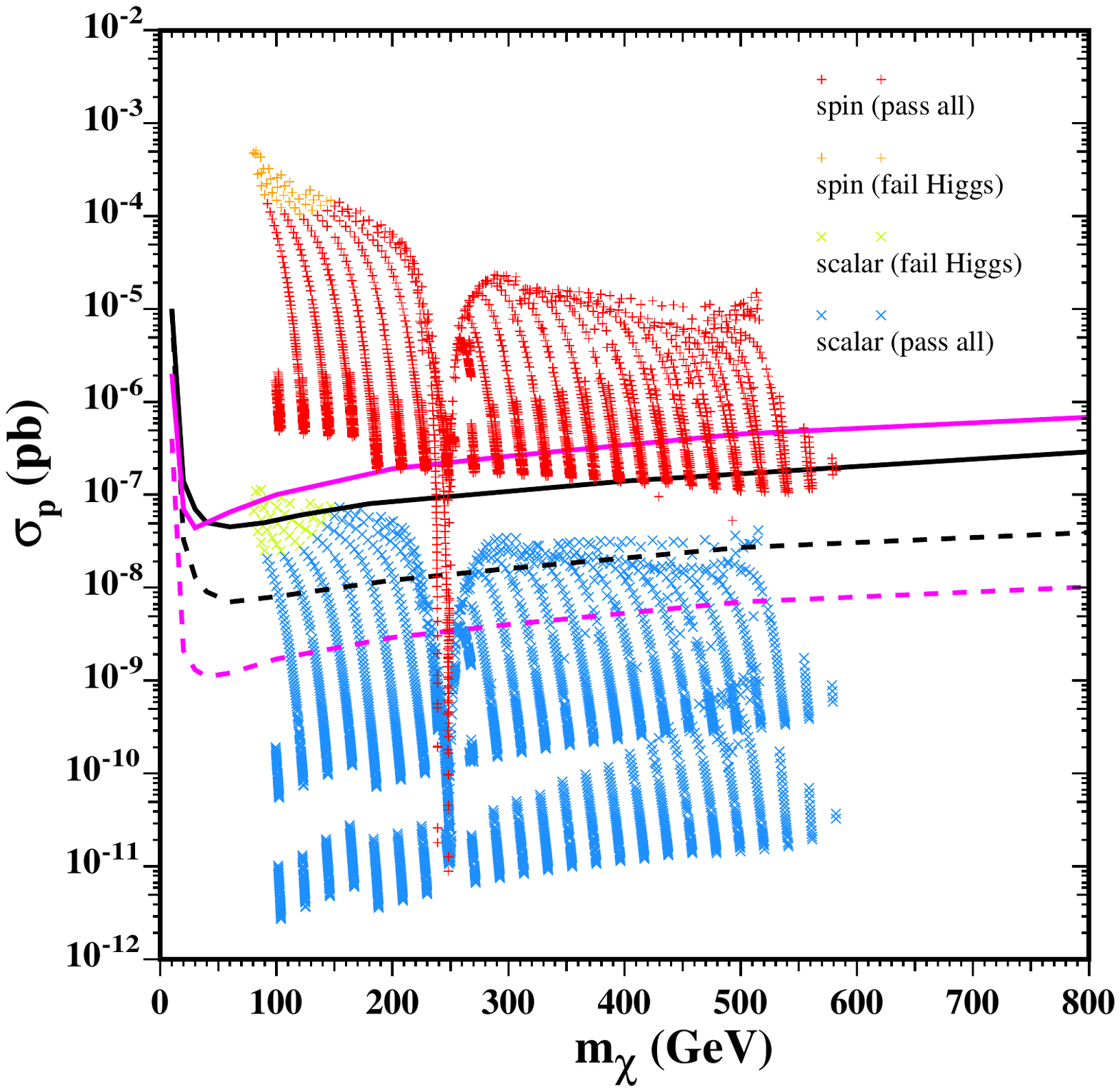}}
\end{center}
\end{wide}
\caption{\it Panel (a) shows the NUHM2 $(\mu,m_{1/2})$ plane for $m_0=500$ GeV, 
$m_A=500$ GeV and $\tanb=10$. Panel (b) shows the corresponding neutralino-nucleon 
elastic scattering cross sections as functions of neutralino mass.
\label{fig:NUHM2mu500500}}
\end{figure}

Fig.~\ref{fig:NUHM2mu500500}(a) displays the NUHM2 $(\mu,m_{1/2})$ plane 
for $m_0=500$ GeV, $m_A=500$ GeV and $\tanb=10$. Highly visible at large $m_{1/2}$
are regions excluded by the neutralino LSP requirement. 
In the dark (very dark) shaded regions, the LSP is the stau (selectron/smuon).
Much of the remaining $\mu < 0$
half-plane is excluded by $b \to s \gamma$, while the lower part of the $\mu > 0$
half-plane with $m_{1/2} < 300$~GeV has $m_h < 114$~GeV. Apart from a very small
section of a crossover strip near $\mu \sim - 500$~GeV and $m_{1/2} \sim 1200$~GeV,
the WMAP density range is attained only in the $\mu > 0$ half-plane. This occurs in a
rapid-annihilation funnel on either side of the blue line where $m_A = 500$~GeV $= 2 m_\chi$,
and along a crossover strip extending to higher (and somewhat lower) $m_{1/2}$.

Panel (b) of Fig.~\ref{fig:NUHM2mu500500} displays the corresponding dark
matter scattering cross sections. We see that the spin-independent cross section
may be as high as $5 \times 10^{-8}$~pb for $m_\chi \sim 500$~GeV, to be
compared with a maximum of $\sim 10^{-9}$~pb in the CMSSM: this possibility
occurs for points towards the top of the crossover strip. At small $m_\chi$,
there are a few points with $m_h < 114$~GeV
that are excluded by the XENON10 and CDMS~II experiments, which occur at
the bottom end of the crossover strip. In between, for $m_\chi \sim 250$~GeV,
there is a suppression of the maximum spin-independent cross section,
corresponding to the rapid-annihilation funnel extending to large $\mu$.
Foreseen experiments should be able to cover all the WMAP strip except a
small portion of this funnel. On the other hand, there are many points with
lower effective cross sections, suppressed by the low relic density. These include
some at low $m_\chi < 200$~GeV with lower cross sections than those found in the
CMSSM.

\begin{figure}[htb]
\begin{wide}{-1in}{-1in}
\begin{center}
\vskip -1.4in
\hskip .6in
\resizebox{0.55\textwidth}{!}{\includegraphics{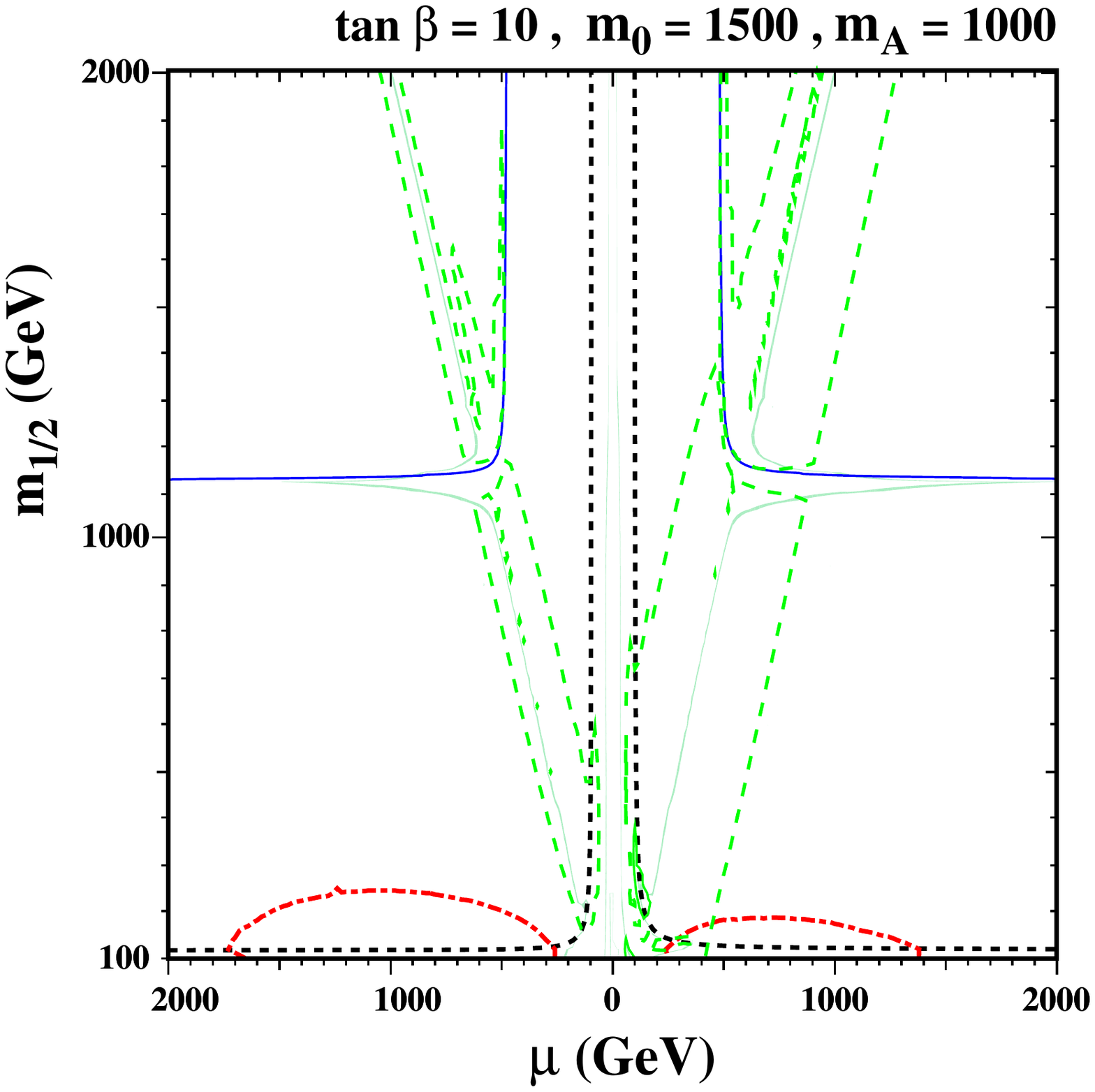}}
\hskip -.6in
\resizebox{0.55\textwidth}{!}{\includegraphics{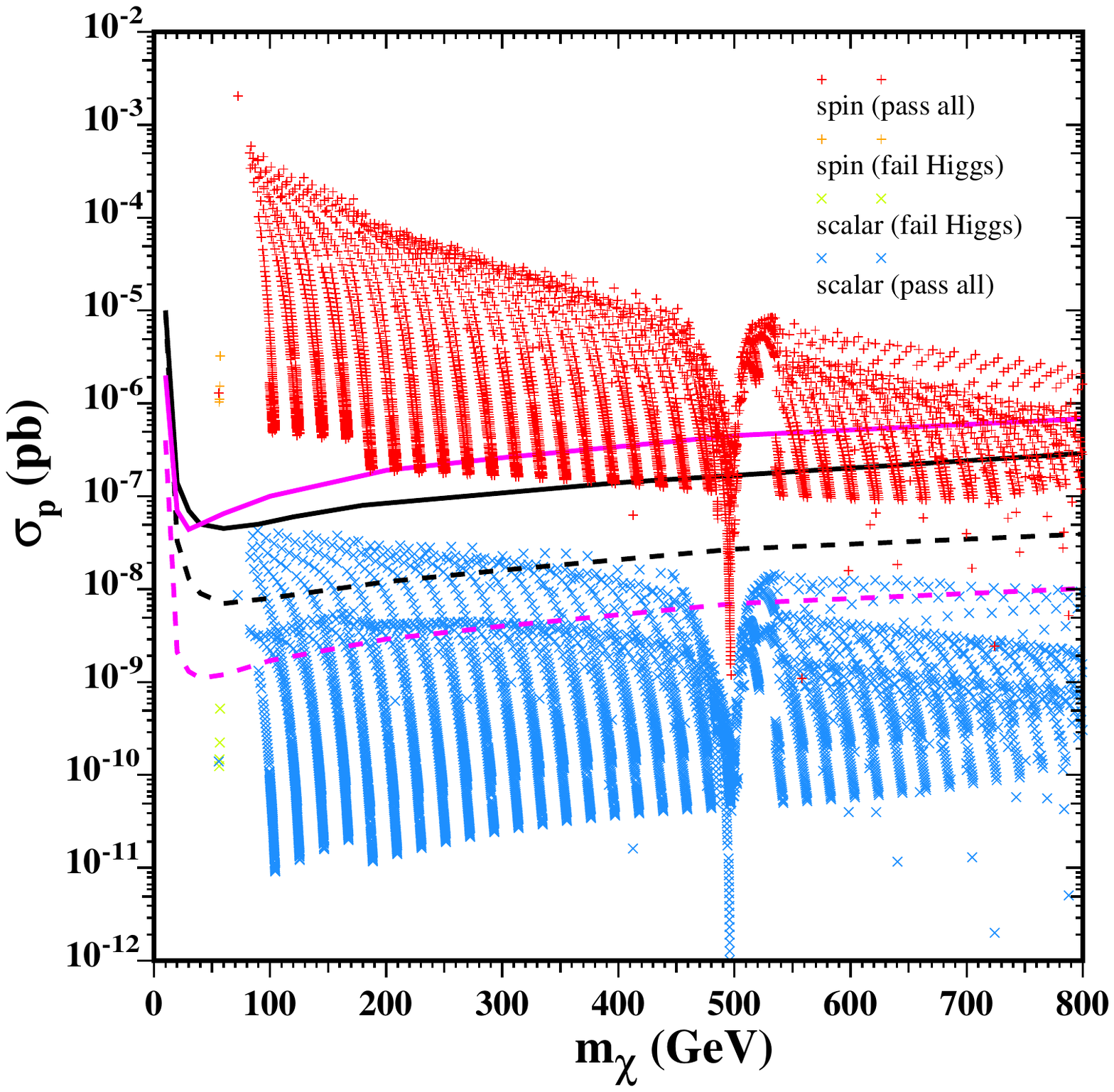}}
\end{center}
\end{wide}
\caption{\it Panel (a)  shows the NUHM2 $(\mu,m_{1/2})$ planes for $m_0=1500$ GeV, 
$m_A=1000$ GeV and $\tanb=10$. Panel (b) shows the corresponding neutralino-nucleon 
elastic scattering cross sections as functions of neutralino mass.
\label{fig:NUHM2mu15001000}}
\end{figure}

Fig.~\ref{fig:NUHM2mu15001000} displays similar NUHM2 $(\mu,m_{1/2})$ planes
for $m_0=1500$ GeV, $m_A=1000$ GeV and $\tanb=10$. In this case, 
we see in panel (a) that the neutralino
LSP and $b \to s \gamma$ constraints have no effect. Almost all the half-planes
for both signs of $\mu$ have $m_h > 114$~GeV, and the rapid-annihilation funnel has
risen to $m_{1/2} \sim 1100$~GeV, corresponding to the higher value $m_A = 1000$~GeV.
In addition to the rapid-annihilation funnels, almost all the crossover strips are allowed
for both signs of $\mu$. Panel (b) has features rather similar to those in panel (b)
of Fig.~\ref{fig:NUHM2mu500500}, with relatively large spin-independent cross sections
$\sim 10^{-8}$~pb possible for all $m_\chi < 800$~GeV, along the crossover strips,
and dips around $m_\chi = 500$~GeV, corresponding to the rapid-annihilation
funnels. As in the previous figure, low effective cross sections are again possible for points
with low $m_\chi$, if they have a suppressed relic density, as occurs between the two
crossover strips for the different signs of $\mu$.

\subsection{Summary}

\begin{figure}[htb]
\begin{wide}{-1in}{-1in}
\begin{center}
\vskip -1.4in
\hskip .6in
\resizebox{0.55\textwidth}{!}{\includegraphics{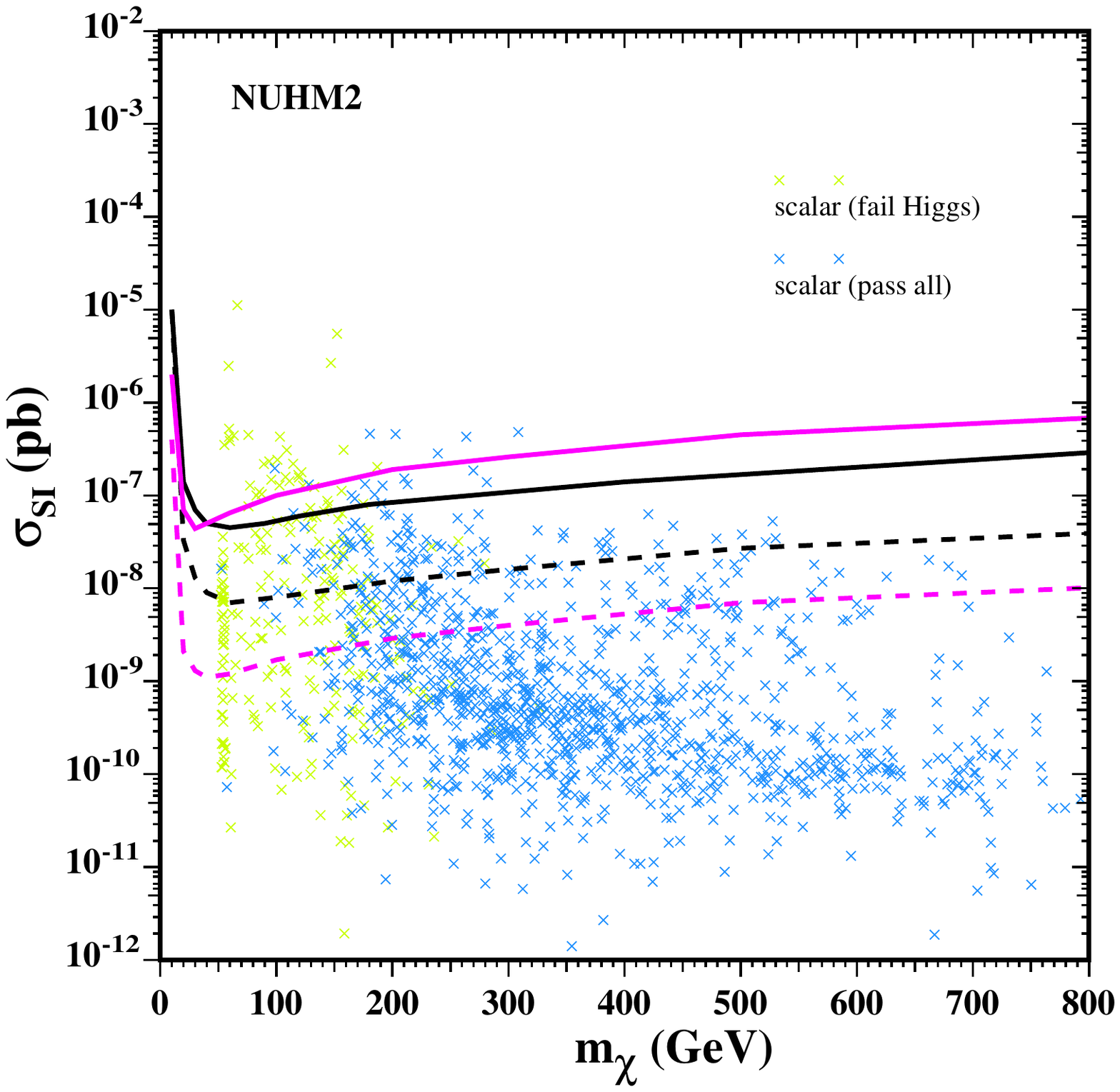}}
\hskip -.6in
\resizebox{0.55\textwidth}{!}{\includegraphics{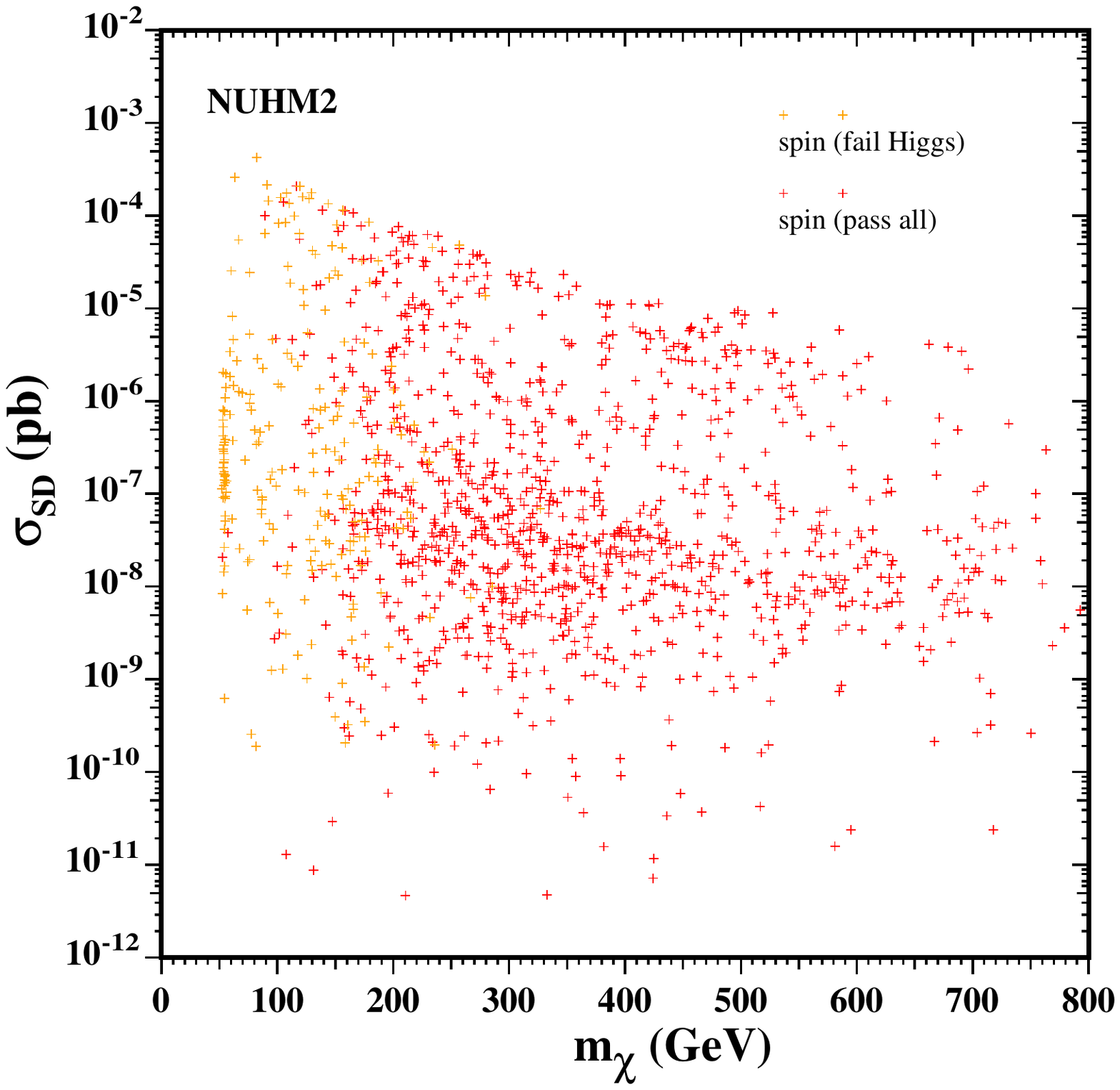}}
\end{center}
\end{wide}
\caption{\it Panels (a) and (b) show the entire potential range of neutralino-nucleon cross sections 
as functions of neutralino mass for the NUHM2 for the scalar and spin dependent cross sections
respectively. In both scans, we scan $5 \leq \tanb \leq 55$, 0 $\leq m_{1/2} \leq 2000$ GeV, 
100 GeV $\leq m_0 \leq 2000$ GeV, $-3 m_{1/2} \leq A_0 \leq 3 m_{1/2}$, and the GUT-scale 
values of $m_1$ and $m_2$ are each in the range $(-2000,2000)$ GeV. 
\label{fig:NUHM2summary}}
\end{figure}

Fig.~\ref{fig:NUHM2summary} displays the potential ranges
of (a) the spin-independent and (b) the spin-dependent dark matter scattering rates in
the NUHM2. Comparing with the corresponding plots for the CMSSM in
Fig.~\ref{fig:cmssm}, we note that the spin-independent cross section in the NUHM1 may be
up to an order of magnitude larger for $m_\chi > 300$~GeV. As discussed in the
previous subsections, the neutralino LSPs at these points typically have large Higgsino
components, despite their large masses, and lie along crossover strips. This feature is
common with the corresponding scatter plots for the NUHM1 shown in 
Fig.~\ref{fig:NUHM1summary}. We also note the
appearance of NUHM2 points with $m_\chi < 200$~GeV and low spin-independent
cross sections $\sim 10^{-10}$~pb. These points are typically in regions between the
crossover strips, in regions with a very suppressed relic density.
Similar features are present for the spin-dependent
cross sections: in the NUHM2 this may even be $\sim 10^{-5}$~pb for $m_\chi > 500$~GeV,
whereas values in the CMSSM are over an order of magnitude lower. Also, the NUHM2
allows the possibility of much lower spin-dependent effective cross sections for 
$m_\chi < 300$~GeV than are attained in the CMSSM. These features are again similar to
those found in the NUHM1.

\section{Conclusions}

We have explored in this paper the possible ranges of dark matter scattering
rates in the NUHM1 and NUHM2, and compared them with the ranges
attainable in the CMSSM. We have seen that the ranges that could be found in
the NUHM1 and NUHM2 are both significantly broader than is possible in the
CMSSM. 
In addition to the expectations for direct detection from broad scans,
we have displayed calculated cross sections in specific slices of the NUHM1,2
parameter spaces to highlight the physical processes in $\chi-p$ elastic scattering.
Larger cross sections are possible at large $m_\chi$, and smaller 
cross sections are possible at small $m_\chi$. The high-mass points with the
largest cross sections occur typically for points in the crossover regions with
relatively large Higgsino components. Mixed Bino-Higgsino states with the
a relic density in the WMAP range may occur at much larger masses than
those found in the focus-point region of the CMSSM. Several mechanisms
yield points with small effective cross sections. Some points have a very
suppressed relic density, e.g., points with even larger Higgsino components
that are located between crossover strips. In other cases, the cross section
may be suppressed because the point is in a rapid-annihilation
funnel region at larger $m_0$ and smaller $m_{1/2}$ than is 
possible in the CMSSM, or in a crossover region at larger $m_0$, or
at larger $m_A$ or $\mu$.

Present direct dark matter searches for spin-independent scattering, 
in particular the XENON10 and CDMS~II experiments,
are beginning to chip away at the parameter spaces of the CMSSM, NUHM1 and
NUHM2. However, most of the points excluded so far have small 
$m_\chi$ and $m_h < 114$~GeV.
On the other hand, prospective searches for spin-independent scattering
could be sensitive to NUHM1 or NUHM2 models with LSP masses as large as
800~GeV, the largest we have sampled, whereas they would be sensitive only to
$m_\chi < 350$~GeV within the CMSSM~\footnote{Unfortunately,
searches for spin-dependent dark matter scattering are still some distance away
achieving the sensitivity needed to probe the classes of models discussed in
this paper.}. Thus, observation of spin-independent
scattering by a heavy LSP would be a good diagnostic for a more complicated
supersymmetric model than the CMSSM, and perhaps a hint for a neutralino with
a significant Higgsino component in one of the crossover regions. Conversely,
if some other experiment, e.g., at the LHC, establishes the existence of a light
LSP, but its spin-independent scattering is {\it not} seen, this could also be a hint
for some model more complicated than the CMSSM in which the relic density
is suppressed, e.g., by one of the mechanisms described in the previous
paragraph that operate in the NUHM1 and NUHM2.

Neither the LHC experiments nor direct dark matter searches can, by themselves,
tell us all we would like to know about the manner in which supersymmetry is
broken. However, as we have illustrated in this paper, the combination of
LHC experiments and direct dark matter searches could provide some
interesting hints.

\section{Acknowledgements}
The work of K.A.O. is supported in part by DOE grant DE-FG02-94ER-40823 at the 
University of Minnesota.
The work of P.S. is supported in part by the National Science Foundation under Grant
No. PHY-0455649.

\end{document}